\documentclass[showpacs,aps,superscriptaddress,letterpaper,nofootinbib]{revtex4}

\usepackage{graphicx}
\usepackage{epsfig}
\usepackage{float}
\usepackage{wasysym}

\newcommand{\be}{\begin{equation}}
\newcommand{\ee}{\end{equation}}
\newcommand{\ba}{\begin{eqnarray}}
\newcommand{\ea}{\end{eqnarray}}

\def\sss{\scriptscriptstyle}

\begin{document}

\begin{flushright}
\vbox{
\begin{tabular}{l}
{ SNOW13-00159, FERMILAB-PUB-13-386-PPD, arXiv:1309.4819 [hep-ph]}
\end{tabular}
}
\end{flushright}

\vspace{0.6cm}

\title{Constraining anomalous $HVV$ interactions at proton and lepton colliders}%
\author{Ian Anderson\thanks{e-mail:  ianderso@pha.jhu.edu}}
\affiliation{Department of Physics and  Astronomy, Johns Hopkins University, Baltimore, MD 21218, USA}
\author{Sara Bolognesi \thanks{e-mail:  sbologne@pha.jhu.edu}}
\affiliation{Department of Physics and  Astronomy, Johns Hopkins University, Baltimore, MD 21218, USA}
\author{Fabrizio Caola  \thanks{e-mail:  caola@pha.jhu.edu}}
\affiliation{Department of Physics and  Astronomy, Johns Hopkins University, Baltimore, MD 21218, USA}
\author{Yanyan Gao \thanks{e-mail:  ygao@fnal.gov}}
\affiliation{Fermi National Accelerator Laboratory (FNAL), Batavia, IL 60510, USA}
\author{Andrei V. Gritsan \thanks{e-mail:  gritsan@pha.jhu.edu}}
\affiliation{Department of Physics and  Astronomy, Johns Hopkins University, Baltimore, MD 21218, USA}
\author{Christopher B. Martin \thanks{e-mail:  cmartin@pha.jhu.edu}}
\affiliation{Department of Physics and  Astronomy, Johns Hopkins University, Baltimore, MD 21218, USA}
\author{Kirill Melnikov \thanks{e-mail:  melnikov@pha.jhu.edu}}
\affiliation{Department of Physics and  Astronomy, Johns Hopkins University, Baltimore, MD 21218, USA}
\author{Markus Schulze \thanks{e-mail:  markus.schulze@anl.gov}}
\affiliation{Argonne National Laboratory (ANL), Lemont, IL 60439, USA}
\author{Nhan V. Tran  \thanks{e-mail:  ntran@fnal.gov}}
\affiliation{Fermi National Accelerator Laboratory (FNAL), Batavia, IL 60510, USA}
\author{Andrew Whitbeck  \thanks{e-mail:  whitbeck@pha.jhu.edu}}
\affiliation{Department of Physics and  Astronomy, Johns Hopkins University, Baltimore, MD 21218, USA}
\author{Yaofu Zhou  \thanks{e-mail:  yzhou49@pha.jhu.edu}}
\affiliation{Department of Physics and  Astronomy, Johns Hopkins University, Baltimore, MD 21218, USA}

\date{September 19, 2013}

\begin{abstract}
\vspace{2mm}
In this paper, we study the extent to which $C\!P$ parity  of a Higgs boson, and more generally
its anomalous couplings to gauge bosons,  can be measured at the LHC and 
a future electron-positron collider.
We consider several processes, including Higgs boson production 
in gluon and weak boson fusion and  production of a Higgs boson 
in association with an electroweak  gauge  boson. We consider decays of a Higgs boson 
including $ZZ, WW, \gamma \gamma$, and $Z \gamma$.
A matrix element approach to three production and decay topologies is developed and applied in the analysis. 
A complete Monte Carlo simulation of the above processes at  proton and $e^+e^-$ colliders 
is performed and verified by comparing it to an analytic calculation.
Prospects for  measuring various tensor couplings at
existing and proposed facilities are  compared.
\end{abstract}


\maketitle

\thispagestyle{empty}


\section{Introduction}
\label{sect:intro}

The existence of a Higgs boson with the mass around 125 GeV has now been 
firmly established by the ATLAS and CMS
experiments at the Large Hadron Collider~\cite{discovery-atlas, discovery-cms}
with supporting evidence from the Tevatron experiments~\cite{evidence-tev}.
However, detailed understanding of the properties of this particle will require 
an array of precision measurements  of Higgs boson production and decay processes. 
The purpose of this paper is to
present a coherent framework for studying  anomalous couplings  of a Higgs boson in processes 
which involve its interactions with weak vector bosons, photons, and gluons. 
We develop tools for measuring  the anomalous couplings and compare the expected sensitivity in 
different modes at existing and planned experimental facilities. 

Several facts about Higgs boson spin, parity, and its couplings have already been established. 
The new boson cannot have spin one because it decays to two on-shell photons \cite{landau}.
The spin-one assignment is also strongly disfavored by the measurement of angular distributions 
in  $H \to ZZ$ decays~\cite{properties-cms, properties-atlas}.
Under the assumption of minimal coupling to vector bosons or fermions, 
the new boson is unlikely to be a spin-two particle~\cite{properties-cms, properties-atlas}.
The spin-zero, negative parity hypothesis is also 
strongly disfavored~\cite{properties-cms, properties-atlas}.  Therefore, the new particle 
appears to be predominantly a $J^{\,C\!P} = 0^{++}$ state whose couplings to gauge bosons 
may, however, have small anomalous components. Constraining and possibly measuring these anomalous couplings 
will require  an extensive experimental program.

\begin{figure}[t]
\centerline{
\setlength{\epsfxsize}{0.33\linewidth}\leavevmode\epsfbox{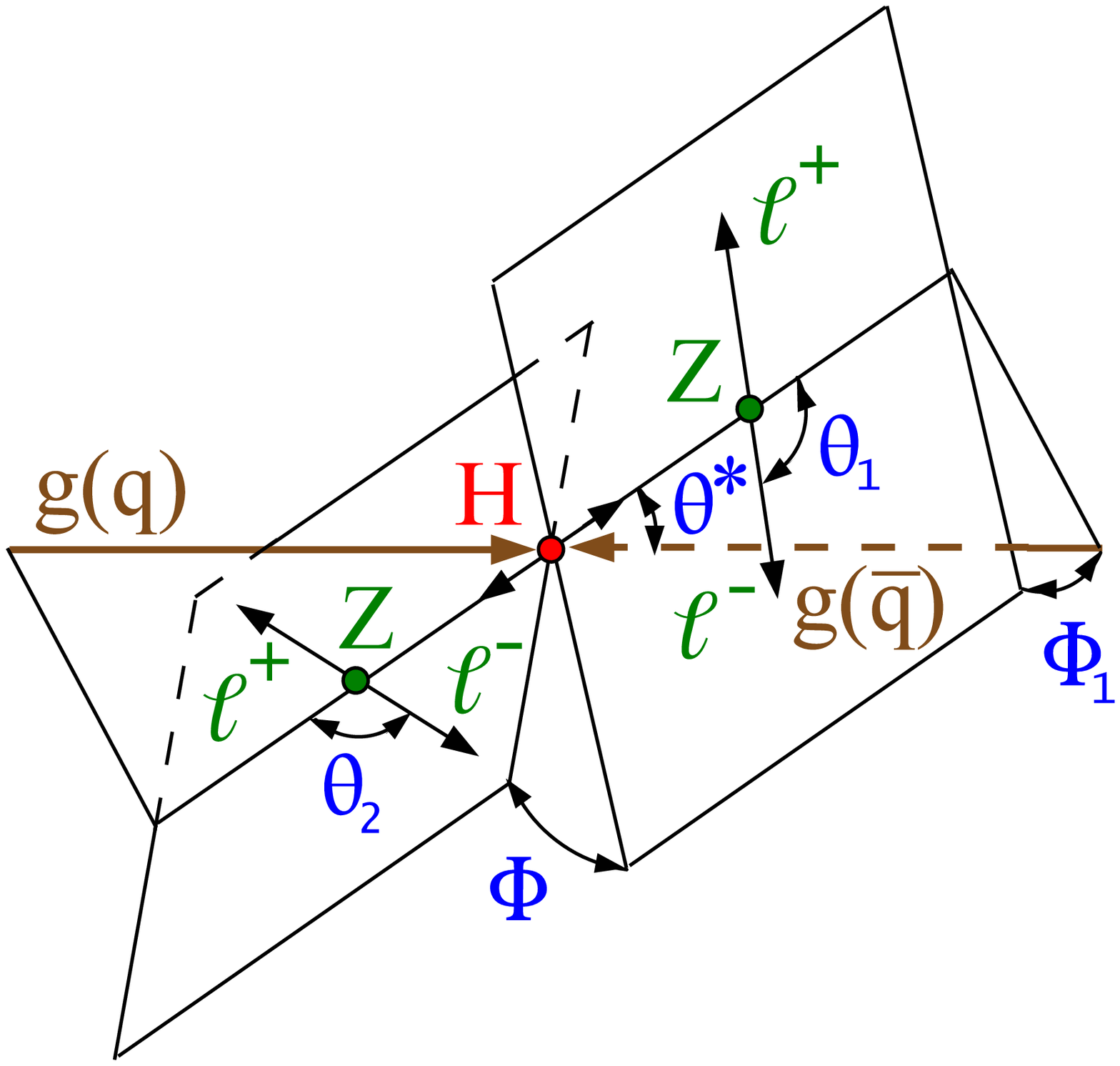}
\setlength{\epsfxsize}{0.33\linewidth}\leavevmode\epsfbox{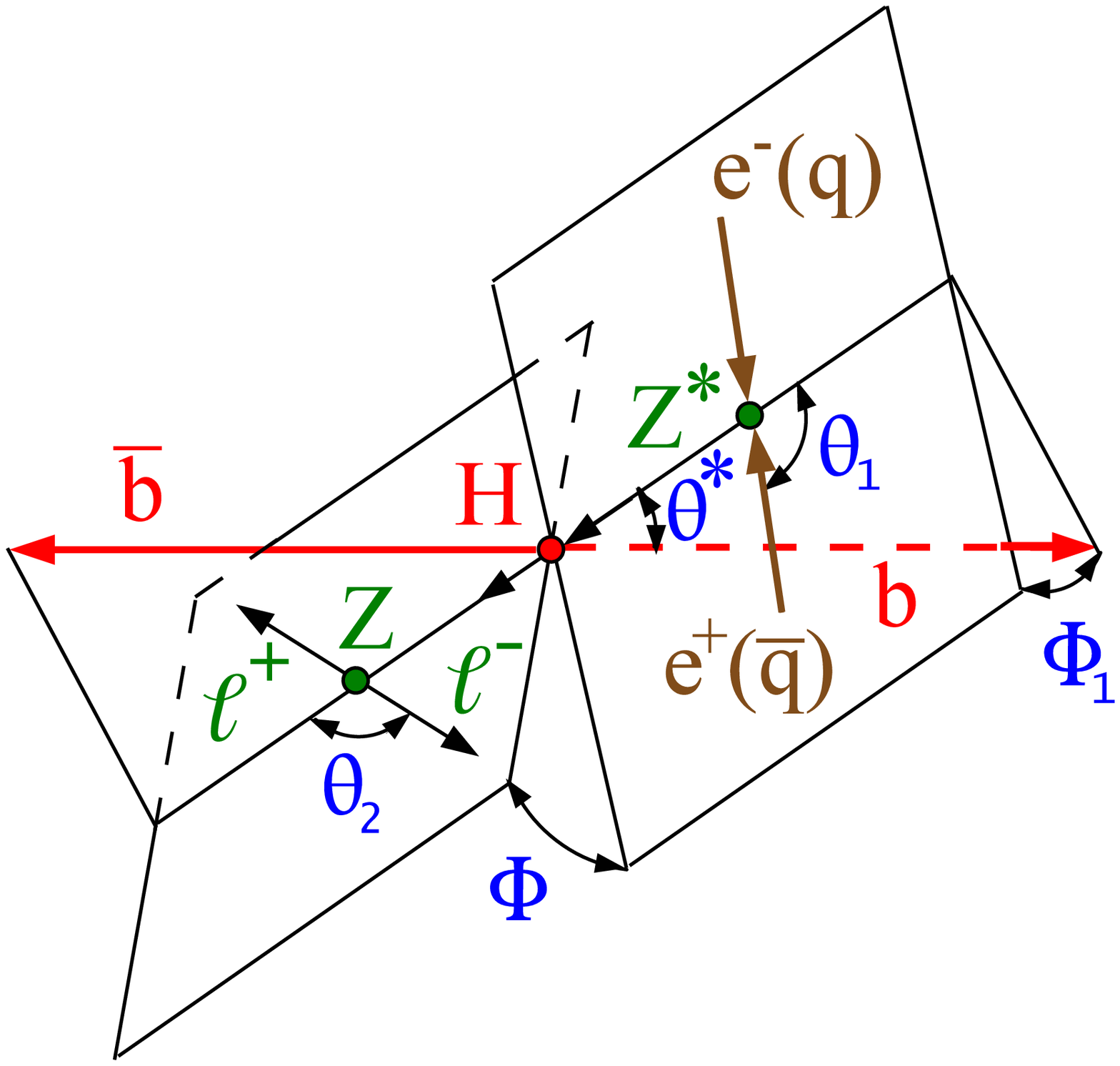}
\setlength{\epsfxsize}{0.33\linewidth}\leavevmode\epsfbox{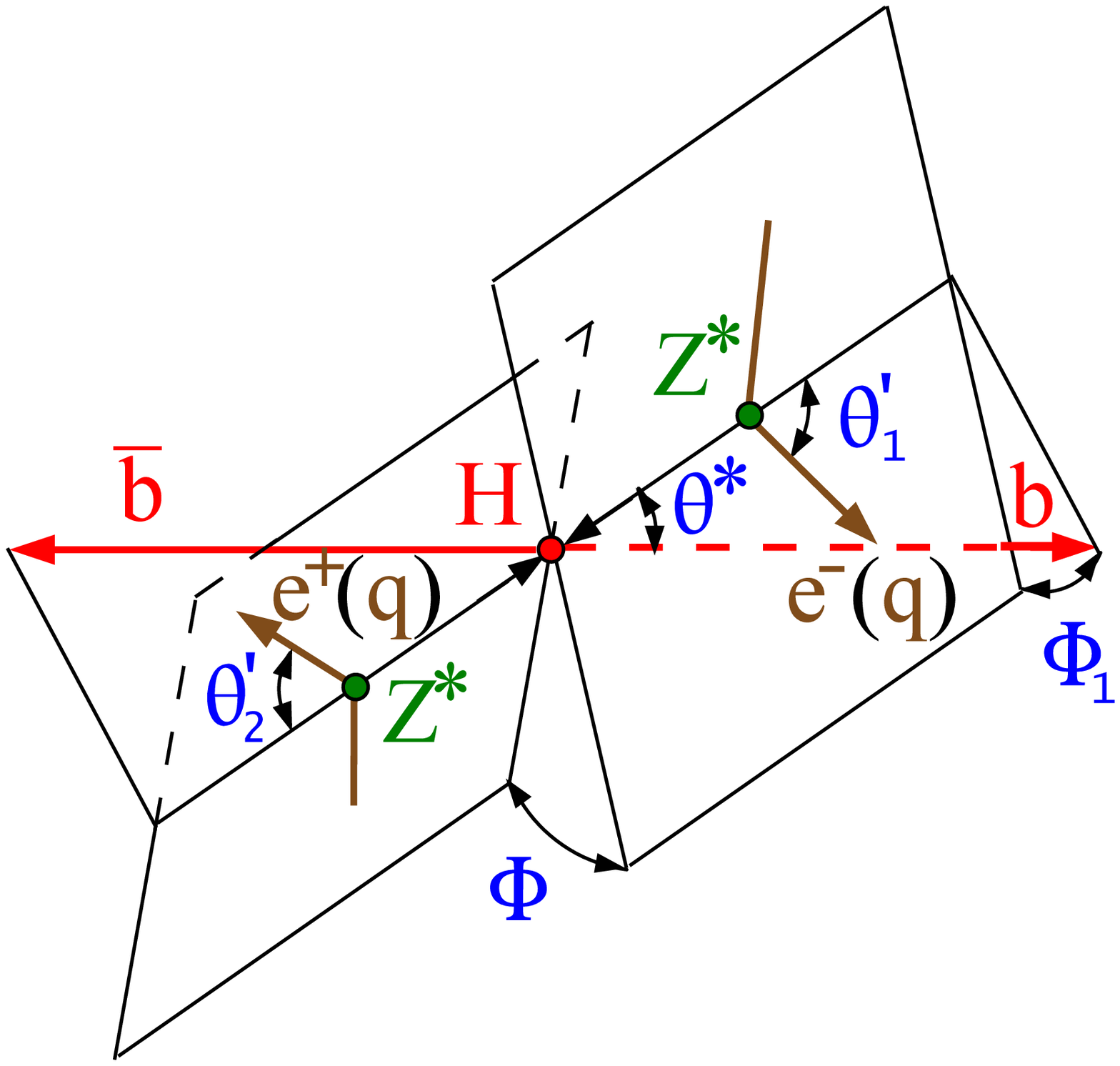}
}
\caption{
Illustrations of $H$ particle production and decay in $pp$ or $e^+e^-$  collision $gg/q\bar{q}\to H\to ZZ\to 4\ell^\pm$ (left),
$e^+e^-(q\bar{q})\to Z^*\to ZH\to\ell^+\ell^-b\bar{b}$ (middle), or 
$e^+e^-(q{q^\prime})\to e^+e^-(q{q^\prime})H \to e^+e^-(q{q^\prime}) b\bar{b}$ (right).
The $H\to b\bar{b}$ decay and $HZZ$ coupling are shown as examples, so that  $Z$ can be substituted by other vector bosons.
Five angles fully characterize the orientation of the production and decay chain and are defined 
in the suitable rest frames.  
}
\label{fig:decay}
\vspace{0.4cm}
\centerline{
  \setlength{\epsfxsize}{0.18\linewidth}\leavevmode\epsfbox{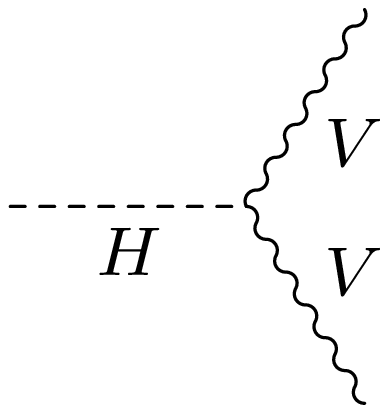}
~~~~~~~~~~~~~~~~~~~
\setlength{\epsfxsize}{0.18\linewidth}\leavevmode\epsfbox{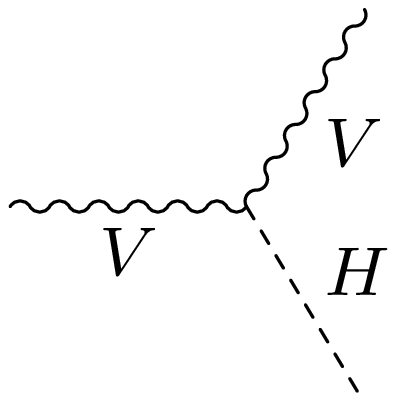}
~~~~~~~~~~~~~~~~~~~~~~~~~~~
\setlength{\epsfxsize}{0.18\linewidth}\leavevmode\epsfbox{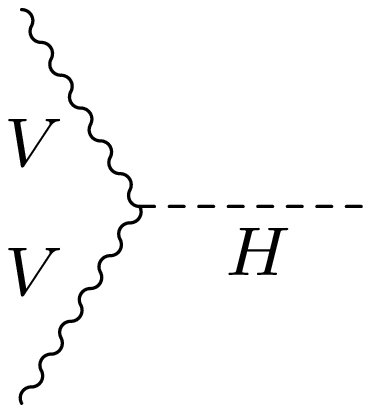}
}
\caption{
Illustration of an effective $HVV$ coupling, where $V=Z,W,\gamma,g$ with
$H$ decay to two vector bosons (left),
associated $H$ production with a vector boson (middle),
and vector boson fusion (right).
}
\label{fig:decay-feyn}
\end{figure}

The basic idea behind any spin-parity measurement is that different spin-parity assignments restrict the allowed 
types of interactions between the Higgs boson and other particles. This feature manifests itself in various kinematic
distributions of either the decay products of the Higgs particle or particles produced in association with it. 
There are three processes that can be used to determine the Lorentz structure of the  $HVV$ interaction vertex,
where $V$ stands for a vector boson $Z, W, \gamma, g$, cf.  Figs.~\ref{fig:decay},~\ref{fig:decay-feyn}. They are
\begin{itemize} 
\item production of a Higgs boson (in any process) followed by its decay to two vector bosons followed by a decay to fermions,
such as $H\to ZZ, WW\to 4f$,  $H\to Z\gamma\to 2f\gamma$, see left panels in Figs.~\ref{fig:decay},~\ref{fig:decay-feyn},
where definition of kinematic observables through the particle momenta can be found in Refs.~\cite{Gao:2010qx,Bolognesi:2012mm};
\item production of $Z^*(W^*)$ followed by its decay into $Z$ or $W$ and  a Higgs boson.
The Higgs boson then decays into any final state, see middle panels in  Figs.~\ref{fig:decay},~\ref{fig:decay-feyn};
\item production of a Higgs boson in association with two jets in weak boson fusion or gluon fusion,
followed by the Higgs boson decay into any final state, see right panels in Figs.~\ref{fig:decay},~\ref{fig:decay-feyn}.
\end{itemize}
Many of these processes were already studied from the point of view of 
spin-parity determination~\cite{Gao:2010qx,Bolognesi:2012mm,Accomando:2006ga,Heinemeyer:2013tqa,
Barger:1993wt,Mahlon:2006zc,DelDuca:2006hk,Hagiwara:2009wt,
Artoisenet:2013puc, sp1,sp2,Boughezal:2012tz,sp3,sp4,sp5,sp6,sp7,sp8,sp9,sp10,sp11,sp12,sp13,sp14,sp15,sp16,
Gainer:2011aa, Chen:2012jy, Gainer:2013rxa,Sun:2013yra}. 
The goal of this paper is to combine all these studies into a single framework and estimate the ultimate sensitivity 
to anomalous couplings that can be reached at the LHC and future lepton colliders. 

We build upon our previous analysis of this problem described in Refs.~\cite{Gao:2010qx,Bolognesi:2012mm}.
Techniques developed there are well-suited for measuring  $HVV$ anomalous couplings since these  couplings affect 
angular and mass distributions and can  be constrained  by fitting observed distributions to theory predictions.  
However,  such multi-parameter fits require large samples of signal events that  are currently not available.   
Nevertheless, it is interesting to study  the ultimate precision on anomalous 
couplings that can be achieved at the LHC and a future lepton  collider since the 
expected number of events can be easily estimated.

We organize the rest of the paper  as follows. In Sec.~\ref{sect:ampl}  we briefly review parameterization of the $HVV$ vertex.  
In Sec.~\ref{sect:mc} we discuss Monte Carlo (MC) and likelihood techniques, since they  provide the necessary tools for the experimental studies.  
In Sec.~\ref{sect:measure} we explore various approaches to anomalous couplings measurements
and summarize the precision that is achievable at different facilities.  We conclude in Sec.~\ref{sect:summary}. 
Additional details, including discussion of the matrix element method and methodology of the analysis, can be found in Appendices.


\section{Parametrization of the scattering amplitudes}
\label{sect:ampl}

Studies of spin, parity, and couplings of a Higgs boson employ generic parameterizations of scattering amplitudes. 
Such parameterizations contain all possible tensor structures consistent with assumed symmetries and Lorentz invariance.  
We  follow the notation of Refs.~\cite{Gao:2010qx,Bolognesi:2012mm}
and write the general scattering amplitude that describes  interactions of a spin-zero boson 
with the gauge bosons, such as $ZZ$, $WW$, $Z\gamma$, $\gamma \gamma$, or $gg$
\begin{equation}
A(X_{J=0} \to VV) = \frac{1}{v} \left ( 
  g^{}_{ 1} m_{\sss V}^2 \epsilon_1^* \epsilon_2^* 
+ g^{}_{ 2} f_{\mu \nu}^{*(1)}f^{*(2),\mu \nu}
+ g^{}_{ 4}  f^{*(1)}_{\mu \nu} {\tilde f}^{*(2),\mu  \nu}
\right )\, .
\label{eq:fullampl-spin0} 
\end{equation}
In Eq.~(\ref{eq:fullampl-spin0}),  $f^{(i),{\mu \nu}} = \epsilon_i^{\mu}q_i^{\nu} - \epsilon_{i}^\nu q_i^{\mu} $ 
is the field strength tensor of a gauge boson 
with momentum $q_i$ and polarization vector $\epsilon_i$;
${\tilde f}^{(i),\mu \nu} = 1/2 \epsilon^{\mu \nu \alpha \beta} f_{\alpha \beta}$  
is the  conjugate field strength tensor.
Parity-conserving interactions of a scalar (pseudo-scalar)  are 
parameterized by the couplings $g^{}_{1,2}$($g_4$), respectively. 
In the Standard Model (SM), the only non-vanishing coupling of the Higgs  to $ZZ$ or $WW$ bosons
at tree-level is $g_1=2i$, while $g^{}_{ 2}$   is generated through radiative corrections. 
For final states with at least one massless gauge boson, such as $\gamma \gamma$, $gg$ or 
$Z\gamma$, the SM interactions with the Higgs boson are loop-induced; 
these interactions are described by the coupling $g_2$.  

In Refs.~\cite{Gao:2010qx,Bolognesi:2012mm} it was shown that an additional $g_{3}$ term in Eq.~(\ref{eq:fullampl-spin0})
can be absorbed into the ``constant'' $g_2$ if the coupling constants in Eq.~(\ref{eq:fullampl-spin0})  are treated
as momentum-dependent form factors.  This is a general feature and we illustrate it with examples shown below.
Consider the following addition to the amplitude\footnote{
A ``derivative operator''  introduced in Ref.~\cite{Artoisenet:2013puc}  is equivalent to the $g_{35}$ and $g_{36}$ terms
in Eq.~(\ref{eq:fullampl-spin0-extra}).}
$A(X_{J=0} \to VV)$
\begin{eqnarray}
\frac{1}{v \, \Lambda^2} f^{*(1),\mu \nu} f^{*(2)}_{\mu \alpha} \left (
g_{ 3} q_{2\nu} q_1^{\alpha}
+ g_{ 32} q_{1\nu} q_1^{\alpha}
+ g_{ 33} q_{2\nu} q_2^{\alpha}
+ g_{ 34} q_{1\nu} q_2^{\alpha}
\right )
+  \frac{1}{v} \left(
g_{ 35} f^{*(1),\mu \nu}  q_{1\mu} \epsilon_{2\nu}^*
+g_{ 36} f^{*(2),\mu \nu}  q_{2\mu} \epsilon_{1\nu}^*
\right )\,,
\label{eq:fullampl-spin0-extra}
\end{eqnarray}
where for identical vector bosons $g_{32} = g_{33}$ and $g_{35} = g_{36}$.
Using the definition of the field strength tensor and $\epsilon_i \cdot q_i = 0$,
we find that all terms in Eq.~(\ref{eq:fullampl-spin0-extra}) can be desribed by Lorentz structures in
Eq.~(\ref{eq:fullampl-spin0}) provided that $g_{1}$ and $g_{2}$ are modified as
\begin{eqnarray}
&& g_{1} \rightarrow
\left(
g_{1} - g_{ 35} \frac{m_1^2}{m_{\sss V}^2} - g_{ 36} \frac{m_2^2}{m_{\sss V}^2}  + g_{ 34} \frac{m_1^2m_2^2}{m_{\sss V}^2\Lambda^2}
\right)\,,~~~~~~
g_{2} \rightarrow
\left(
g_{2} + g_{3} \frac{m_{\sss X}^2-m_1^2-m_2^2}{4\Lambda^2} + g_{ 32} \frac{m_1^2}{2\Lambda^2}+ g_{ 33} \frac{m_2^2}{2\Lambda^2}
\right)\,.
\label{eq-equaiv}
\end{eqnarray}

In this paper, we  focus on the determination of anomalous couplings of the predominantly 
$J^{\,C\!P}=0^{++}$ Higgs-like boson to SM gauge bosons since existing experimental data 
already disfavors other exotic spin-parity assignments~\cite{properties-cms, properties-atlas}.  
For $HZZ$ or $HWW$ vertices, we therefore assume that the coupling constants satisfy a hierarchical 
relation $g_1 \gg g_{2,4}$  and that non-standard couplings {\it always} provide small modifications of the SM contributions. 

It is convenient to express the results of the measurement of the anomalous couplings 
in terms of physical quantities. To this end, we consider three independent, and generally complex,
couplings $g^{}_{ 1}$, $g^{}_{ 2}$, and $g^{}_{ 4}$  for each of the vector bosons $Z,\gamma,W,g$.
Assuming no $q^2$-dependence,
five independent numbers are needed to parameterize the couplings since one overall complex phase is not measurable.  
We take one of these numbers to be the $H \to VV$ decay rate; the remaining four real numbers parameterize ratios of couplings 
and their relative phases.  We find it convenient to use effective fractions of events defined as
%
\begin{eqnarray}
&& f_{gi} =  \frac{|g^{}_{i}|^2\sigma_i}{|g^{}_{1}|^2\sigma_1+|g^{}_{2}|^2\sigma_2+|g^{}_{4}|^2\sigma_4},
~~~~~~~~~
\label{eq:fractions}
\end{eqnarray}
%
to parameterize coupling ratios. The phases are defined as  $\phi_{gi} = \arg\left(g_i/g_1\right)$.
For real couplings, $\phi_{gi} = 0$ or $\pi$. Complex couplings may appear if light particles contribute
to the loops, as very small anomalous complex couplings in fact may appear in the Standard Model. 
Even under assumption of real constant couplings, as in an Effective Lagrangian framework, 
it is of interest to test consistency of the model by relaxing both real and momentum-independent
requirements on the couplings. 

We note that $\sigma_i$ in Eq.~(\ref{eq:fractions}) 
is the cross section for the process $H \to VV$, $V^*\to VH$, or  $V^*V^*\to H$
that corresponds to $g^{}_{ i}=1, g_{j \ne i}=0$. The advantage of introducing fractions $f_{gi}$ 
is that, for fixed tensorial structure of the $HVV$ vertex, they  are invariant under independent re-scalings 
of all couplings. They may also be 
interpreted  as  fractions of event yields corresponding to each anomalous coupling independently.
Contributions that originate from interferences of different amplitudes can be 
described using  parameterization introduced above; for this, both fractions $f_{gi}$ and phases $\phi_{gi}$ are required. 
Once fractions $f_{gi}$ are  measured, one can extract the coupling constants in a straightforward way by inverting Eq.~(\ref{eq:fractions}),
e.g. $|g_i/g_1|=  ( f_{gi}/(1-\sum_k f_{gk}) )^{1/2}\times ({\sigma_1}/{\sigma_i})^{1/2}$.
The parameter $f_{g4}$ is equivalent to the parameter $f_{a3}$ as introduced by the CMS collaboration~\cite{properties-cms}
under the assumption $g_2=0$; it is the fraction of a $C\!P$-odd contribution to the 
total production cross section of a Higgs boson.  For the ease of comparison with earlier CMS studies, we will use 
$f_{a2}$ and $f_{a3}$ instead of $f_{g2}$ and $f_{g4}$, respectively, to denote event fractions throughout the paper.  
The $f_{a2}^{\rm dec}$ and $f_{a3}^{\rm dec}$ values correspond to cross sections defined in decay $H \to VV$. 

The above discussion is well-suited in case when  effective couplings can be treated 
as  $q^2$-independent  constants.   However, it may also be desirable to treat these 
couplings as functions of invariant masses of gauge bosons $q_{1,2}^2$ and we show how to do this 
in the next section.  However, we do not pursue such a general analysis 
in this paper. Instead, we focus  on the lowest-order modification to $HVV$ interaction 
vertex caused by each of the anomalous couplings. 
This means that we treat $g_2$ and $g_4$ as $q^2$-independent constants but 
account for  $q^2$-dependent correction to $g_1$.  The parametrization of this correction 
is described in detail below; here, we just mention that the new contribution 
is treated as yet another anomalous coupling to which the construction of effecitve 
event fractions Eq.~(\ref{eq:fractions}) is applied.

\bigskip


\section{Analysis tools}
\label{sect:mc}

Analyses reported in this paper require a simulation program to describe production of resonances 
in hadron-hadron or $e^+e^-$ collisions, followed by their subsequent decays.
Anomalous couplings to vector bosons must be included. 
The simulation program is supplemented by both analytical and numerical calculations of the
likelihood distributions based on the matrix element method. 
These analysis tools are described in this section. Additional details can be found in Appendices.
 
Events are simulated with the {\sf JHU} generator~\cite{Gao:2010qx,Bolognesi:2012mm,support},
a dedicated  Monte Carlo program, that features implementations  of the processes 
$gg / q\bar q \to X\to ZZ (WW) \to 4f $ as well as  $gg / q\bar q\to X\to \gamma\gamma$.  
The {\sf JHU}  generator incorporates all spin correlations, interference of all contributing amplitudes, 
and the general couplings of the $X$ particle to gluons and quarks in production and to vector bosons in decay.  
New features of the {\sf JHU} generator, implemented since the last release, are summarized below. 

The {\sf JHU} generator has been extended to include new processes:
associated production of a Higgs boson in either proton or electron collisions
$q\bar{q}^\prime\to V^*\to VH$, $e^+e^- \to Z^*\to ZH$, and associated
production with two jets from either gluon fusion $gg \to H+2$ jets
or weak boson fusion $q{q^\prime}\to q{q^\prime} V^*V^* \to H q{q^\prime} $,  where $V=Z, W$. 
In all cases, parameterization of the $HVV$ vertex with all anomalous couplings as in Eq.~(\ref{eq:fullampl-spin0}) is included.  
Extension to other spin assignments of an exotic boson following formalism in Refs.~\cite{Gao:2010qx,Bolognesi:2012mm} 
is also available for some of these processes, but it is not the focus of the study presented here. 
We also introduce the decay mode $H\to Z\gamma$.
In both $H\to ZZ$ and $Z\gamma$ decays we allow $Z^*/\gamma^*$ interference 
covering the intermediate states  $H\to Z^*Z^* / Z^*\gamma^* / \gamma^*\gamma^* \to 4f$ and 
$H\to Z^*\gamma/\gamma^* \gamma \to 2f\gamma$.

Another feature of the generator implemented recently concerns the 
dependence of the effective coupling constants $g_{1...4}$ on the virtualities of 
two vector bosons, cf. Eq.~(\ref{eq:fullampl-spin0}). 
To describe this effect, we parameterize the couplings as  
%
\begin{eqnarray}
&&  g_i(q_1^2, q_2^2) = g^{\rm SM}_i + g^\prime_i \times \frac{\Lambda_i^4}{(\Lambda_i^2 + |q_1^2|)\,(\Lambda_i^2 + |q_2^2|))} \,,
\label{eq:formfactors}
\end{eqnarray}
%
where $\Lambda_i$ is the energy  scale that is correlated with  
 masses of new, yet unobserved,  particles that  contribute to $HVV$ interaction vertex 
and $g^{\rm SM}_i=g_1\cdot \delta_{i1}$ appears at tree level in the coupling 
of a Higgs boson to  weak vector bosons in the Standard Model.  Although we do not use this 
feature of the generator in the current paper, we expect that it will be helpful for checking 
the sensitivity of various observables employed for spin-parity analysis to high-energy or high invariant-mass
tails of kinematic distributions that may be affected by poorly controlled form-factor effects. 
In case when form-factor scales $\Lambda_i$ are much higher than any of the kinematic invariants 
in the physics process of interest, the form-factors can be expanded into series 
of $q^2/\Lambda_i^2$, enabling a connection to the effective field theory approach to Higgs 
couplings determination.   The option to describe effective couplings as series in $q^2/\Lambda^2$ is 
available in JHU generator as an alternative to Eq.(\ref{eq:formfactors}).  We illustrate the usefulness 
of this feature by considering modifications of the $g_1$ coupling,
 $g_1\to (g_1 - g_1^{\prime\prime}\times(q_1^2+q_2^2)/\Lambda_1^2)$. 

The generator program can be interfaced to parton shower simulation as well as full detector simulation 
through the Les Houches Event (LHE) file format~\cite{Alwall:2006yp}.  
The {\sf JHU} generator now also allows interfacing the decay of a spin-zero particle with the production 
simulated by other MC programs, or by the {\sf JHU} generator itself, through the LHE file format.
This option allows us to combine modeling of the next-to-leading-order (NLO) QCD effects in the production of a 
$0^+$ particle with the description of its decays that includes   both anomalous couplings 
and interference effects of  identical fermions in the final state.

Apart from simulating events, our analysis requires the construction of various  likelihood functions 
to distinguish between different hypotheses about the Lorentz structure of the $HVV$ interaction vertex.  
As described in Appendix~\ref{sect:me}, the likelihood functions are obtained from kinematic probability 
distributions that can be either computed analytically or numerically. 
Analytical parameterizations are currently available for the $H\to VV$, $pp\to VH$, and $e^+e^-\to ZH$ processes,
see Appendix~\ref{sect:me} and Refs.~\cite{Gao:2010qx,Bolognesi:2012mm}. 
Numerical computations of matrix elements are provided by the {\sf JHU} generator.
These matrix elements are also needed to compute cross sections and kinematics distributions.
The matrix elements are implemented in the {\sf JHU} generator as separate functions~\cite{Bolognesi:2012mm} 
and can be accessed by an end-user directly. We provide the necessary codes to compute the likelihood 
functions using both analytic and numerical parameterizations of the matrix elements~\cite{support}.

\begin{figure}[t]
\centerline{
\setlength{\epsfxsize}{0.33\linewidth}\leavevmode\epsfbox{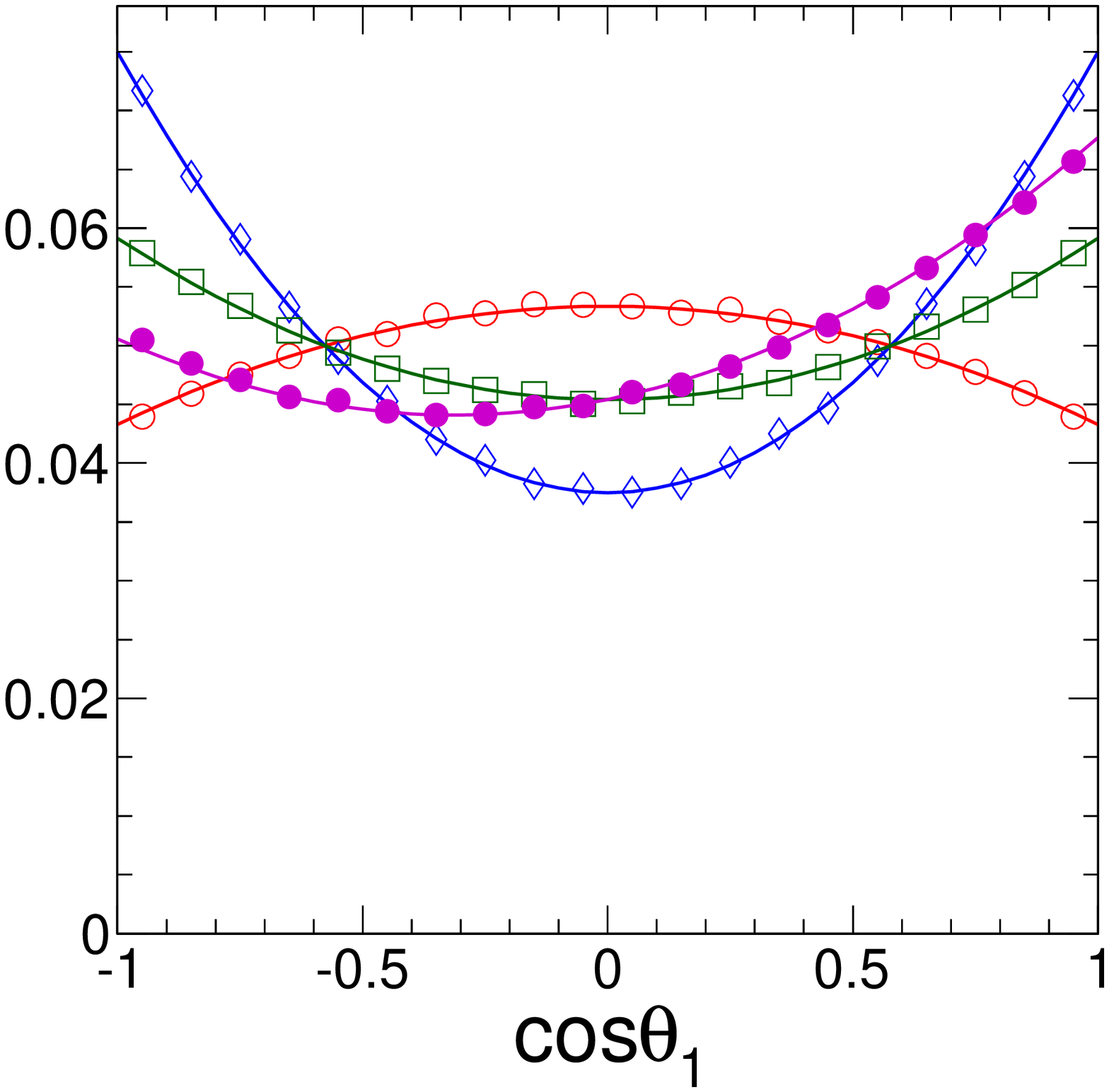}
\setlength{\epsfxsize}{0.33\linewidth}\leavevmode\epsfbox{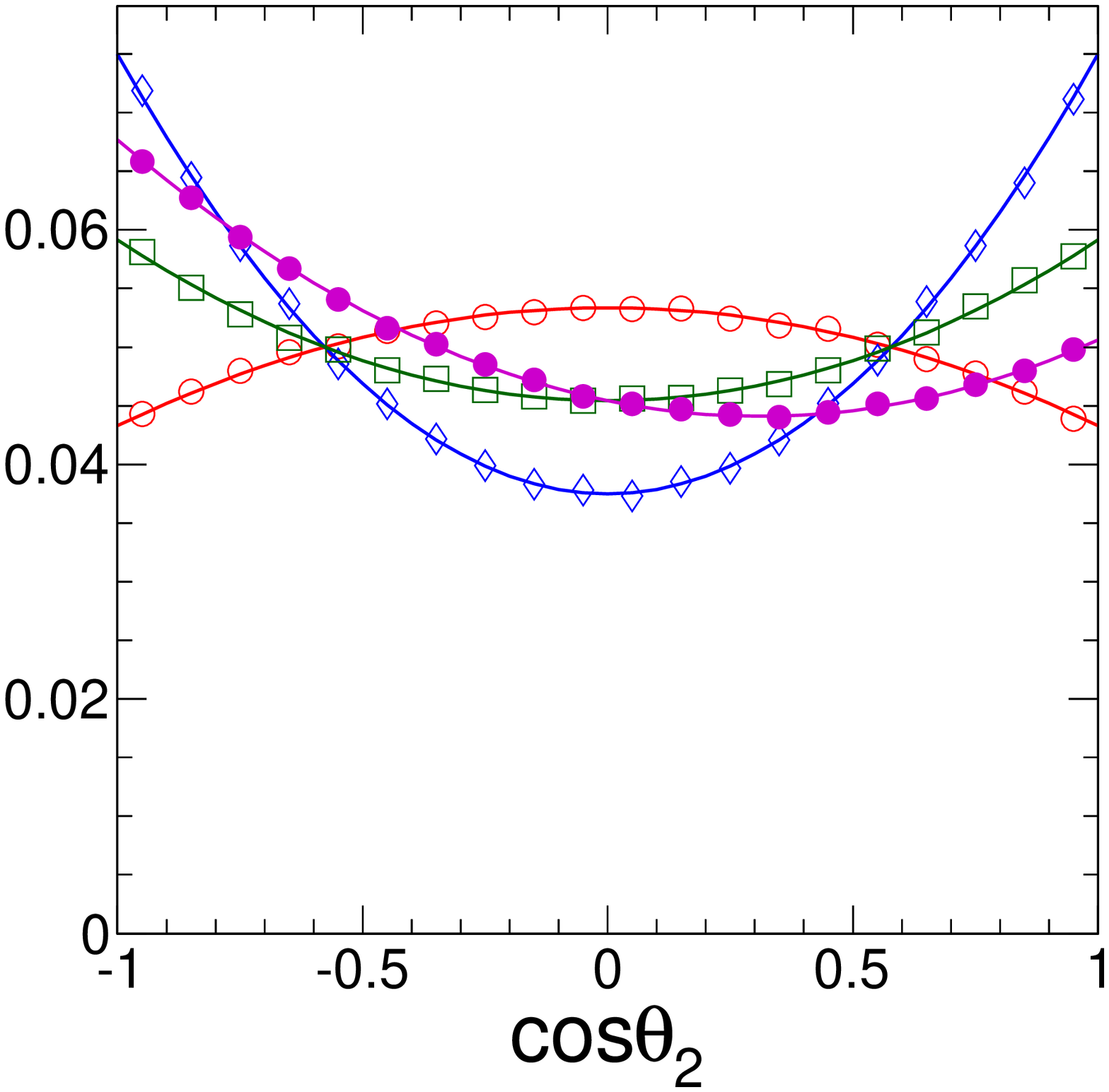}
\setlength{\epsfxsize}{0.33\linewidth}\leavevmode\epsfbox{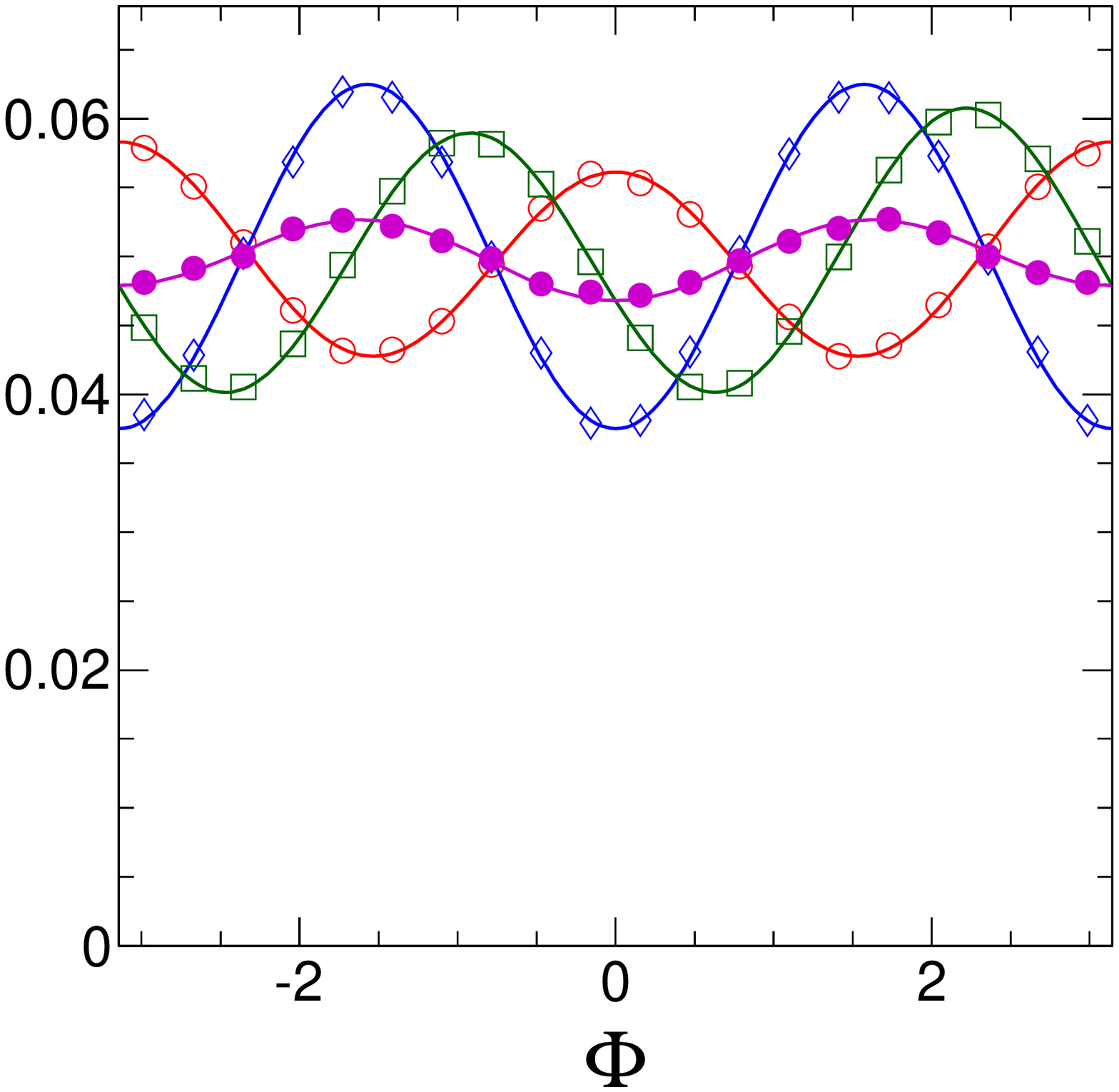}
}
\caption{
Distributions of the observables in the $e^+e^-\to ZH\to (\ell^+\ell^-)H$ analysis at $\sqrt{s}=250$ GeV, 
from left to right: $\cos\theta_1$, $\cos\theta_2$, and $\Phi$.
Points show simulated events and lines show projections of analytical distributions.
Four scenarios are shown: SM scalar ($0^+$, red open circles), pseudoscalar ($0^-$, blue diamonds), 
and two mixed states corresponding to $f_{a3}=0.5$ with $\phi_{a3}=0$ (green squares)
and $\pi/2$ (magenta points).  In all cases we choose $f_{a2} = 0$. 
}
\label{fig:ilc_angles}
\end{figure}

The availability of the two methods allows independent validation of the same analysis. 
Results presented in this paper employ analytic parameterization of the probabilities when
available. This allows analytic normalization of the probability distributions  to 
facilitate multi-dimensional and multi-parameter fits.  When analytic results are not available, 
 we use numerical computations of the matrix element squared. 

Examples of both analytical and generator distributions are shown in Fig.~\ref{fig:ilc_angles}
for the $e^+e^-\to ZH\to (\ell^+\ell^-)H$ process. More examples are available in 
Refs.~\cite{Gao:2010qx,Bolognesi:2012mm} for decay processes and in Appendix~\ref{sect:me}
for the production and decay processes at the  LHC. 
Examples of analyses based on the implementation of the matrix element techniques 
are given in Appendix~\ref{sect:statistics}.  We will use the distributions in Fig.~\ref{fig:ilc_angles}
to explain some results obtained in the next section. 


\section{Measurements of $HVV$  anomalous couplings}
\label{sect:measure}

In this section we describe prospects for measuring the anomalous $HVV$ couplings 
both at the LHC and at a future $e^+e^-$ collider. 
We consider all types of processes that allow such measurements, 
including gluon fusion at LHC (SBF), weak boson fusion (WBF), and $VH$ production.
For the analysis of the Higgs boson decay $H\to VV$, all production mechanisms can be combined. 
The cleanest and most significant SM Higgs boson decay mode at the LHC 
is  $H\to ZZ^*\to 4\ell$ and we consider this mode in the following analysis~\cite{properties-cms,properties-atlas}.
The decay $H\to WW^*\to 2\ell2\nu$ can also be used for anomalous coupling measurements, as demonstrated 
in Ref.~\cite{Bolognesi:2012mm}, but precision of spin-zero measurements is lower. 
Inclusion of other decay modes will only improve estimated precision and we examine
such examples as well ($H\to \gamma\gamma$ in VBF and $H\to b\bar{b}$ in $VH$ production).
At an $e^+e^-$ collider, we consider the dominant decay mode $H \to b \bar b$, 
but  other final states could be considered  as well.

\begin{figure}[t]
\centerline{
\setlength{\epsfxsize}{0.5\linewidth}\leavevmode\epsfbox{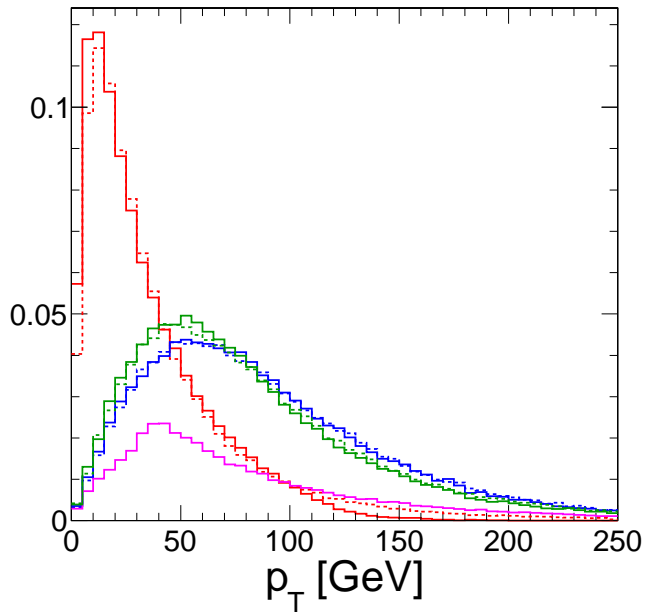}
}
\caption{
Comparison of transverse momentum $p_T$ distribution of a SM Higgs boson with 
$m_H=125$ GeV in MC simulation of 14~TeV $pp$ collisions at the LHC. 
Higgs production in the gluon fusion is generated by {\sf JHU} generator combined with {\sf Pythia} parton 
shower (solid red) and  by {\sf POWHEG} (dashed red) where NLO QCD approximation 
is matched to parton shower. 
The decay  $H\to ZZ\to 4\ell$ is simulated using the {\sf JHU} generator in both cases.
Also shown in the order of decreasing peak position:
$VH$ production (solid green), WBF production (solid blue),
and gluon fusion $H+2$ jets production (solid magenta) with the {\sf JHU} generator.
For $VH$ and WBF production, parton shower is included and comparison
with NLO QCD {\sf POWHEG} simulation (dashed distributions) is shown.
All distributions are normalized to unit area except for $H+2$ jets, which is normalized
with respect to inclusive gluon fusion production according to its relative cross section
with selection requirements on jets $p_T>15$~GeV and $\Delta R_{jj}>0.5$
as discussed in text.
}
\label{fig:pT}
\end{figure}

We now discuss details of event simulation and selection. 
In  this paper, signal events were simulated with the {\sf JHU} generator.
Background events were  generated with 
 {\sf POWHEG}~\cite{powheg}  ($q\bar{q}\to ZZ^{(*)} / Z\gamma^{(*)}$ + jets) and 
 {\sf MadGraph}~\cite{Alwall:2011uj} ($q\bar{q}\to ZZ^{(*)} / Z\gamma^{(*)} /\gamma\gamma$ + 0 or 2 jets, $e^+e^-\to ZZ$).
When backgrounds from other processes are expected, their effective contribution is included
by rescaling the expected event yields of the aforementioned processes. 
The vector boson fusion (VBF) and $VH$ topology of the SM Higgs boson production has been tested against 
{\sf POWHEG}, see Fig.~\ref{fig:pT}, as well as against 
{\sf VBF@NLO}~\cite{Arnold:2008rz,Arnold:2011wj,Arnold:2012xn} and {\sf MadGraph} simulation, respectively. 

To properly simulate recoil of the final state particles caused by QCD radiation, we interface 
the {\sf JHU} generator with  parton shower in {\sf Pythia}~\cite{pythia},
or, alternatively, simulate the decay of the Higgs boson with the {\sf JHU } generator 
and production of the Higgs boson through NLO QCD accuracy with 
 {\sf POWHEG}. We point out  that  this way of interfacing {\sf POWHEG} and {\sf JHU} generator 
is exact  for spin-zero particle production since no spin correlations connect initial and final 
states. 
We note that  quality of the approximation with {\sf Pythia} parton showering is surprisingly high 
as can be seen in Fig.~\ref{fig:pT} where   we  compare the transverse momentum distribution of 
a Standard Model  Higgs boson  obtained within this framework with the NLO QCD computation 
of the same distribution  as implemented in {\sf POWHEG}.
Effects of beyond-the-standard-model (BSM) couplings in gluon fusion production on recoil of the final state particles caused 
by the QCD radiation have been tested explicitly in the $pp\to H+2$ jets process; we found that 
their impact on recoil kinematics is negligible for the analysis of Higgs boson decays.
We conclude that parton shower description of QCD effects is sufficient at the current 
level of analysis but further refinements of such an approach, for example by means 
of dedicated NLO QCD computations, are certainly possible, see e.g.~Ref.~\cite{Artoisenet:2013puc}.

In this paper, we employ a simplified detector simulation similar to our earlier 
studies~\cite{Gao:2010qx,Bolognesi:2012mm}. Lepton momenta are smeared with an rms 
$\Delta p/p = 0.014$ for 90\% of events and a broader smearing for the remaining 10\%.
Hadronic jets are smeared with an rms $\Delta p/p = 0.1$. Events are selected in which leptons 
have $|\eta|<2.4$, and transverse momentum $p_T>5$\,GeV; jets, defined with anti-$k_\perp$ algorithm, 
 have $\Delta R_{jj}>0.5$,  $p_T >30$\,GeV, and $|\eta_j|<4.7$.
The jet $p_T$ threshold is raised to 50 GeV to study the effects of pileup when we consider the high luminosity LHC scenario.
The invariant mass of the di-lepton pairs from a $Z^{(*)}$ decay is required to exceed 12 GeV.
These selection criteria are  chosen to be as close as possible to existing 
LHC  analyses~\cite{properties-cms,properties-atlas}
and we assume that similar selection criteria will be also adopted for a future $e^+e^-$ collider.
The estimated number of reconstructed events in Table~\ref{table:XS} is scaled down from the number 
of produced events by 30\% and 80\% at $pp$ and $e^+e^-$ colliders, respectively. 
The $ZH$ channel at a $pp$ collider with $H\to b\bar{b}$ accounts for tighter selection requirements discussed in text. 

The expected statistical precision of the analysis depends on the number of Higgs bosons 
produced at each collider which is proportional to collider's integrated luminosity. 
To estimate the number of Higgs bosons expected at  the LHC and at a future $e^+e^-$ collider 
we note that each of the two LHC experiments will collect  300~fb$^{-1}$ of integrated luminosity at $pp$
collision energy of about 14 TeV. 
Beyond that, a high-luminosity upgrade is planned where 3000~fb$^{-1}$ per experiment 
are expected to be collected~\cite{ATLAS:2013hta,CMS:2013xfa,Dawson:2013bba}.
Among future facilities, an $e^+e^-$ collider operating at the center-of-mass energies of 250~GeV and above 
with either linear~\cite{Behnke:2013xla} or circular~\cite{Koratzinos:2013ncw} design 
could deliver a luminosity that ranges  from several hundred to several thousand fb$^{-1}$. 
At an $e^+e^-$ collider the $ZH$ production dominates  at lower energies 
while at  higher energies $WW$ or $ZZ$ fusion dominates.  
However, although  $e^+e^-\to\nu\bar\nu W^*W^*\to \nu\bar\nu H$ 
cross section exceeds the cross section for  $e^+e^-\to e^+e^-Z^*Z^*\to e^+e^-H$
by about an order of magnitude, no angular analysis  is possible in 
final states with neutrinos. 
The process $e^+e^-\to e^+e^-Z^*Z^*\to e^+e^-H$ would dominate over the $ZH$ production at high 
$e^+e^-$ energies, as evident from Table~\ref{table:XS}, but it does not provide enhanced sensitivity 
to anomalous couplings with increased $e^+e^-$ energy, as discussed below.

\begin{table}[t]
\begin{center}
\caption{
Summary of collider options considered for the production of a Higgs boson with the mass of 125 GeV.
Collider center-of-mass energy, integrated luminosity, cross sections for relevant  production modes and 
decay channels are shown.  Reconstructed efficiencies are estimated using selection criteria described in the text
and relate the number of produced and reconstructed events ($N_{\rm prod}$ and $N_{\rm reco}$). 
In several cases we also show fractions $f_{\rm jet}$ of events with two associated jets with 
$p_T>30$~GeV and $\Delta R_{jj}>0.5$.
}
\begin{tabular}{|lccccccccc|}
\hline\hline
collider & energy & $\int{\cal L}dt$ (fb$^{-1}$) & production &  $\sigma$ (fb)  & decay &  $\sigma\times{\cal B}$ (fb)  & $N_{\rm prod}$ & $N_{\rm reco}$  & $f_{\rm jet}$ \\
\hline\hline
$pp$ & 14 TeV & 3000 & $gg\to H$ & 49850 & $H\to ZZ^*\to 4\ell$ & $6.23$ & 18694 & 5608 & 0.1 \\ 
$pp$ & 14 TeV & 3000 & $V^*V^*\to H$           & 4180 & $H\to ZZ^*\to 4\ell$ & $0.52$ &  1568 &  470 & 0.6 \\ 
$pp$ & 14 TeV & 3000 & $W^*\to WH$       &   1504 & $H\to ZZ^*\to 4\ell$ & $0.19$&     564 &  169 & 0.5 \\ 
$pp$ & 14 TeV & 3000 & $Z^*\to ZH$         &   883 & $H\to ZZ^*\to 4\ell$ & $0.11$ &     331 &    99 & 0.5 \\ 
$pp$ & 14 TeV & 3000 & $t\bar{t}\to t\bar{t}H$&  611 & $H\to ZZ^*\to 4\ell$ & $0.08$ &     229 &    69 & 1.0 \\ 
\hline
$pp$ & 14 TeV & 3000 & $V^*V^*\to H$           & 4180 & $H\to \gamma\gamma$ & $9.53$ &  28591 &  8577 & 0.6 \\ 
$pp$ & 14 TeV & 3000 & $Z^*\to ZH$        &   883 & $H\to b\bar{b}$, $Z\to\ell\ell$ & $34.3$ &  102891 &  690 & -- \\ 
\hline
$e^+e^-$ & 250 GeV & 250 & $Z^*\to ZH$ & 240 & $H\to b\bar{b}$, $Z\to\ell\ell$ & 9.35 & 2337 & 1870 & -- \\  
$e^+e^-$ & 350 GeV & 350 & $Z^*\to ZH$ & 129 & $H\to b\bar{b}$, $Z\to\ell\ell$ & 5.03 & 1760 & 1408 & -- \\  
$e^+e^-$ & 500 GeV & 500 & $Z^*\to ZH$ &  57 & $H\to b\bar{b}$, $Z\to\ell\ell$ & 2.22 & 1110 &   888 & -- \\  
$e^+e^-$ & 1 TeV     & 1000 & $Z^*\to ZH$ & 13 & $H\to b\bar{b}$, $Z\to\ell\ell$ & 0.51 &  505 &   404 & -- \\  
\hline
$e^+e^-$ & 250 GeV & 250 & $Z^*Z^*\to H$ & $0.7$ & $H\to b\bar{b}$ & $0.4$ & 108 & 86 & -- \\
$e^+e^-$ & 350 GeV & 350 & $Z^*Z^*\to H$ & $3$ & $H\to b\bar{b}$ & $1.7$ & 587 & 470 & -- \\
$e^+e^-$ & 500 GeV & 500 & $Z^*Z^*\to H$ & $7$ & $H\to b\bar{b}$ & $4.1$ & 2059 & 1647 & -- \\
$e^+e^-$ & 1 TeV & 1000 & $Z^*Z^*\to H$    & $21$  & $H\to b\bar{b}$ & $12.2$ & 12244 & 9795 & -- \\
\hline\hline
\end{tabular}
\label{table:XS}
\end{center}
\end{table}

The resulting numbers of a $125~{\rm GeV}$ Standard Model Higgs bosons expected at the 
LHC and at an $e^+e^-$ collider are summarized in Table~\ref{table:XS}.
We calculate the number of produced signal events $N_{\rm prod}$ using SM Higgs boson cross sections 
and branching fractions from Ref.~\cite{Heinemeyer:2013tqa}. 
The cross sections at an $e^+e^-$ collider are calculated  with the {\sf JHU} generator
for $e^+e^-\to ZH$ process and {\sf MadGraph} for $e^+e^-\to e^+e^-H$ VBF-only process.
The selection criteria described above are used to find the number of reconstructed Higgs bosons $N_{\rm reco}$.
We assume only small contributions of anomalous couplings which would not change this number significantly. 
The LHC experiments are expected to collect sufficient statistics to study $HVV$ tensor 
structure both in production and in decay of a Higgs boson. At the same time, the $e^+e^-$ machines are 
in a much better position to study the $HVV$ tensor structure in production, especially at high energy. 
However, considerations  based entirely on  event yields are insufficient 
since  both kinematics and relative importance of various tensor structures' contributions change depending 
on the process and collision energies. To illustrate this, in Table~\ref{table:sigmas} we show examples where 
cross sections $\sigma_i$, defined below  Eq.~(\ref{eq:fractions}), are computed for several processes. 

\begin{table}[ht]
\begin{center}
\caption{
Description of processes used for $HVV$ tensor structure measurements with the corresponding 
cross sections ratios, where 
$\sigma_1$$, \sigma_2$, or $\sigma_4$ corresponds to $g_1=1$,  $g_2=1$, or $g_4=1$, respectively,
and $\sigma_+=\sigma_1$  ($g_+=g_1$) for all processes except couplings to massless vector bosons
($Z\gamma, \gamma\gamma, gg$) where $\sigma_+=\sigma_2$ ($g_+=g_2$).
MC simulation parameters used in studies are shown, where the generated coupling $g_i$ values
correspond to certain $f_{a2}$ and $f_{a3}$ values.
The expected precision on the $ f_{a2}$ and $f_{a3}$ parameters are quoted for 
300~fb$^{-1}$ (first row) and 3000~fb$^{-1}$ (second row) scenarios on LHC
and four energy scenarios on an $e^+e^-$ machine, as discussed in Table~\ref{table:XS}.
This expected precision corresponds to about 3$\sigma$ deviation from zero of the MC simulated values.
The $f_{a2}^{\rm dec}$ and $f_{a3}^{\rm dec}$ values correspond to cross sections defined in decay. 
}
\begin{tabular}{|lcccc || cccccc || cccc |}
\hline
\hline
 \multicolumn{5}{|c||}{process description}  &  \multicolumn{6}{c||}{MC simulation parameters}  &  \multicolumn{4}{c|}{~~expected precision~~}  \\
\hline
collider & energy &  mode  &  $\sigma_2/\sigma_1$ &  $\sigma_4/\sigma_{+}$ & $|g_2/g_1|$ &  $|g_4/g_{+}|$  & ~~$f_{a2}$ &  $f_{a2}^{\rm dec}$ &  $f_{a3}$ &  $f_{a3}^{\rm dec}$
 & $\delta f_{a2}$  & $\delta f_{a2}^{\rm dec}$ & $\delta f_{a3}$  & $\delta f_{a3}^{\rm dec}$ \\
\hline\hline
any   & any       &  $H\to ZZ^*$ & 0.362 &  0.153                      & 0 & 1.20 & \multicolumn{2}{c}{0} & \multicolumn{2}{c||}{0.18}  & \multicolumn{2}{c}{--}  & \multicolumn{2}{c|}{ 0.06 }  \\
          &               &                                            &            &              &  0 & 0.67 & \multicolumn{2}{c}{0} & \multicolumn{2}{c||}{0.06} & \multicolumn{2}{c}{--}  & \multicolumn{2}{c|}{ 0.02 }  \\
          &               &                                            &            &              & 0.78 & 0& \multicolumn{2}{c}{0.18} & \multicolumn{2}{c||}{0} & \multicolumn{2}{c}{ 0.088 }  & \multicolumn{2}{c|}{ -- }  \\
          &               &                                            &            &              &  0.42 & 0 & \multicolumn{2}{c}{0.06} & \multicolumn{2}{c||}{0} & \multicolumn{2}{c}{ 0.014  }  & \multicolumn{2}{c|}{ -- } \\
\hline
any   & any       &  $H\to WW^*$ & 0.776 &  0.322                      & 0 & 1.76 & \multicolumn{2}{c}{0} & \multicolumn{2}{c||}{0.50}  & \multicolumn{2}{c}{--}  & \multicolumn{2}{c|}{ -- }  \\
          &               &                                            &            &              & 1.13 & 0& \multicolumn{2}{c}{0.50} & \multicolumn{2}{c||}{0} & \multicolumn{2}{c}{ -- }  & \multicolumn{2}{c|}{ -- }  \\
\hline
any   & any       &  $H\to\gamma\gamma, gg$ & N/A &  1.0                     & N/A & 1.0 & \multicolumn{2}{c}{0} & \multicolumn{2}{c||}{0.50}  & \multicolumn{2}{c}{--}  & \multicolumn{2}{c|}{ -- }  \\
any   & any       &  $H\to Z\gamma$ & N/A &  1.0                     & N/A  & 1.0 & \multicolumn{2}{c}{0} & \multicolumn{2}{c||}{0.50}  & \multicolumn{2}{c}{--}  & \multicolumn{2}{c|}{ -- }  \\
\hline
\hline
$pp$ & 14 TeV &  $gg\to H$                        & N/A        &  1.0       &   N/A & 1.0 &  0 & 0 & 0.50 & 0.50 & \multicolumn{2}{c}{ --  } &  0.50 & 0.50 \\
          &               &   ($H\to ZZ^*$)              &            &               &  N/A & 1.0 &  0 & 0  & 0.50 & 0.50 & \multicolumn{2}{c}{ --  } & 0.16 & 0.16 \\
\hline
$pp$ & 14 TeV &  $V^*V^*\to H$                & 14.0     &  11.3    &  0 & 0.299      &  0 & 0 & 0.50 & 0.013  & \multicolumn{2}{c}{ --  } & 0.190 &$7\!\times\!10^{-3}$ \\
          &               &   ($H\to ZZ^*$)                &            &              &   0 & 0.109 &  0 & 0  & 0.12 & 0.0018 & \multicolumn{2}{c}{ --  } & 0.036 & $6\!\times\!10^{-4}$ \\
\hline
$pp$ & 14 TeV &  $V^*V^*\to H$                            & 14.0     &  11.3       & 0 & 0.109 &  0 & 0  & 0.12 & 0.0018 & \multicolumn{2}{c}{ --  }  & 0.04 & $7\!\times\!10^{-4}$  \\
          &               &   ($H\to\gamma\gamma$)             &           &                & 0 & 0.052 &  0 & 0  & 0.030 & 0.0004 & \multicolumn{2}{c}{ --  } &  0.009 & $1.3\!\times\!10^{-4}$\\ 
\hline
$pp$ & 14 TeV &  $V^*\to VH$                    &  76.1     &  46.8      & 0 & 0.145   & 0 & 0  & 0.50 & 0.0032  & \multicolumn{2}{c}{ --  }  & 0.32 & $3\!\times\!10^{-3}$  \\
          &              \multicolumn{2}{c}{($V\to q\bar{q}^\prime, H\to ZZ^*$)}   
                                                                            &           &                & 0  & 0.095      &  0 & 0  & 0.30 & 0.0014 & \multicolumn{2}{c}{ --  } & 0.10 & $6\!\times\!10^{-4}$  \\ 
\hline
$pp$ & 14 TeV &  $V^*\to VH$                                                                   &  76.1   &  46.8      & 0 & 0.061 & 0 & 0  & 0.15 & 0.0006 & \multicolumn{2}{c}{ --  }  & 0.09  & $4\!\times\!10^{-4}$  \\
         &            \multicolumn{2}{c}{($V\to \ell^+\ell^-, H\to b\bar{b}$)}   &           &                 &  0 & 0.049 &  0 & 0  & 0.10 &  0.0004 & \multicolumn{2}{c}{ --  } & 0.029 & $1.2\!\times\!10^{-4}$  \\ 
\hline\hline
$e^+e^-$ & 250 GeV & $Z^*\to ZH$ & 34.1 &  8.07 & 0  & 0.117 & 0 & 0  & 0.10 & $2\times10^{-3}$ & \multicolumn{2}{c}{ --  } & 0.032 & $7\!\times\!10^{-4}$ \\
         &               &                            &         &       &  0.057 &  0 & 0.10 & $1.2\times10^{-3}$ &  0 & 0  & 0.033 & $4\!\times\!10^{-4}$ & \multicolumn{2}{c|}{ --  }  \\
\hline
$e^+e^-$ & 350 GeV & $Z^*\to ZH$ & 84.2 &  50.6 &  0 & 0.0469 & 0 & 0  & 0.10  & $3\times10^{-4}$ & \multicolumn{2}{c}{ --  } & 0.031 &  $1.1\!\times\!10^{-4}$ \\
        &               &                            &         &       &  0.025  &  0 & 0.05 & $2\times10^{-4}$ &  0 & 0  & 0.015 & $7\!\times\!10^{-5}$   & \multicolumn{2}{c|}{ --  }   \\
\hline
$e^+e^-$ & 500 GeV & $Z^*\to ZH$ & 200.8 &  161.1 &  0 & 0.0263 & 0 & 0  & 0.10 & $1.1\times10^{-4}$ & \multicolumn{2}{c}{ --  } & 0.034 &  $4\!\times\!10^{-5}$ \\
        &               &                            &         &       &   0.024 &  0 & 0.10 & $2\times10^{-4}$  &  0 & 0  & 0.033 &  $7\!\times\!10^{-5}$ & \multicolumn{2}{c|}{ --  }   \\
\hline
$e^+e^-$ & 1 TeV     & $Z^*\to ZH$ & 916.5 &  870.8 &   0 & 0.0113 & 0 & 0 & 0.10 & $2\times10^{-5}$ & \multicolumn{2}{c}{ --  } & 0.037 &  $8\!\times\!10^{-6}$ \\
        &               &                            &                        &       &  0.014 &  0 & 0.15 & $7\times10^{-5}$  &  0 & 0  & 0.049 & $3\!\times\!10^{-5}$ & \multicolumn{2}{c|}{ --  }  \\
\hline
\hline
\end{tabular}
\label{table:sigmas}
\end{center}
\end{table}

As evident from Table~\ref{table:sigmas}, relative cross sections corresponding to 
scalar ($g_1$) and pseudoscalar ($g_4$) couplings are different in various $HVV$ processes. 
For example the ratio $\sigma_4/\sigma_1$ is 0.153 in the $H\to ZZ$ decay, 
8.07 in $e^+e^-\to ZH$ production at $\sqrt{s}=250$ GeV and grows linearly with increasing $\sqrt{s}$. 
This is caused by the  different dependence of the scalar and pseudoscalar tensor couplings in 
Eq.~(\ref{eq:fullampl-spin0}) on the off-shellness of the vector boson, which leads to an 
asymptomatically energy-independent $e^+e^-$ cross section in case of $C\!P$-odd higher-dimensional operator.   
This feature means  that, for a fixed ratio of coupling constants 
$|g_4/g_1| $, it is beneficial to go to highest available energy where the production cross section 
due to $g_4$ is  kinematically enhanced \cite{sp12}. 
Therefore, the same fraction of events for $C\!P$-odd contributions at different collider energies 
translates into different  sensitivities  for effective couplings $g_i$. To compare different cases, we express 
the results of the analysis  
in terms of $f_{a3}^{\rm dec}$, defined for the  Higgs boson decay to two vector bosons 
 since in this case the kinematics are entirely fixed 
and this choice determines the ratio of the coupling constants uniquely. 

\begin{figure}[t]
\centerline{
\setlength{\epsfxsize}{0.45\linewidth}\leavevmode\epsfbox{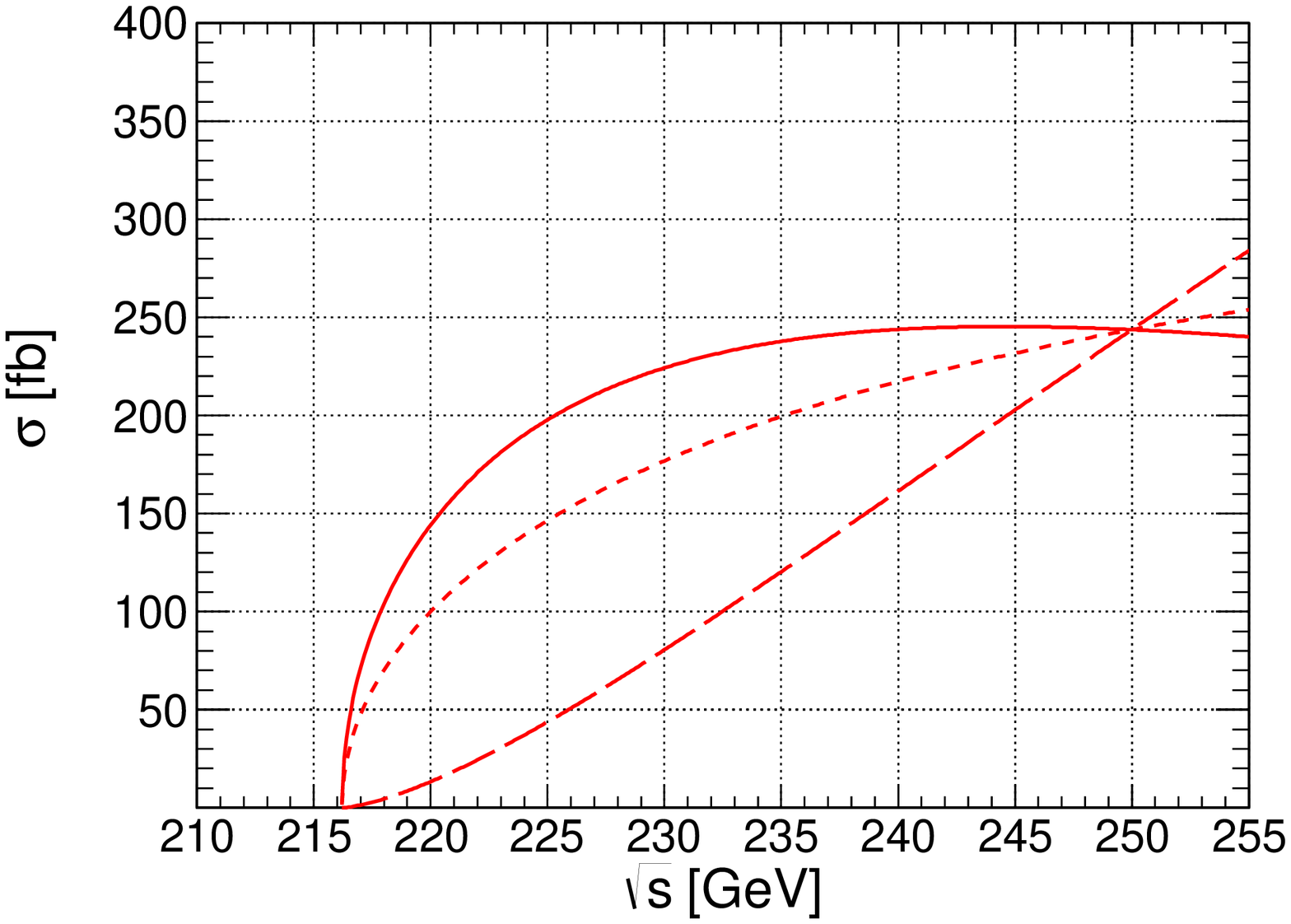}
\setlength{\epsfxsize}{0.45\linewidth}\leavevmode\epsfbox{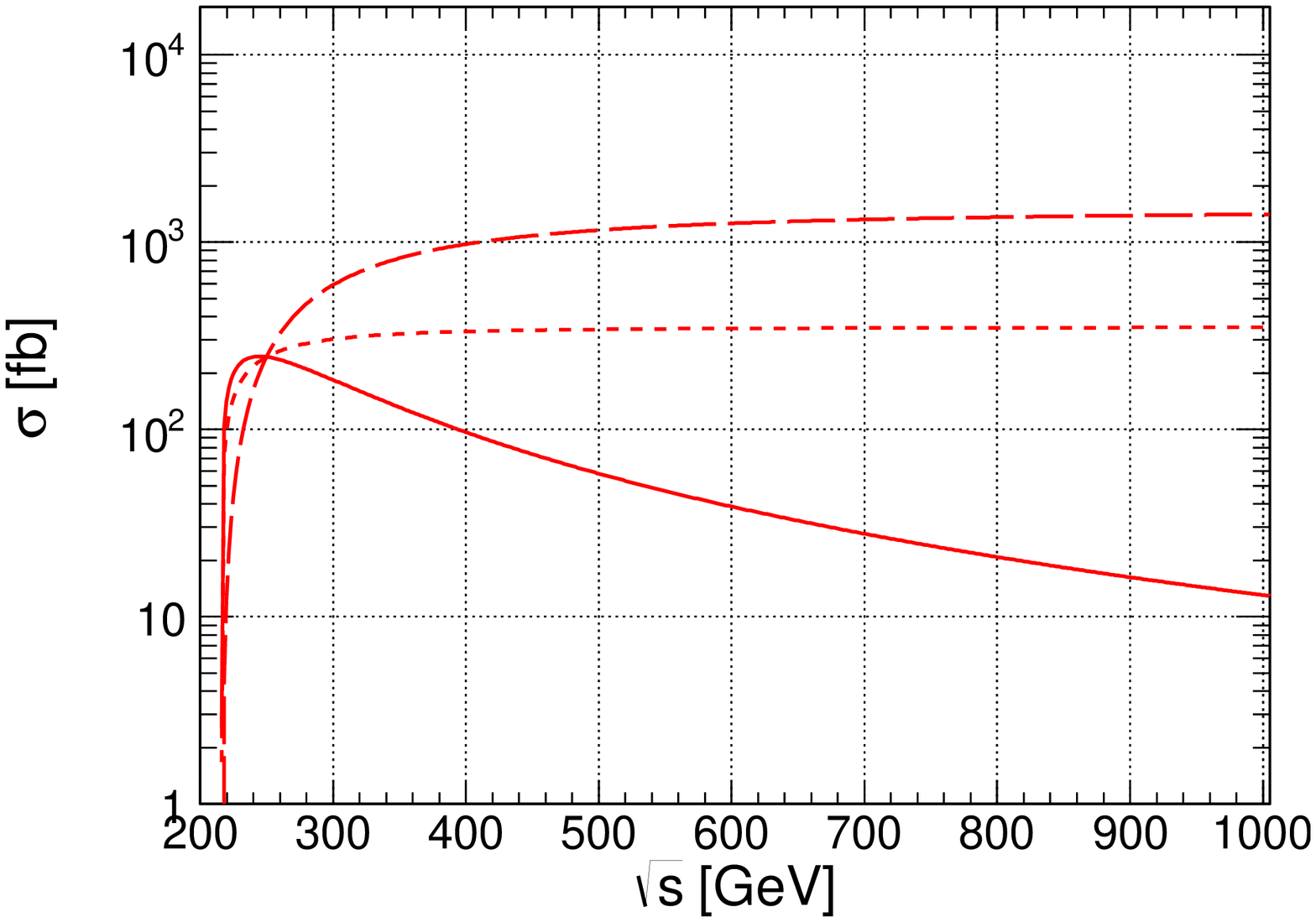}
}
\caption{
Cross sections for  $e^+e^-\to Z^*\to ZX$ process as a function of $\sqrt{s}$ for three models:
SM Higgs boson ($0^+$, solid), 
scalar with higher-dimension operators ($0^+_h$, short-dashed), 
and pseudoscalar ($0^-$, long-dashed).
All cross sections are normalized to SM value at $\sqrt{s}=250$ GeV. Different high-energy behavior 
of cross sections related to point-like interactions (solid) and higher-dimensional non-renormalizable 
operators (dashed) is apparent from the right panel. 
}
\label{fig:eeXS}
\end{figure}

To illustrate this point further, we examine the energy dependence of the $e^+e^-\to Z^*\to ZH$ cross section
for various tensor couplings. 
In Fig.~\ref{fig:eeXS}, cross section dependence on  $\sqrt{s}$ is shown for the ratio of the coupling constants 
chosen in such a way that cross sections for all tensor structures at $\sqrt{s}=250$ GeV are equal to the SM 
$e^+e^- \to Z^* \to ZH$ cross section.
The threshold behavior for $\sqrt{s}<250$~GeV of the cross sections $e^+e^- \to Z^* \to XZ$ 
has been suggested as a useful observable to determine the spin of the new boson~\cite{sp1}. 
Similarly, in a mixed $C\!P$-case, the dependence of $e^+e^- \to ZH $ cross section on the 
energy of the collision will differ from a pure $J^{\,C\!P}=0^{++}$ case; therefore, a measurement 
of the cross section at several different energies will give us useful information 
about anomalous $HVV$ couplings. 
For example, if the $e^+e^-\to Z^*\to ZH$ cross section is
first measured at the center of mass energy 
$\sqrt{s}=250$~GeV, the  scan of cross sections at 350, 500, and 1000~GeV 
will lead to a measurement of $f_{a3}$ with precision $0.035, 0.041$, and $0.055$, respectively,  
using the expected signal yields reported in Table~\ref{table:XS}. 
This would translate to precision on $f_{a3}^{\rm dec}$ of $10^{-4}, 4\times10^{-5}$, and $10^{-5}$, 
respectively, as defined in the decay $H\to ZZ^*$. 

As we have already mentioned, 
the reason for the significantly improved precision on $f_{a3}^{\rm dec}$ that appears to be achievable 
at higher energy $e^+e^-$ colliders is the energy-independence of the  cross section 
for pseudoscalar couplings,  caused by the 
non-renormalizable nature of the operator $Z_{\mu \nu} {\tilde Z}^{\mu \nu}$.
Of course,  this feature  cannot continue forever and, 
in  any  theory, the coupling ``constant''  $g_4$  should eventually become 
a $q^2$-dependent form factor, which will provide 
suppression for the cross section  at higher energies.  The energy scale where the $q^2$-dependence 
of the effective couplings  
can no longer be ignored is denoted as $\Lambda_4$ in Eq.~(\ref{eq:formfactors}) and we do not know this 
scale a priori.  For this reason, we ignore the $q^2$-dependent form-factors in this paper, but we
note that results presented above can be changed to incorporate possible reduction 
of the coupling constants with energy. 
Studies  of experimental data should, ideally,  include  tests of different  values of the 
form-factor scales $\Lambda_i$.

We conclude this general discussion by pointing out 
that three types of  observables  can be used to measure tensor couplings 
of the Higgs bosons in general and $f_{a3}$ in particular. They are 
\begin{enumerate} 
\item cross sections, especially  their dependences on virtualities of weak bosons \cite{sp3, sp12,Boughezal:2012tz}.
Examples  are shown  in Fig.~\ref{fig:eeXS} 
for the $e^+e^-\to Z^*\to ZH$ process and in Fig.~\ref{fig:lhc_angles}
for the decay $H\to ZZ^*$.  We note that while measurements of cross sections 
in different kinematic regimes appear to be a 
powerful tool to study  anomalous couplings, it relies on our understanding 
of {\it dynamics}, rather than kinematics, and therefore may be sensitive to poorly 
understood form-factor effects or breakdown of effective field-theoretic description. 
\item Angular distributions particular  
to scalar and pseudoscalar $HVV$ interactions or, more generally, to different types of tensor couplings. 
Examples of such distributions are shown in 
Figs.~\ref{fig:ilc_angles}, \ref{fig:lhc_angles}, \ref{fig:vbf_angles}, \ref{fig:h2j_angles}.
\item Angular distributions  or other observables 
particular  to interferences between $C\!P$-even and $C\!P$-odd couplings.
Examples include forward-backward asymmetry with respect to $\cos\theta_1$ or $\cos\theta_2$ 
and non-trivial phase in the $\Phi$ distributions shown in Figs.~\ref{fig:ilc_angles} and~\ref{fig:lhc_angles}.
Such asymmetries require undefined $C\!P$ to appear; 
as the result,  $C\!P$ violation would follow as an unambiguous interpretation e.g. once the forward-backward asymmetry is observed. 
\end{enumerate} 
In order to measure or set a limit on $f_{a3}$, it is important  to employ 
all types of observables described above and not limit 
oneself to $C\!P$-specific ones, such as interferences.
In particular, if only a limit is set on $f_{a3}$, the phase
of $C\!P$-odd contribution $\phi_{a3}$ is generally unknown and one cannot predict the 
forward-backward asymmetry in $\cos\theta_1$ nor the non-trivial phase in $\Phi$,
as shown in Figs.~\ref{fig:ilc_angles} and \ref{fig:lhc_angles}. 
For example, even under the assumption of real coupling constants, $\phi_{a3}$ ambiguity between
$0$ and $\pi$ needs to be resolved.
In principle, model-dependent assumptions can be made about such phases and tighter constraints on $f_{a3}$ 
can be obtained, but it is important to pursue  coupling measurements that are as model-independent as possible. 
On the other hand, once a non-zero value of $f_{a3}$ is observed, its phase $\phi_{a3}$ can be measured
directly from the data,  as we illustrate below. 
While we focus on the measurement of the $C\!P$-odd contribution $f_{a3}$, we also illustrate 
measurements of $f_{a2}$ and $f_{\Lambda1}$, which can be performed with a similar precision. 
Here $f_{\Lambda1}$ is defined as in  Eq.~(\ref{eq:fractions}); it provides the cross section fraction 
that is induced by  $-g_1^{\prime\prime}\times(q_1^2+q_2^2)/\Lambda_1^2$ anomalous coupling. 


\subsection{The $e^+e^-\to ZH$ process}
\label{sect:VH_ILC}

To illustrate the above points, we considered $e^+e^-\to ZH$ process, 
with $Z\to \ell^+\ell^-$ and $H\to b\bar{b}$.
The number of signal events is estimated in Table~\ref{table:XS} for four energies 
$\sqrt{s}=$ 250, 350, 500, 1000 GeV, that are under discussion for an electron-positron collider,
and are rounded to 2000, 1500, 1000, 500 events, respectively.
The effective number of background events is estimated to be 10\%
of the number of signal events and is modeled with the $e^+e^-\to ZZ\to \ell^+\ell^- b\bar{b}$ process. 
Cross sections for several simulated signal samples are displayed  in Table~\ref{table:sigmas}. 
We assume that the signal can be reconstructed inclusively by tagging  $Z\to \ell^+\ell^-$
decay and using energy-momentum  constraints, but further improvements can be achieved through
the analysis of the Higgs boson decay products and by considering other $Z$ decay final states.
In view of this, our estimates of expected sensitivities are conservative. 

Our analysis techniques are identical to what has been used earlier to study Higgs spin and parity in the 
$pp \to H \to ZZ$ process at the LHC~\cite{Gao:2010qx,Bolognesi:2012mm}. 
For this channel and the channels in the following subsections,
the details of the analyses are explained in Appendix~\ref{sect:statistics}. 
We employ either the dedicated discriminants 
$D_{0^-}$ and $D_{C\!P}$, or the multi-dimensional probability distribution.
Several thousand statistically-independent experiments are generated and fitted using 
different approaches. Detector effects and backgrounds are included 
either with direct parameterization of one- or two-dimensional distributions or by exploiting certain 
approximations of a multidimensional model, as explained in Appendix~\ref{sect:statistics}. 

\begin{figure}[t]
\centerline{
\setlength{\epsfxsize}{0.33\linewidth}\leavevmode\epsfbox{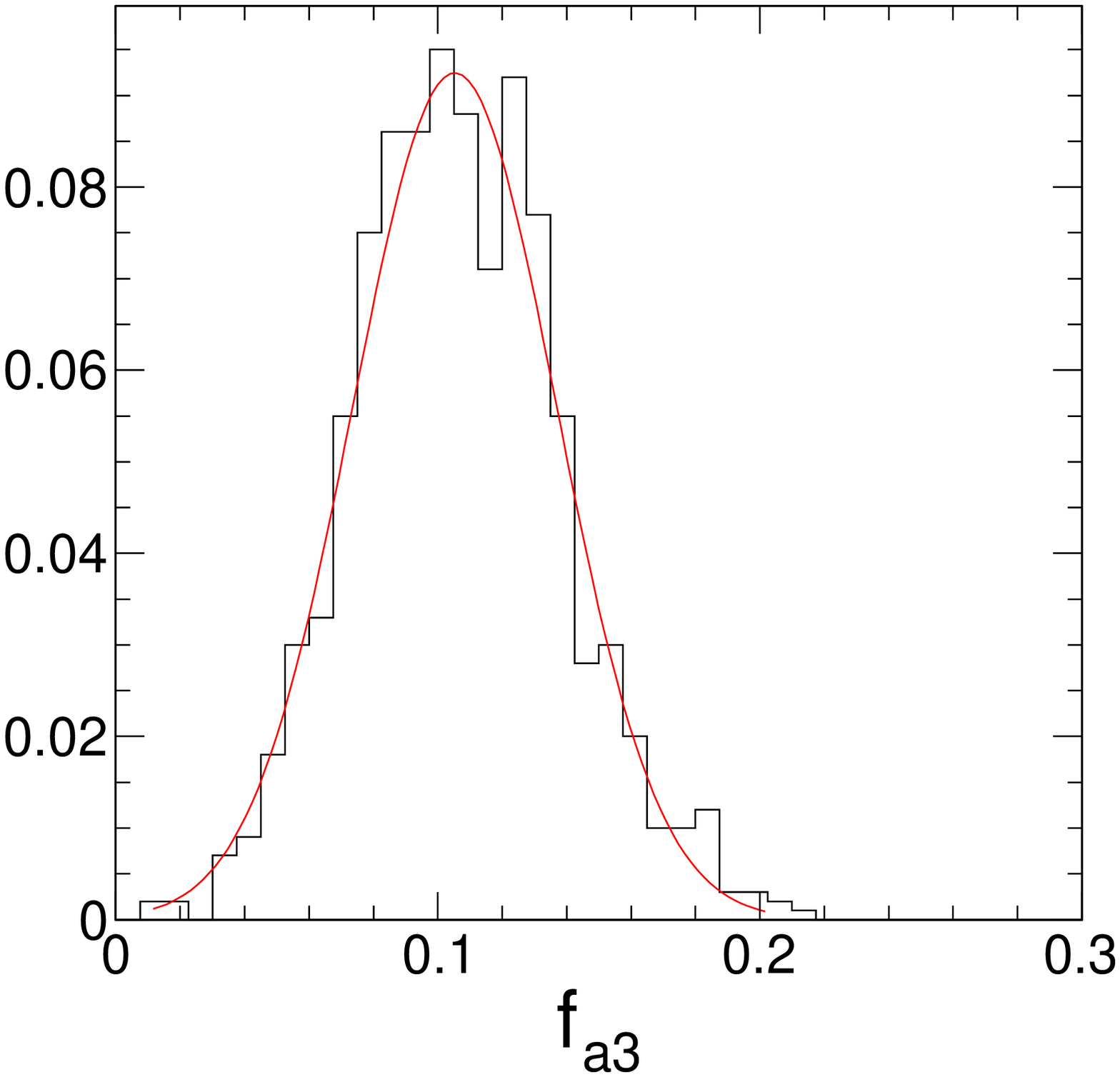}
\setlength{\epsfxsize}{0.33\linewidth}\leavevmode\epsfbox{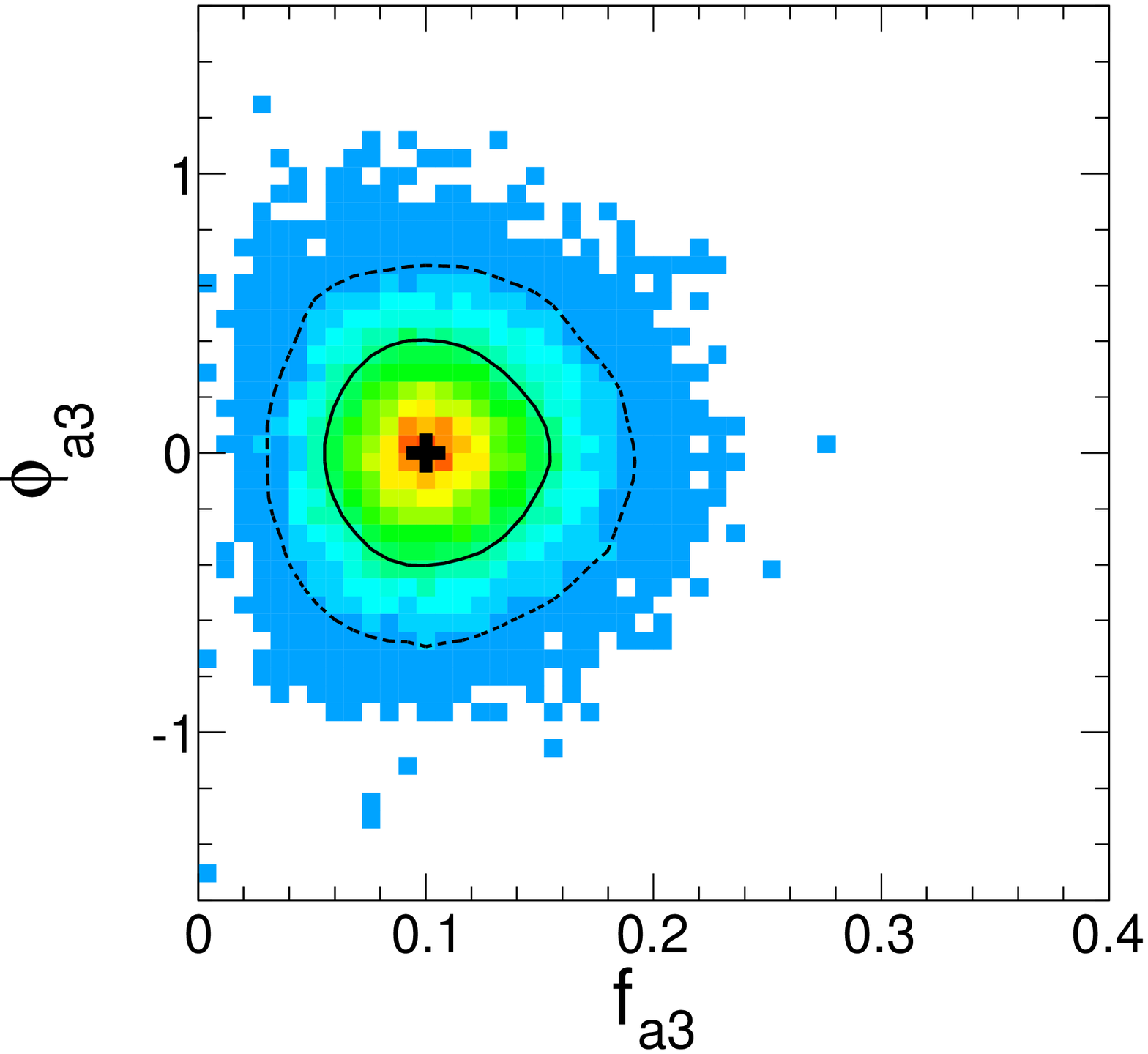}
\setlength{\epsfxsize}{0.33\linewidth}\leavevmode\epsfbox{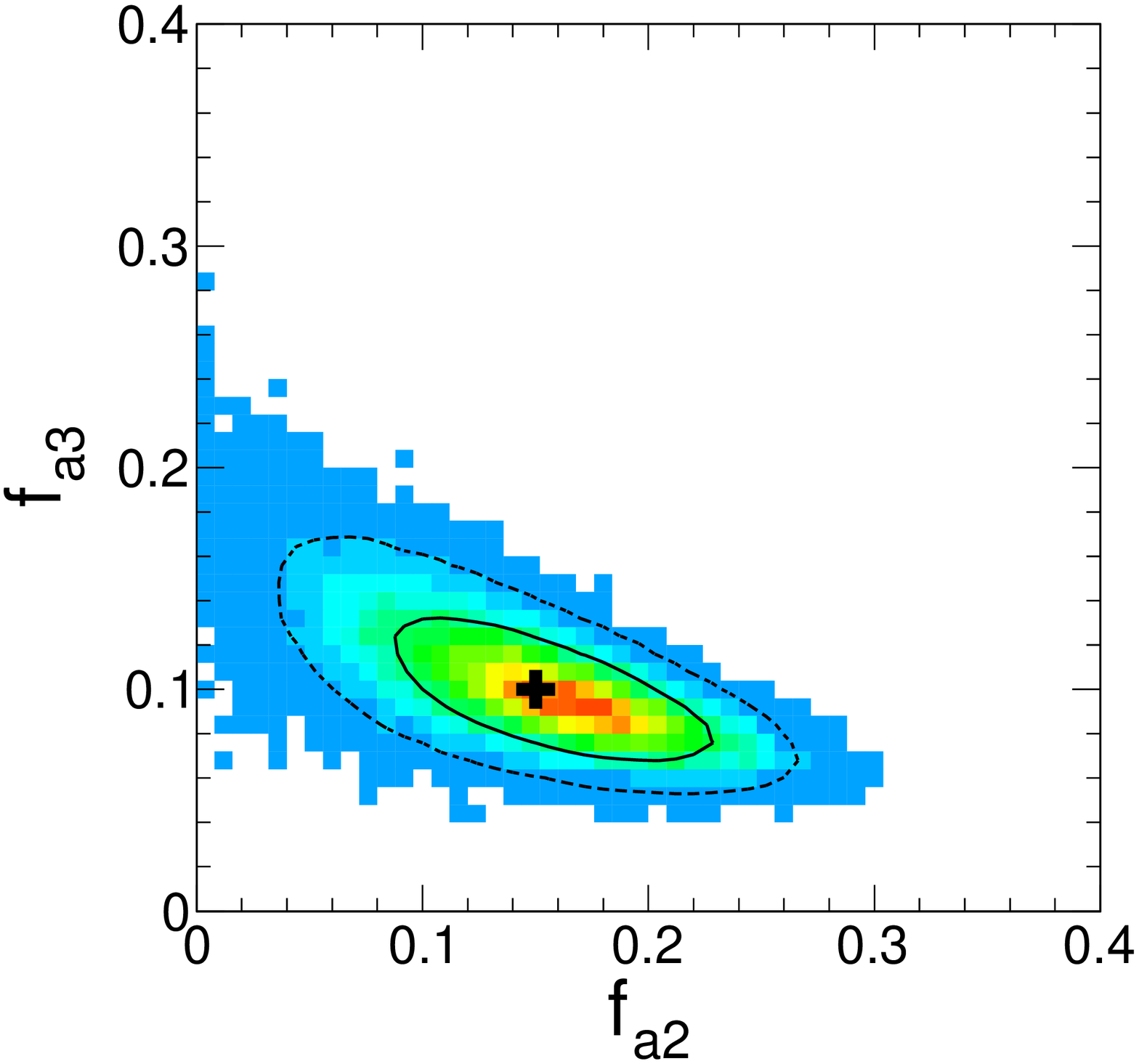}
}
\caption{
Distribution of fitted values of $f_{a3}$,  $\phi_{a3}$, and $f_{a2}$ in a large number of generated 
experiments in $e^+e^-\to ZH$ process at $\sqrt{s}=250$~GeV.
Left plot: $f_{a3}$ results from simultaneous fit of $f_{a3}$ and $\phi_{a3}$.
Middle and right plots: simultaneous fit of $f_{a3}$ and $\phi_{a3}$ or $f_{a3}$ and $f_{a2}$,
with 68\% and 95\% confidence level contours shown.
}
\label{fig:ilc_fa3}
\end{figure}

For the $e^+e^-$ case discussed in this section, 
we first obtained results for the sensitivity to the fractions $f_{a2,a3}$ at fixed collider energy and 
then expressed these constraints in terms of the parameters $f_{a2,a3}^{\rm dec}$.
Figure~\ref{fig:ilc_fa3} shows precision on $f_{a3}$ and $f_{a2}$ obtained with generated experiments that 
include background.
Expected precisions of $f_{a2,a3}$ measurements are shown  in Table~\ref{table:sigmas}. As can be seen there, 
the expected precision on $f_{a3}$ is in the range $0.03-0.04$, independent of the $e^+e^-$ collision energy.
This translates to very different constraints on $f_{a3}^{\rm dec}$ that range from  $7\times10^{-4}$ to $8\times10^{-6}$; 
as we already explained, measuring a similar fraction of events caused by the pseudoscalar anomalous couplings 
at higher energy means a sensitivity to a smaller value of $g_4$. 
The  expected precision is therefore similar  to what can be 
achieved from  cross section measurements at different energies, but 
in this case it relies on kinematic observables rather than dynamic ones that can be subject to form-factor effects. 
The expected precision of $f_{a2}^{\rm dec}$ is comparable to that of $f_{a3}^{\rm dec}$. 
We also confirm that precision on $f_{a3}$ does not change significantly if $\phi_{a3}$ is either floated or kept 
fixed  provided that the measured value of $f_{a3}$ is at least 3$\sigma$ away from zero. 

\begin{figure}[t]
\centerline{
\setlength{\epsfxsize}{0.33\linewidth}\leavevmode\epsfbox{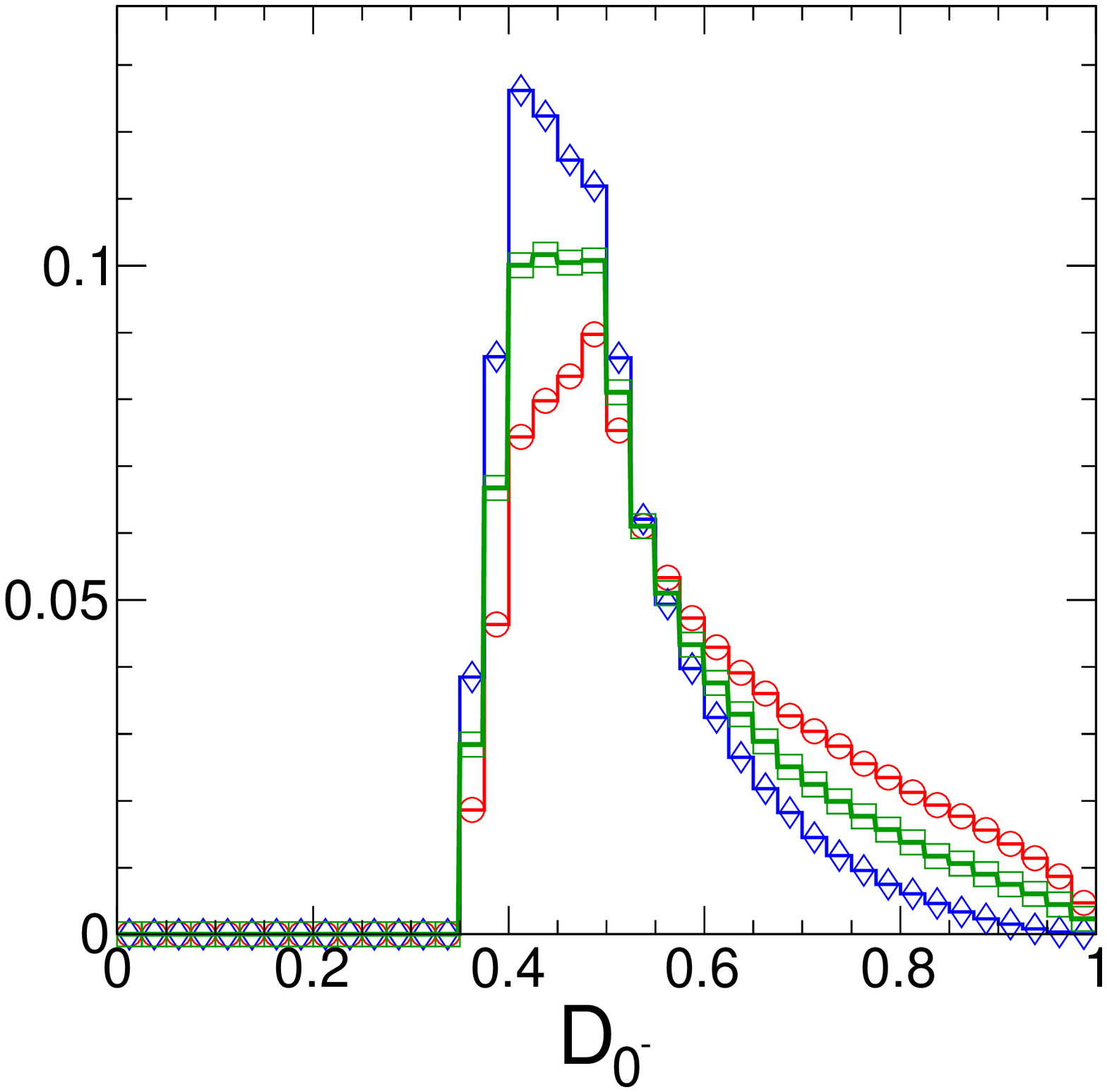}
\setlength{\epsfxsize}{0.33\linewidth}\leavevmode\epsfbox{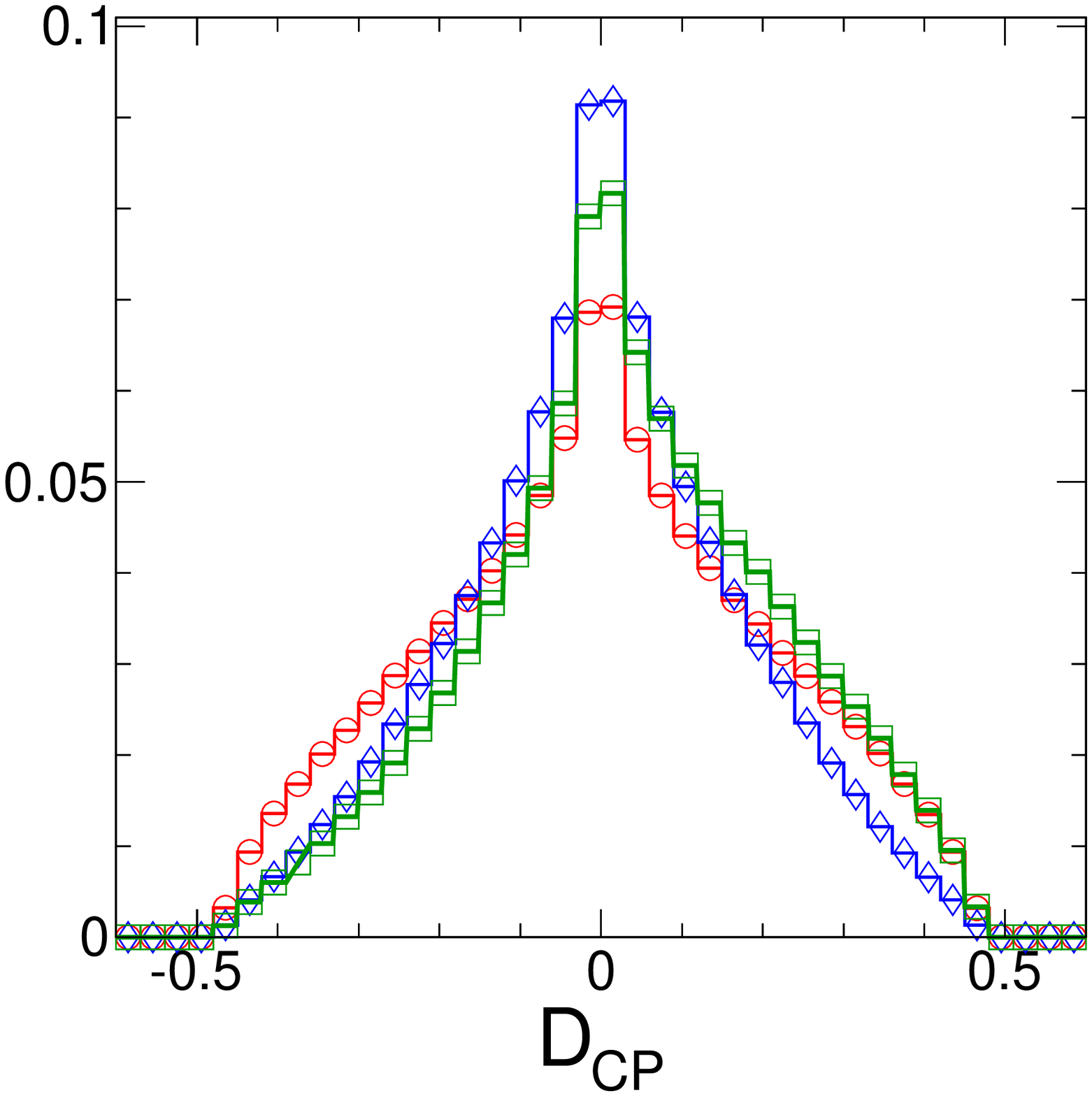}
\setlength{\epsfxsize}{0.33\linewidth}\leavevmode\epsfbox{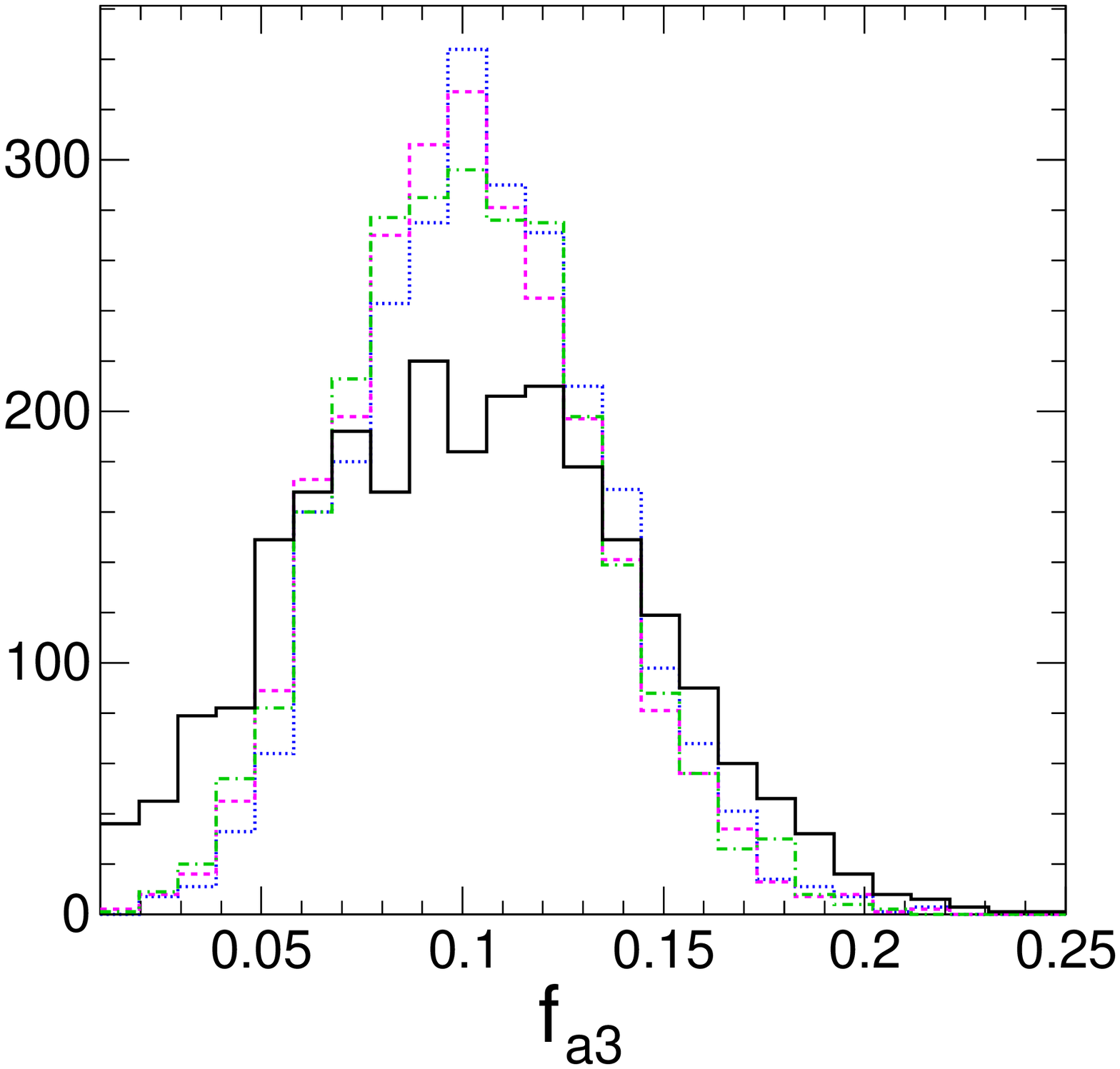}
}
\caption{
Distribution of $D_{0^-}$ and  $D_{C\!P}$ for generated events $e^+e^-\to ZH$ at $\sqrt{s}=250$~GeV.
Three processes  are shown: SM ($0^+$, red open circles), pseudoscalar ($0^-$, blue diamonds), 
and a mixed state corresponding to $f_{a3}=0.5$ with $\phi_{a3}=0$ (green squares). 
Right plot: $f_{a3}$ results without considering background and detector effects:
1D fit of $D_{0^-}$ (solid black);
2D fit of $D_{0^-}$ and $D_{C\!P}$  (dot-dashed green);
3D fit with $f_{a3}$ and $\phi_{a3}$ unconstrained (dotted blue);
and 3D fit with $f_{a3}$ only unconstrained (dashed magenta).
}
\label{fig:ilc_fa3_ND}
\end{figure}

The process  $e^+e^-\to ZH\to (\ell^+\ell^-)H$ is relatively simple and the three-dimensional (3D) analysis
is sufficient to extract most information from the multi-parameter fit, as illustrated above. 
Let us discuss this example as an illustration of how $C\!P$-analysis 
can be performed in other, more complicated, channels at both proton and lepton colliders. 
At a given $e^+e^-$ energy, there are no form-factor effects to study and the couplings are constant
and, in general, complex numbers. Therefore, dynamic information sensitive to form factors is contained 
in the $\sqrt{s}$ dependence and can be easily  separated from the rest. 
The other two pieces of information, as we discussed above, can be incorporated in two discriminants
$D_{0^-}$ and $D_{C\!P}$, see Fig.~\ref{fig:ilc_fa3_ND} and Appendix~\ref{sect:statistics}.
The $D_{0^-}$ discriminant is optimal to separate amplitudes squared representing the scalar and pseudoscalar contributions. 
The $D_{C\!P}$ discriminant is optimal to separate interference of the scalar and pseudoscalar amplitudes. 

The $D_{C\!P}$ is particularly interesting as it incorporates the full information about interference in a single
observable which exhibits clear forward-backward asymmetry indicating ${C\!P}$ violation.
There is a built-in assumption about the relative phase of the $g_1$ and $g_4$ terms in
the $D_{C\!P}$ construction. Under the assumption $\phi_{a3}=0$ or $\pi$, which can be justified 
if heavy particles generate the $g_4$ coupling perturbatively, $D_{C\!P}$ exhibits maximal 
forward-backward asymmetry, with the sign changing between $\phi_{a3}=0$ and $\pi$. Should the
phase be between $0$ and $\pi$, the asymmetry is reduced and, eventually,  vanishes at $\phi_{a3} = \pi/2$.
If this happens, it is possible to construct another discriminant $D^\perp_{C\!P}$ that has maximal 
asymmetry at $\phi_{a3} = \pm \pi/2$ and has asymmetry vanishing at $\phi_{a3} = 0,\pi$. 
At any rate, it is straightforward to introduce the two discriminants ($D_{C\!P}$, $D^\perp_{C\!P}$)
that will allow us to measure non-zero interference and the phase $\phi_{a3}$.

We stress that it is  advantageous  to use $D_{0^-}$ and $D_{C\!P}$ discriminants. Indeed, 
they cleanly separate information contained either in the yields 
of $C\!P$-odd and  $C\!P$-even contributions or their interference. 
The same information is present in the angular observables, such as those shown in Fig.~\ref{fig:ilc_angles},
but it is hidden in the multi-dimensional space. For example, forward-backward asymmetry
is also visible in the plots in Fig.~\ref{fig:ilc_angles}, but it is less obvious in some cases. For example, in case of 
$\phi_{a3}=0$ no simple observable exists to  illustrate it. 
It is also hard to describe distributions with larger number of dimensions for 
some of the other processes (e.g. VBF discussed later) or to parameterize both the detector effects and background. 
It is relatively simple to parameterize the one- or two-dimensional distributions of $D_{0^-}$ and $D_{C\!P}$ as we show below. 
Moreover, this approach can be easily extended to measure $f_{a2}$ using the dedicated discriminants 
with the same approach, which includes interference of the $g_1$ and $g_2$ terms. 

Figure~\ref{fig:ilc_fa3_ND} illustrates the results of several measurements using either an optimal 3D analysis,
or a single- or double-discriminant analysis. 
We omit background events in this study to simplify presentation, but this has little effect on the conclusion.
For the discriminant parameterization, we use Eq.~(\ref{eq:fractions-P}) with either 1D or 2D template histograms. 
When $f_{a3}$ is obtained from a one-dimensional fit to  $D_{0^-}$,
which does not contain an  interference between the $C\!P$-odd and $C\!P$-even contributions,
the precision on $f_{a3}$ gets worse by about 65\% with $f_{a3}=0.05$, 37\% with $f_{a3}=0.10$ and 
by 12\% with $f_{a3}=0.50$ at $\sqrt{s}=250$ GeV, with each case corresponding to 3$\sigma$
measurements of $f_{a3}$.
Note that interference scales as $\sqrt{f_{a3}}$ and therefore dominates at small values of $f_{a3}$.
Hence, especially for small event fractions,
the interference effects are important to include when non-zero 
$C\!P$-odd contribution is observed, and they appear to be more important in this mode
than in the $H\to ZZ$ decay, as we will see below, because analysis does not rely on observables sensitive to dynamics.
When $f_{a3}$ is obtained from a two-dimensional fit of $D_{0^-}$ and $D_{C\!P}$, precision of the
full multi-dimensional fit is recovered. 
However, we note that $D_{C\!P}$ or $D^\perp_{C\!P}$ do not  provide additional constraint on $f_{a3}$
without constraints on $\phi_{a3}$.

All the above techniques can be applied to all other channels under consideration, as discussed below.
While we provide the tools to explore all these methods, we often choose the more practical ways
to illustrate expected precision in each channel.


\subsection{The $H\to ZZ^*$ process on LHC}
\label{sect:HZZ_LHC}

In this subsection, 
we study  precision on tensor coupling measurements that can be achieved by exploiting  kinematics of 
$H\to ZZ^*$ process at the LHC.
The signal contributions are  listed in Table~\ref{table:XS}; we consider the sum of all five production 
mechanisms. The effective number of background events is estimated to be 0.4 times the
number of signal events; it is modeled with the $q\bar{q}\to ZZ^*/Z\gamma^*$ process. We compare the 
sensitivity that can be reached when  $300$~fb$^{-1}$ and $3000$~fb$^{-1}$ of integrated  
luminosity is collected at  the LHC.  The number of Higgs events 
at  $300$~fb$^{-1}$ is taken to be 10\% of the $3000$~fb$^{-1}$ yields quoted in Table~\ref{table:XS}.
Cross sections for some of the simulated signal samples are listed in Table~\ref{table:sigmas}. 

\begin{figure}[t]
\centerline{
\setlength{\epsfxsize}{0.33\linewidth}\leavevmode\epsfbox{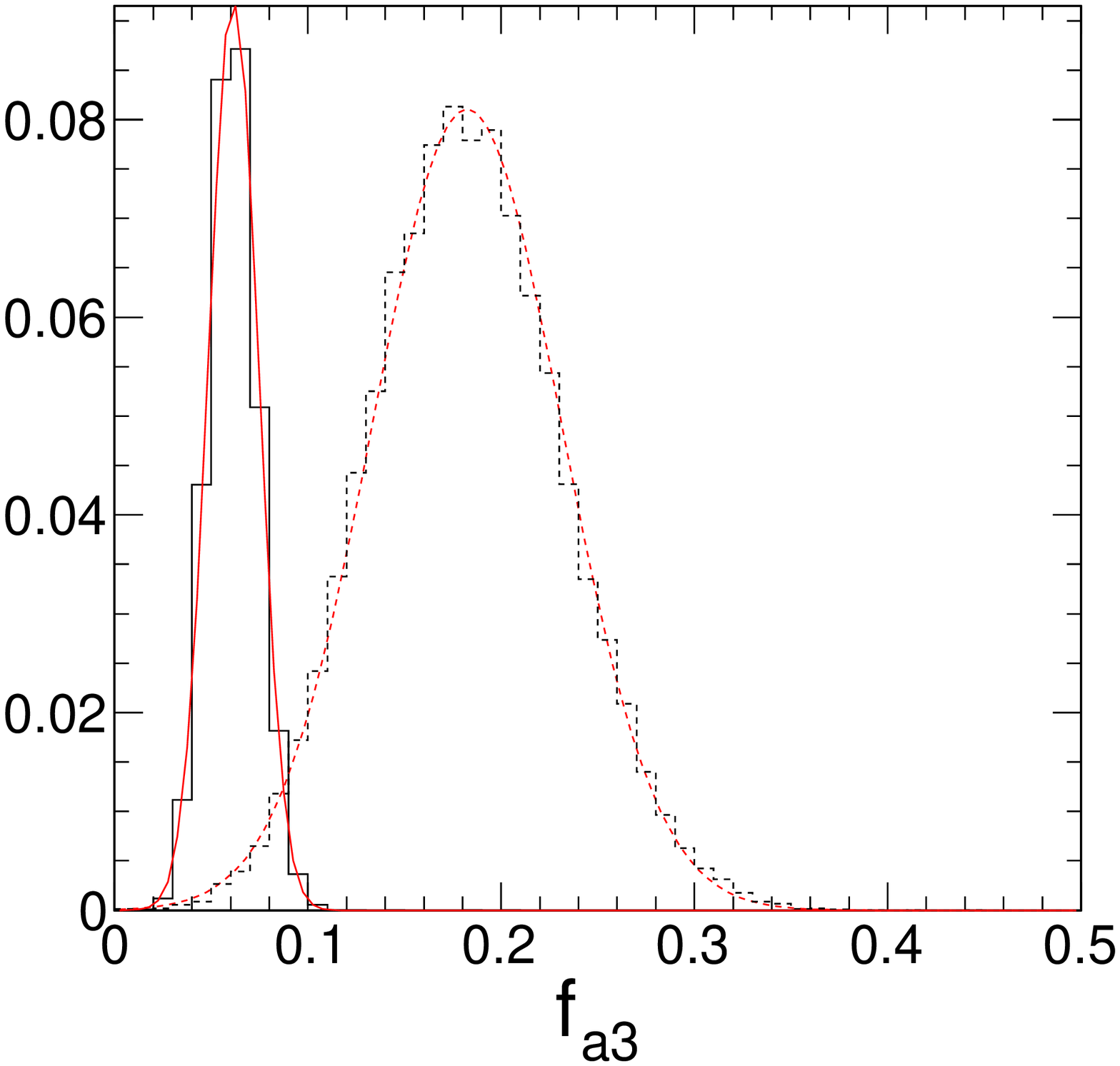}
\setlength{\epsfxsize}{0.33\linewidth}\leavevmode\epsfbox{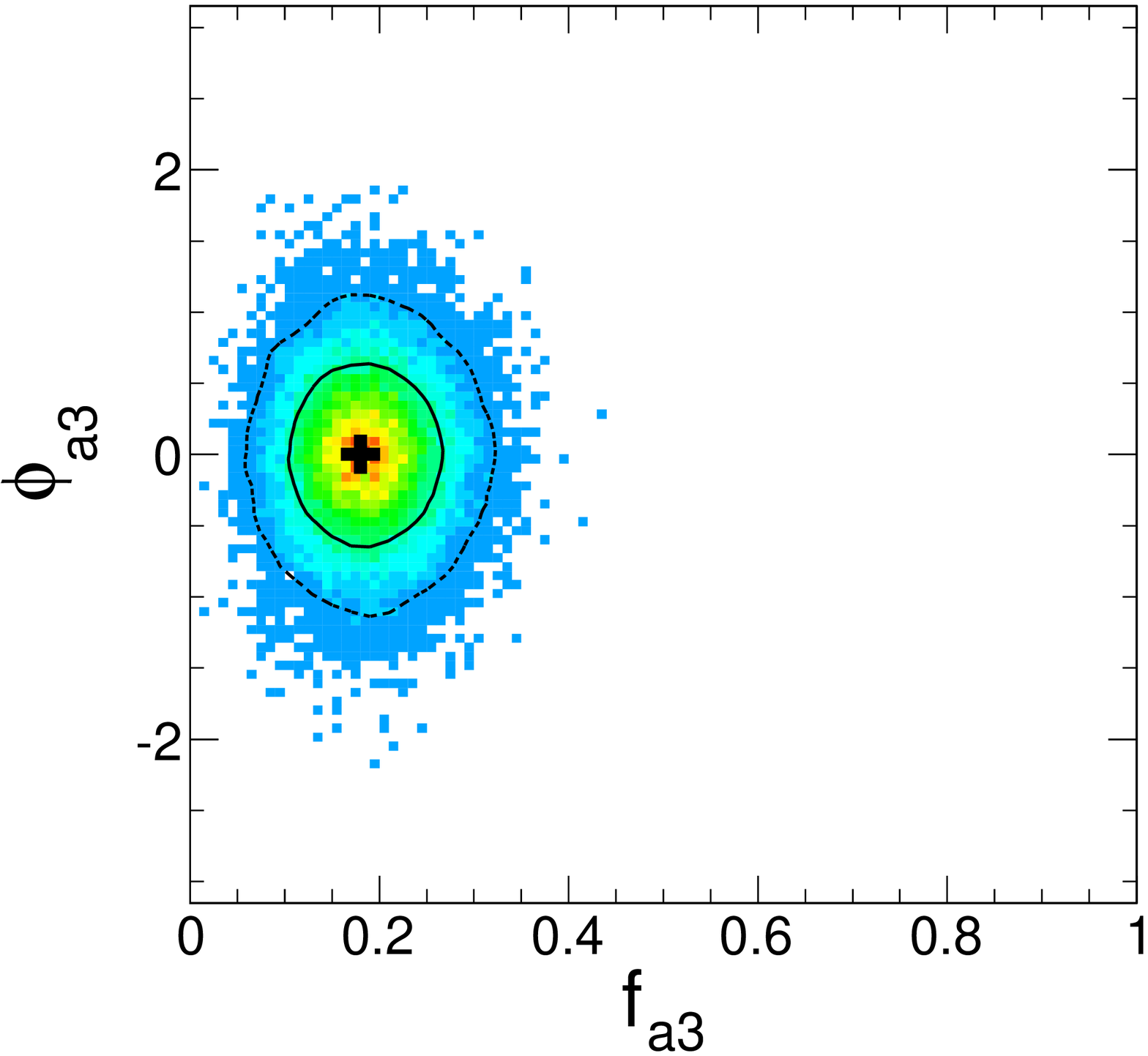}
}
\caption{
Distribution of fitted values of $f_{a3}$,  $\phi_{a3}$, and $f_{a2}$  in a large number of generated 
experiments with a 7D analysis in the $H\to ZZ^*\to 4\ell$ channel with 300 fb$^{-1}$ of data collected at the  LHC.
Left plot: $f_{a3}$ results from simultaneous fit of $f_{a3}$ and $\phi_{a3}$ with 300 fb$^{-1}$ (dotted) and 3000 fb$^{-1}$ (solid).
Right plots: simultaneous fit of $f_{a3}$ and $\phi_{a3}$ with 300 fb$^{-1}$ with 68\% and 95\% confidence level contours shown.
}
\label{fig:lhc_fa3}
\end{figure}

\begin{figure}[t]
\centerline{
\setlength{\epsfxsize}{0.33\linewidth}\leavevmode\epsfbox{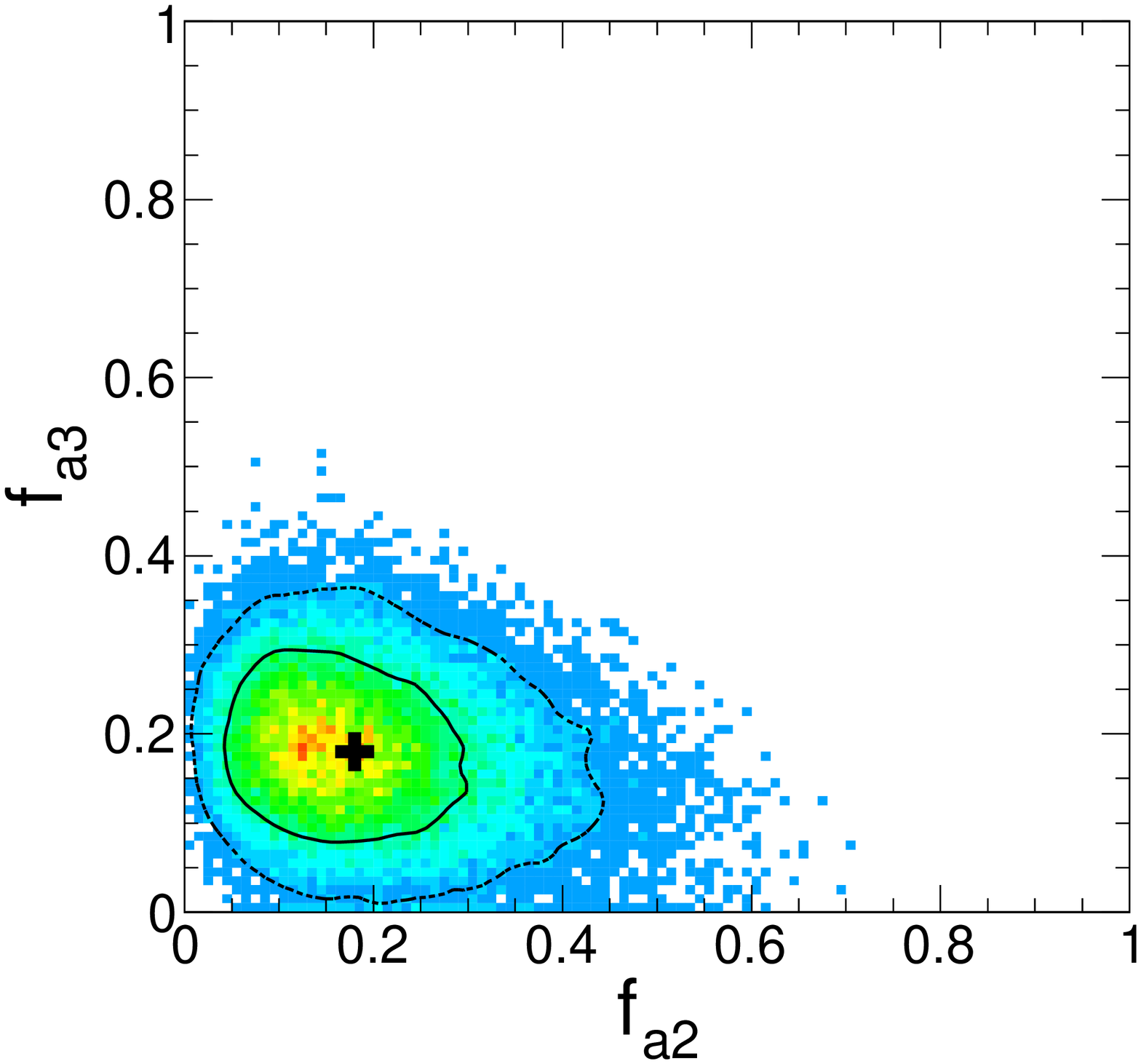}
\setlength{\epsfxsize}{0.33\linewidth}\leavevmode\epsfbox{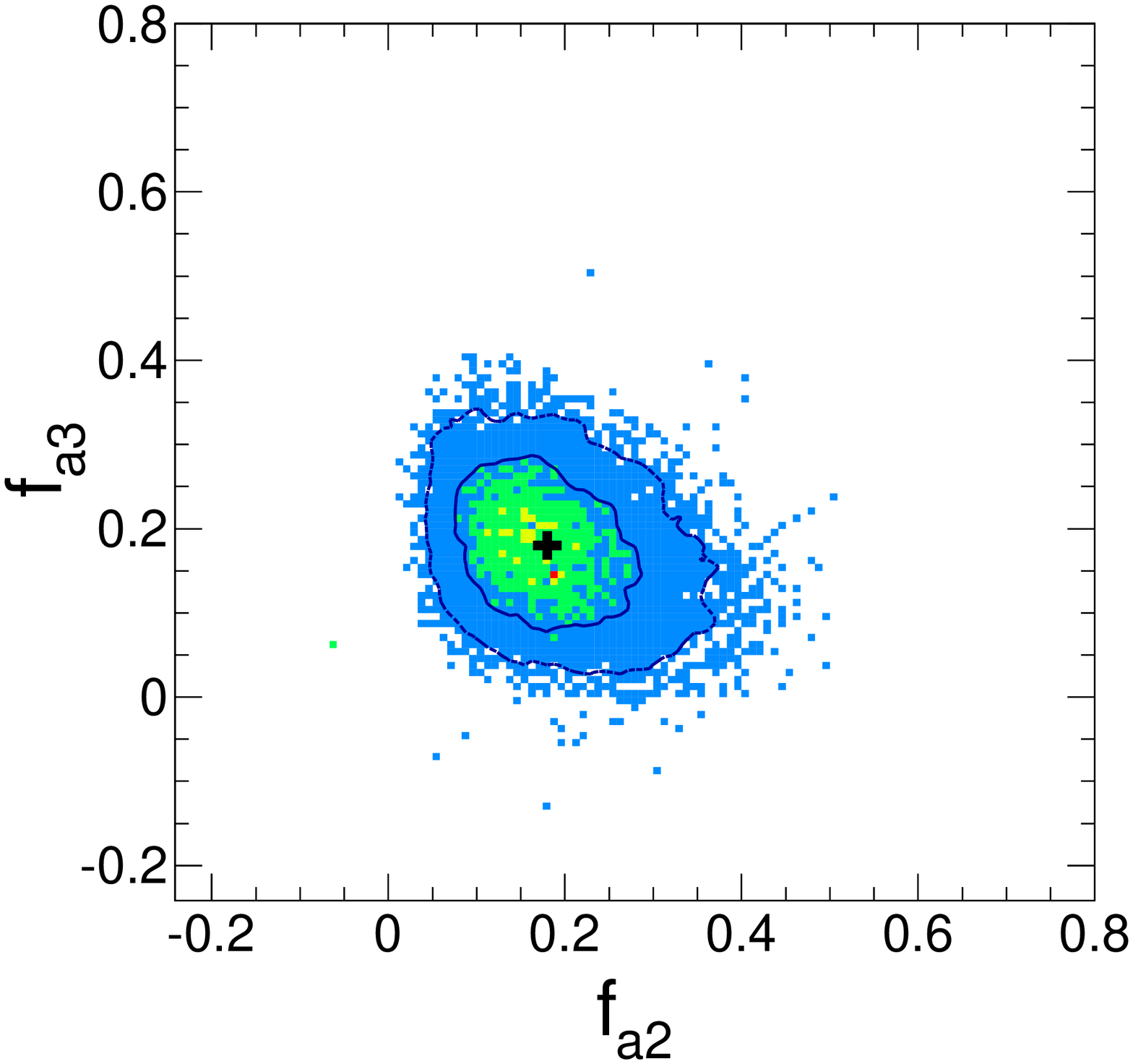}
}
\caption{
Simultaneous fit of $f_{a3}$ and $f_{a2}$ with 68\% and 95\% confidence level contours shown.
Left plot: 7D fit with 300 fb$^{-1}$ scenario.
Right plot: 3D fit with background and detector effects not considered, see text for details. 
Negative values of $f_{a3}$ and $f_{a2}$ correspond to $\phi_{a3}=\pi$ and $\phi_{a2}=\pi$, respectively.
}
\label{fig:lhc_fa3_2}
\end{figure}

Figure~\ref{fig:lhc_fa3} illustrates precision on 
$f_{a3}$ that can be achieved when both $f_{a3}$ and $\phi_{a3}$ are allowed to float in the multi-parameter fit
with seven observables. 
We ignore potential $q^2$-dependence of the couplings in this study due to a small $q^2$ range in $H\to ZZ^*$ process,
but later we examine one such example. 
The generated values for $f_{a3}=0.18$ (0.06) at 300 (3000)~fb$^{-1}$
are about three standard deviations away from zero. 
A similar approach is taken for precision in the $f_{a2}$ measurement, where 
for illustrative purpose we study the $\phi_{a2}=0$ case. These results are summarized in Table~\ref{table:sigmas}. 
We also show that both $f_{a2}$ and $f_{a3}$ could be measured simultaneously, see Fig.~\ref{fig:lhc_fa3_2} (left).
Overall, the expected precision on $f_{a3}$ is 0.06 (0.02) with 300 (3000) fb$^{-1}$ at the  LHC, 
which is in good agreement with similar studies performed by CMS~\cite{CMS:2013xfa}.
The expected precision on $f_{a2}$ is comparable, but it more strongly depends on the phase $\phi_{a2}$
than in the case of $f_{a3}$ measurement.

\begin{figure}[t]
\centerline{
\setlength{\epsfxsize}{0.33\linewidth}\leavevmode\epsfbox{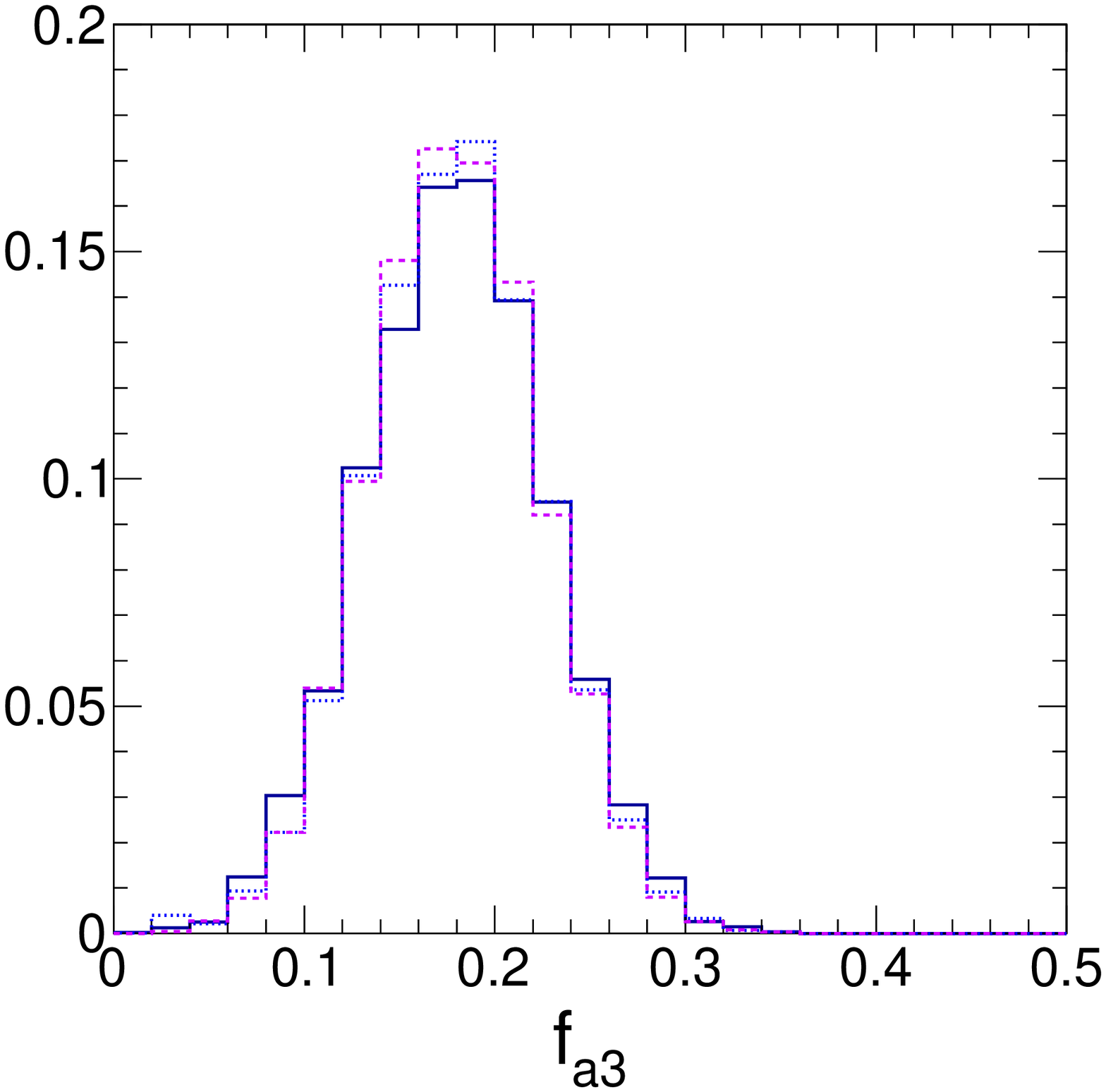}
\setlength{\epsfxsize}{0.33\linewidth}\leavevmode\epsfbox{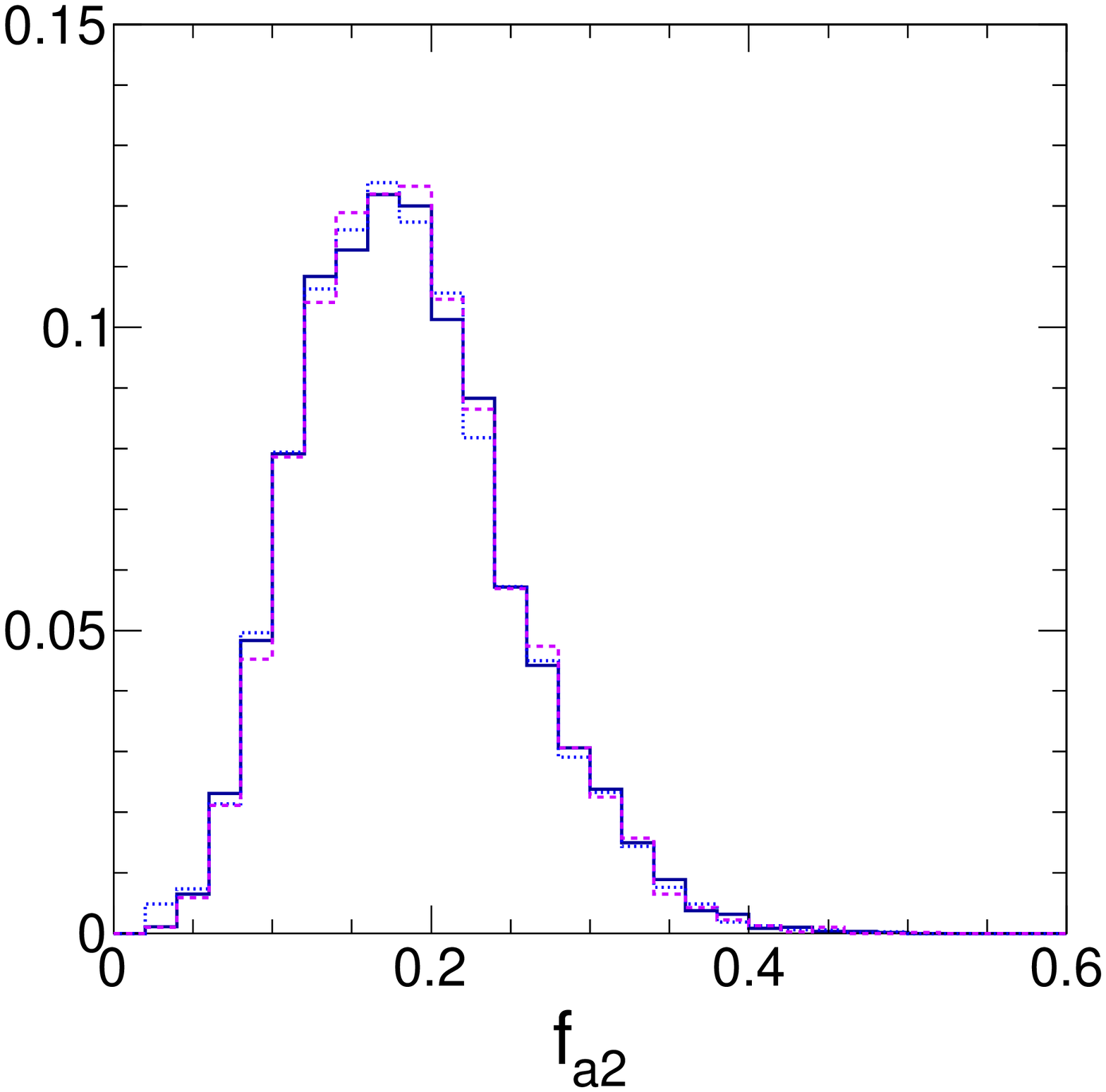}
\setlength{\epsfxsize}{0.33\linewidth}\leavevmode\epsfbox{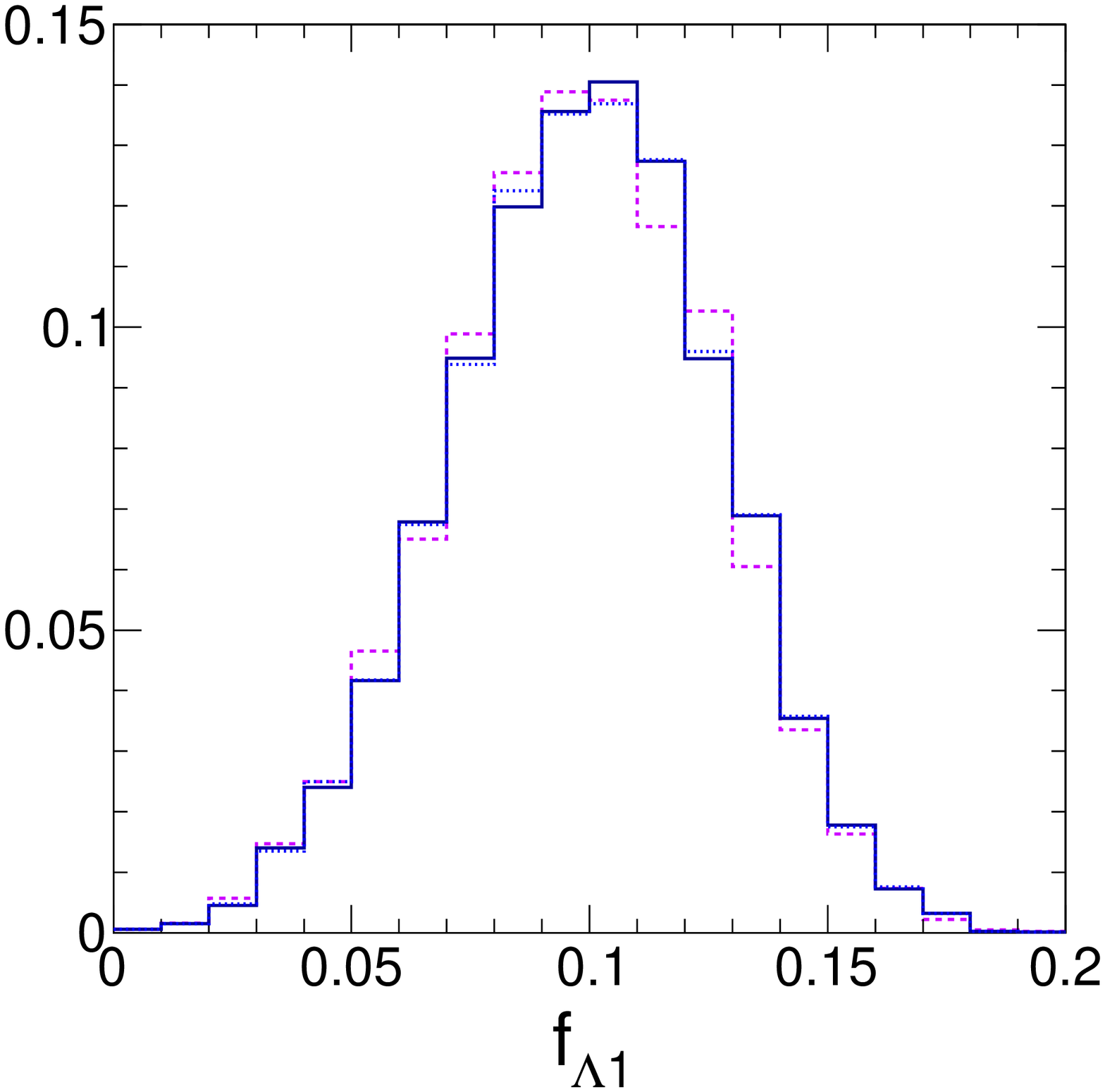}
}
\caption{
Distribution of fitted values  of $f_{a3}$ (left), $f_{a2}$ (middle), and $f_{\Lambda1}$ (right)  in a large number of generated 
experiments in the $H\to ZZ^*\to 4\ell$ channel with 300 fb$^{-1}$ of data collected at the  LHC,
with background and detector effects not considered.
Results of a 1D (solid black), 2D (dotted blue), and 7D (dashed magenta) fits are shown, see text for details. 
}
\label{fig:lhc_fa3_ND}
\end{figure}

To study certain features of the multi-dimensional distributions, no background or 
acceptance effects were included for simplicity of the presentation.  
We do this, in particular, when we show results of the $f_{a3}$ fits obtained in three different ways -- 
one-dimensional fit of $D_{0^-}$, two-dimensional fit of $D_{0^-}$ and $D_{C\!P}$,
and multi-dimensional fit of seven angular and mass observables.
Figure~\ref{fig:lhc_fa3_ND}  shows  results of these fits assuming the 
300 fb$^{-1}$ luminosity at the LHC. The events were generated with $f_{a3}=0.18$.
These studies are performed with a constraint that the coupling phases are real, but
we find $f_{a3}$ precision to be essentially the same if $\phi_{a3}$ is either floated or
constrained in the 7D fit provided, of course, that the number of events is sufficiently high. 
The two-dimensional fit recovers the precision of the 7D fit as the full information relevant for the yield and
interference measurement of the two components is retained.
When the one-dimensional fit of $D_{0^-}$ is employed the  precision of the $f_{a3}$ measurement gets 
worse by about 4\% with $f_{a3}=0.18$ (3$\sigma$ observation at 300 fb$^{-1}$), 
13\% with $f_{a3}=0.06$ (3000 fb$^{-1}$) and  30\% with $f_{a3}=0.02$ (30000 fb$^{-1}$). 
This again illustrates our assessment that interference effects are important to include
when non-zero $C\!P$ contribution is observed {\it but} that they are not the primary 
drivers of the discovery of $C\!P$  violation in $HVV$ interactions with available statistics. 

In Fig.~\ref{fig:lhc_fa3_ND}, a similar study is presented for the measurement of either $f_{a2}$ or $f_{\Lambda1}$. 
In all cases, either a 7D fit is performed, or a 1D fit (with $D_{0_h^+}$ or $D_{\Lambda1}$), or a 2D fit 
(with additional interference discriminant $D_{\rm int}$ optimal for each interference case). 
We find that 1D fits recover the precision of a 7D fit in both of these cases. 
In Fig.~\ref{fig:lhc_fa3_2} (right), we also illustrate the 3D analysis with the discriminants $D_{0^-}, D_{0_h^+}, D_{C\!P}$. 
We find that the three listed discriminants are sufficient to recover precision of the 7D fit with tested statistics. 
In this study we allow negative values of $f_{a2}$ and $f_{a3}$ to incorporate the phase information 
$\phi_{a2,3}=0$ or $\pi$ as $f_{a2}\times\cos(\phi_{a2})$ and $f_{a3}\times\cos(\phi_{a3})$.
The 2D fit with $D_{0^-}, D_{0_h^+}$ is also close in precision to the 7D fit and is not sensitive to $\phi_{a3}$.

We also note that similar techniques can be applied to the decays $H\to WW\to 2\ell2\nu$, 
as demonstrated in Ref.~\cite{Bolognesi:2012mm}, and $H\to Z\gamma\to 2\ell\gamma$, 
as demonstrated in Appendix~\ref{sect:me}. However, only partial polarization information is available 
in those channels. Moreover, any decay mode  can be studied at a  lepton collider.
However, since a typical  lepton collider has the advantage in associated production mode,
only such mode is presented in this study.

\subsection{The VBF process on LHC}
\label{sect:VBF_LHC}

We illustrate analysis of the weak boson fusion process  considering two decays of the Higgs boson, $H\to ZZ^*$ and $H\to\gamma\gamma$. In both cases, 
two high transverse momentum jets are required. Yields of signal events are summarized
in Table~\ref{table:XS}. The $f_{\rm jet}$ parameter indicates the fraction of events with two jets. 
We ignore the $VH$ production of Higgs bosons in this analysis since  
it can be isolated from the WBF events by applying constraints on the invariant mass of the two jets. We discuss 
$VH$ production in the next subsection. 

The gluon  fusion production of a Higgs boson contaminates WBF sample significantly  and is treated 
as a background. As shown below, $C\!P$ properties of  events produced in gluon fusion 
do not affect their kinematics strongly; this allows us to use  
the SM  predictions for $pp \to H+2j$   in the background studies. 
The other background originates from di-boson production with 
associated jets $ZZ(\gamma\gamma)$ + 2 jets and is modeled explicitly in the analysis. Selection requirements
follow closely those suggested by the ATLAS and CMS collaborations~\cite{properties-cms,properties-atlas}. 
In the analysis of the $H\to\gamma\gamma$ channel, additional requirements are applied on the dijet invariant mass 
$m_{jj}>350$ GeV and pseudorapidity difference $\Delta\eta_{jj}>3.5$, to improve the purity of the WBF signal. 
This leads to an additional WBF signal suppression by  a factor 0.6 with respect to 
that quoted in Table~\ref{table:XS}.
The ratio of gluon fusion  and weak boson fusion  events is 0.42 and the ratio of di-boson + 2 jets and weak boson fusion  
events is 4.7 in the $H\to\gamma\gamma$ channel.
The same ratios in the VBF $H\to ZZ^*$ channel are 2.2 and 0.7, respectively. 

\begin{figure}[t]
\centerline{
\setlength{\epsfxsize}{0.33\linewidth}\leavevmode\epsfbox{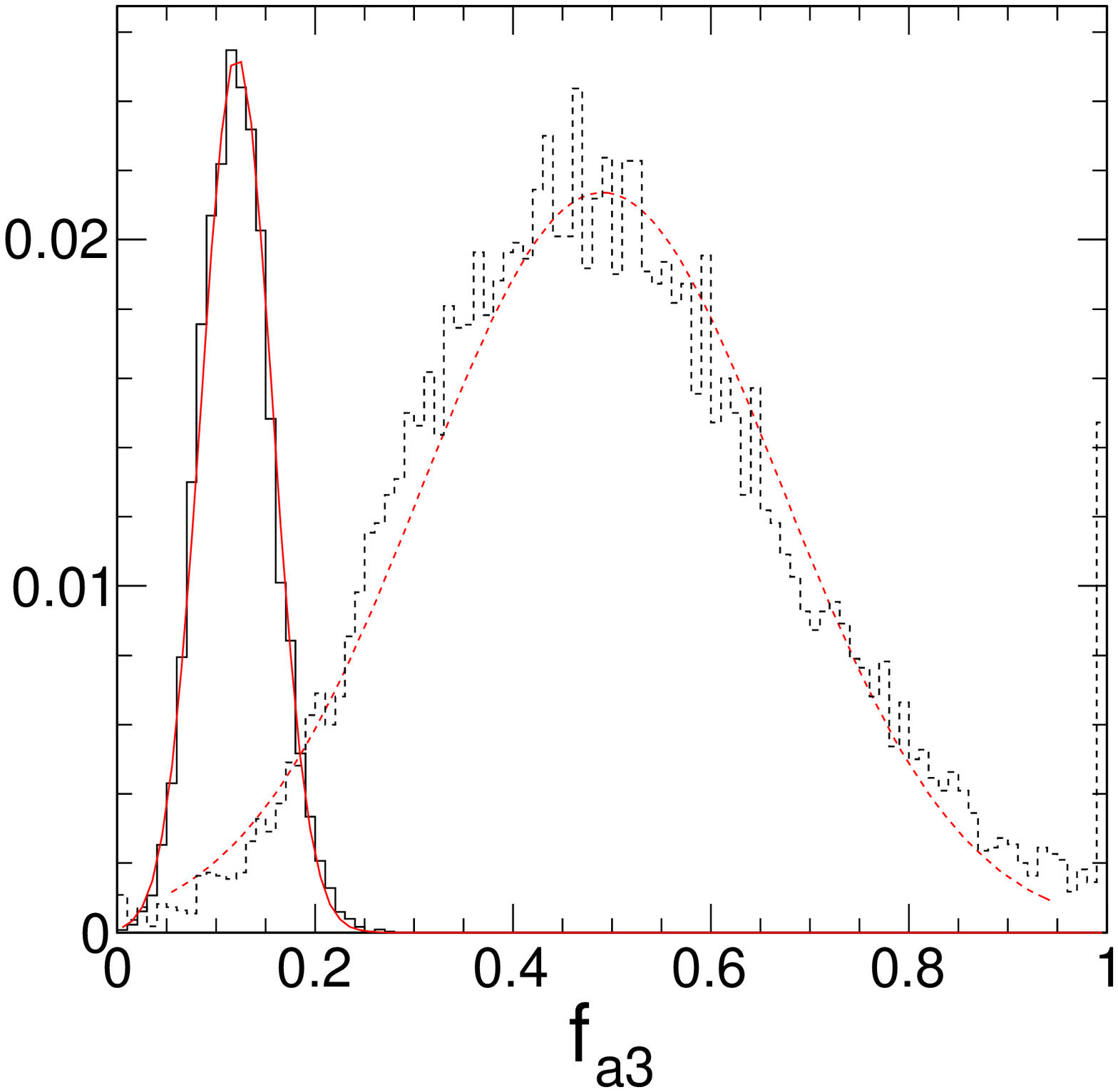}
\setlength{\epsfxsize}{0.33\linewidth}\leavevmode\epsfbox{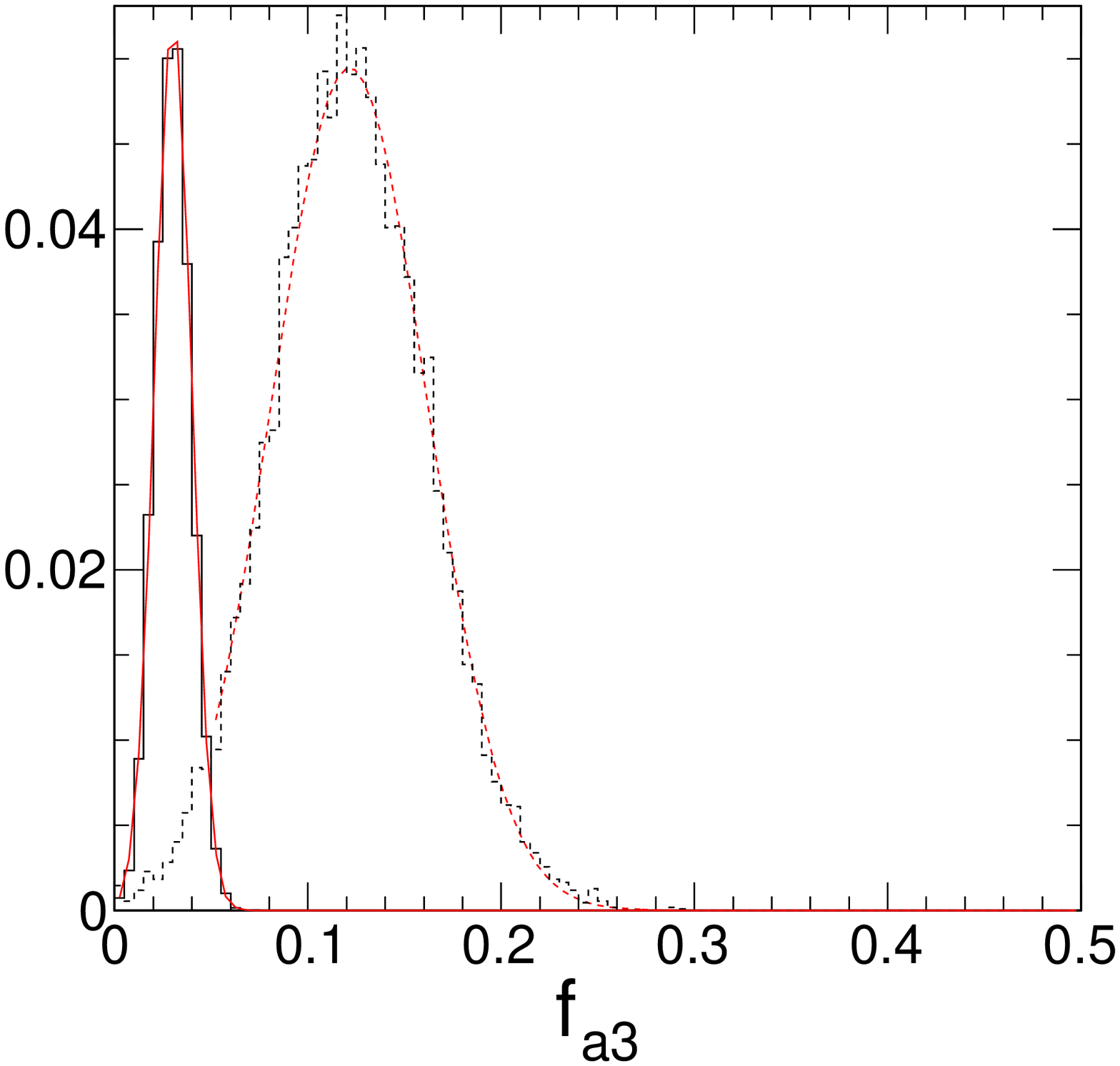}
\setlength{\epsfxsize}{0.33\linewidth}\leavevmode\epsfbox{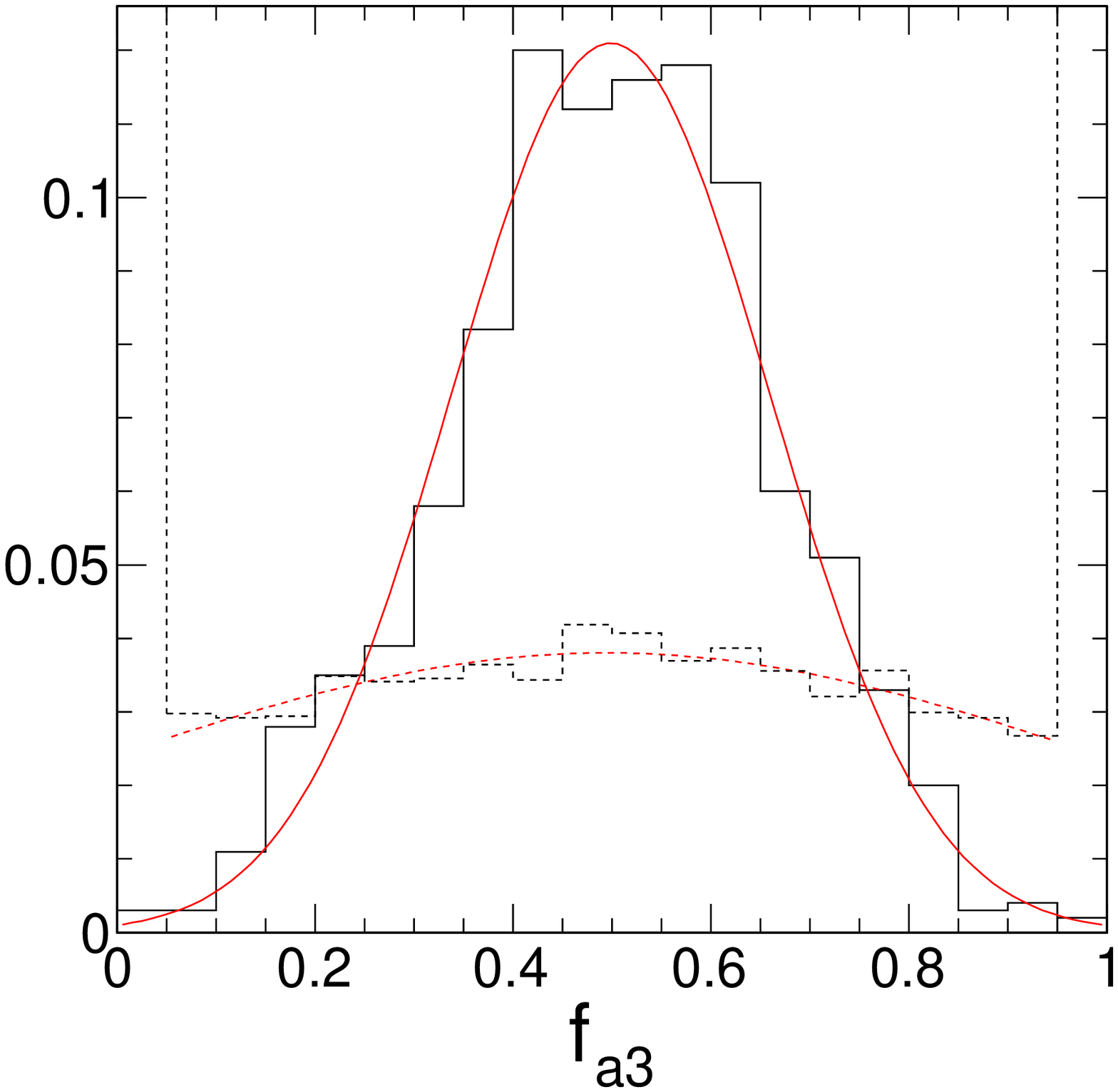}
}
\caption{
Distribution of fitted values  of $f_{a3}$  in a large number of generated experiments
in the weak boson fusion with $H\to ZZ^*$ (left) and $H\to\gamma\gamma$ (middle),  
and strong boson fusion with $H\to ZZ^*$ (right) channels 
with 300 fb$^{-1}$ (dotted) and 3000 fb$^{-1}$ (solid) of data collected at LHC.
}
\label{fig:vbf_fa3}
\end{figure}

Analysis is performed with the 
two discriminants $\vec{x}_{i}=({D}_{0^-}, {D}_{\rm bkg})$, as discussed in Appendix~\ref{sect:statistics}.
The ${D}_{0^-}$ discriminant is sensitive to ratios of 
scalar to pseudoscalar components in the $HVV$ vertex 
and is based on numerical matrix elements for
two types of signals.  The ${D}_{\rm bkg}$ discriminant is constructed 
to facilitate  signal-to-background separation, 
where signal is represented by the scalar
weak boson fusion  matrix element, and background is represented by the scalar $H+2j$  matrix element. 
Results of one-parameter fits of $f_{a3}$ in both topologies are shown in Fig.~\ref{fig:vbf_fa3}
and presented in Table~\ref{table:sigmas}. The $H\to ZZ$ channel is cleaner, but the $H\to\gamma\gamma$ channel
provides higher statistics and, as a result, it has about three times better precision for the same collected luminosity. 
The ultimate precision on $f_{a3}$ is in general comparable to that achieved in $H\to ZZ$ decay. However, due to large off-shell mass
of the $V^*$ in production, this translates to a substantially better precision on $f_{a3}^{\rm dec}$ of
$1.3\times 10^{-4}$ with 3000~fb$^{-1}$.

It is interesting to reverse the analysis and search for $C\!P$ violation in the gluon fusion 
 production process. Since the
selection requirements in the $H\to\gamma\gamma$ channel suppress gluon fusion  production significantly, we investigate the 
feasibility of this measurement in the cleaner $H\to ZZ^*$ channel. The ${D}_{\rm bkg}$ discriminant remains
the same, but it now serves the purpose to separate $H+2$ jets signal from the SM weak boson fusion contamination. 
The ${D}_{0^-}$ discriminant provides  separation between production of scalar and pseudoscalar Higgs 
in gluon fusion events, based on the corresponding
matrix elements. Results of this study are also shown in Fig.~\ref{fig:vbf_fa3} and Table~\ref{table:sigmas}.
With 3000~fb$^{-1}$, the precision on $f_{a3}$ is about 0.16, while with 300~fb$^{-1}$ the precision is about 0.5.

An important consideration in the high-luminosity scenario of the LHC is a very high number of multiple proton-proton
interactions per collision, leading to so-called pileup events. The pileup results 
in a very large number 
of relatively low $p_T$ jets 
from multiple interactions which could fake a 
signal. There are detector design considerations which may improve suppression
of such jets in data analysis. However, for the purpose of this study we mitigate the effects of increased pileup in the
3000~fb$^{-1}$ scenario by imagining that low-$p_T$ jets cannot be reconstructed and 
by increasing $p_T$ threshold   for reconstructed  jets to $50$ GeV. As a consequence, the uncertainty on $f_{a3}$ 
increases by 17\% and 40\% in the channels $V^*V^*\to H$ and $gg\to H+2$~jets with $H\to ZZ^*$, respectively,
while there is no noticeable change in $V^*V^*\to H\to\gamma\gamma$ due to tighter selection requirements. 
The changes are not dramatic and could be offset by other improvements in 
analyses, such as addition of other modes. 

We also note that the VBF process at a lepton collider $e^+e^-\to e^+e^-Z^*Z^*\to e^+e^-H$ can be studied with the 
same techniques as discussed here for the  LHC. This channel will in fact dominate over the $e^+e^-\to ZH\to \ell^+\ell^-H$ 
process at high energies, see Table~\ref{table:XS}. However, the $q^2$ range of the virtual $V^*$ bosons in a VBF
process depends only weakly on the collider energy and therefore we do not expect increased sensitivity to 
$f_{a3}^{\rm dec}$ as observed in the $e^+e^-\to ZH$ production process. We therefore do not study this channel 
in this paper and leave it to future work.


\subsection{The $q\bar{q}^\prime\to VH$ process on LHC}
\label{sect:VH_LHC}

We illustrate analysis of $VH$ events using two processes, $pp\to ZH/WH\to (q\bar{q}^\prime)(ZZ^*)$ 
and  $pp\to ZH\to (\ell\ell)(b\bar{b})$. In the first case, the final state is identical to the one in WBF  analysis,
described  in Sec.~\ref{sect:VBF_LHC}. Discussion of major background contributions  can be found there. 
The distinguishing feature of the $ZH/WH$ signal is the peak in the $Z/W\to2$ jets invariant mass $m_{jj}$ distribution
whose width is dominated by detector resolution. Therefore, we separate the $m_{jj}$ probability distribution 
from  the signal description and parameterize it with an empirical Gaussian function. The rest of the matrix element
squared is parameterized analytically as a function of $(m_{V\!H}, \cos\theta_1, \cos\theta_2, \Phi, Y)$
using Eq.~(\ref{fig:lhc_vhangular}). We find kinematics of the $ZH$ and $WH$ events to be 
essentially identical, except for the small shift in $m_{jj}$. Therefore, the results are obtained by 
combining the $ZH/WH$ channels under a single topology using the $ZH$ model.
Similarly to the VBF case described in the previous subsection, we perform a two-dimensional fit with the 
discriminant  $\vec{x}_{i}=({D}_{0^-}, {D}_{\rm bkg})$.

To discuss  $pp\to ZH\to (\ell\ell)(b\bar{b})$ case, we estimate signal and background yields following 
ATLAS and CMS selection requirements~\cite{properties-cms, properties-atlas}. 
The expected number of signal events is shown  in Table~\ref{table:XS}. 
To suppress otherwise overwhelming background, we require large transverse momentum of the
Higgs boson $p_{T, H} >200$ GeV, see Fig.~\ref{fig:pT}.
This, combined with other selection requirements
of the $Z\to\ell\ell$ and $H\to b\bar{b}$, leads to about 0.7\% reconstruction efficiency. 
The dominant background is from $Z+$jets, which we take to be 5 times the size of signal with the
above selection, but we approximate its shapes with $pp\to ZZ\to (\ell\ell)(b\bar{b})$ simulation.
Approximate modeling of broad kinematic distributions of background does not affect separation between two types of signal. 
Analysis is performed  in a narrow mass window of the $b\bar{b}$ invariant mass with a 
1D $\vec{x}_{i}=({D}_{0^-})$ parameterization using Eq.~(\ref{fig:lhc_vhangular})
for probability calculations. 

\begin{figure}[t]
\centerline{
\setlength{\epsfxsize}{0.33\linewidth}\leavevmode\epsfbox{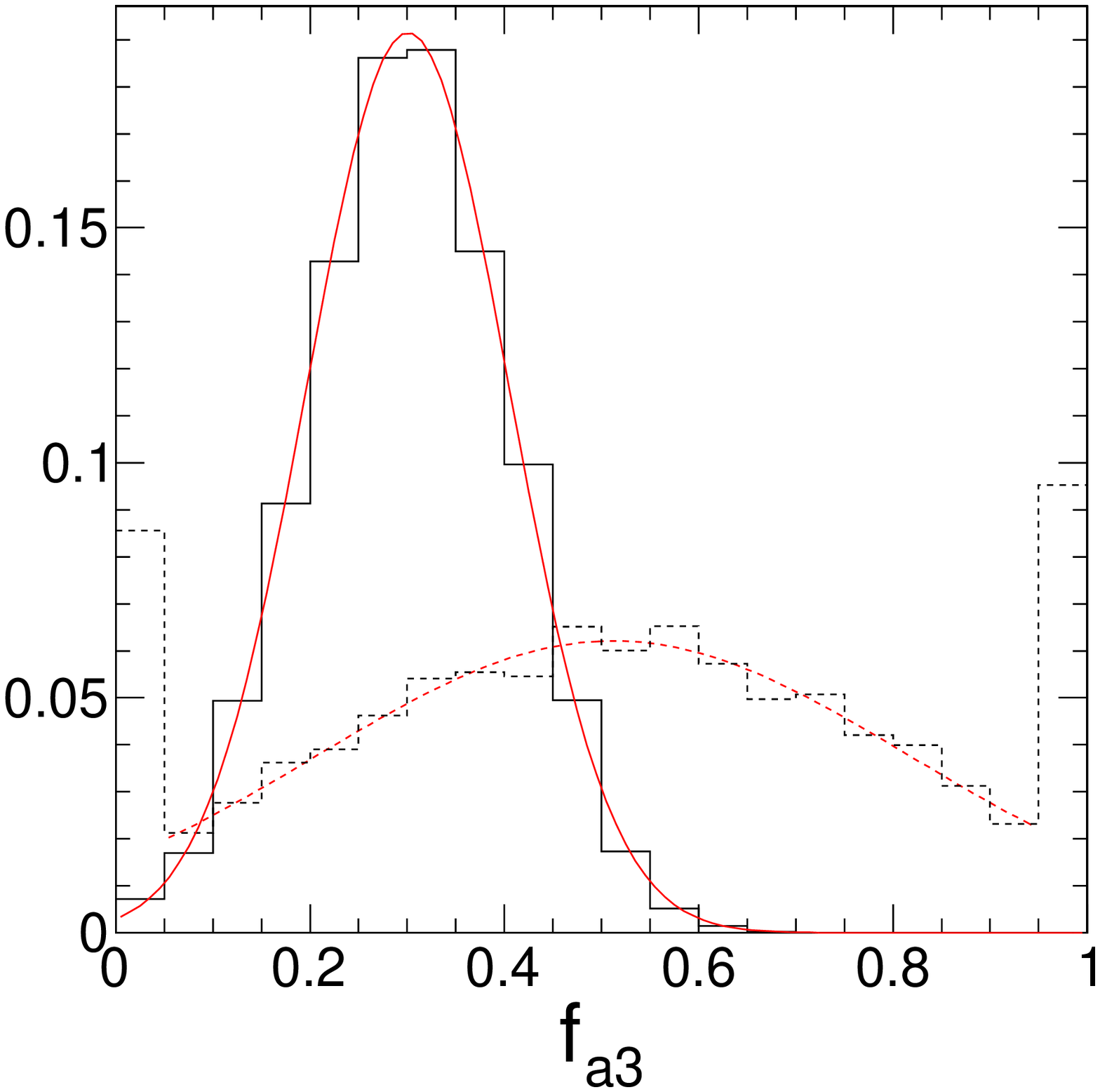}
\setlength{\epsfxsize}{0.33\linewidth}\leavevmode\epsfbox{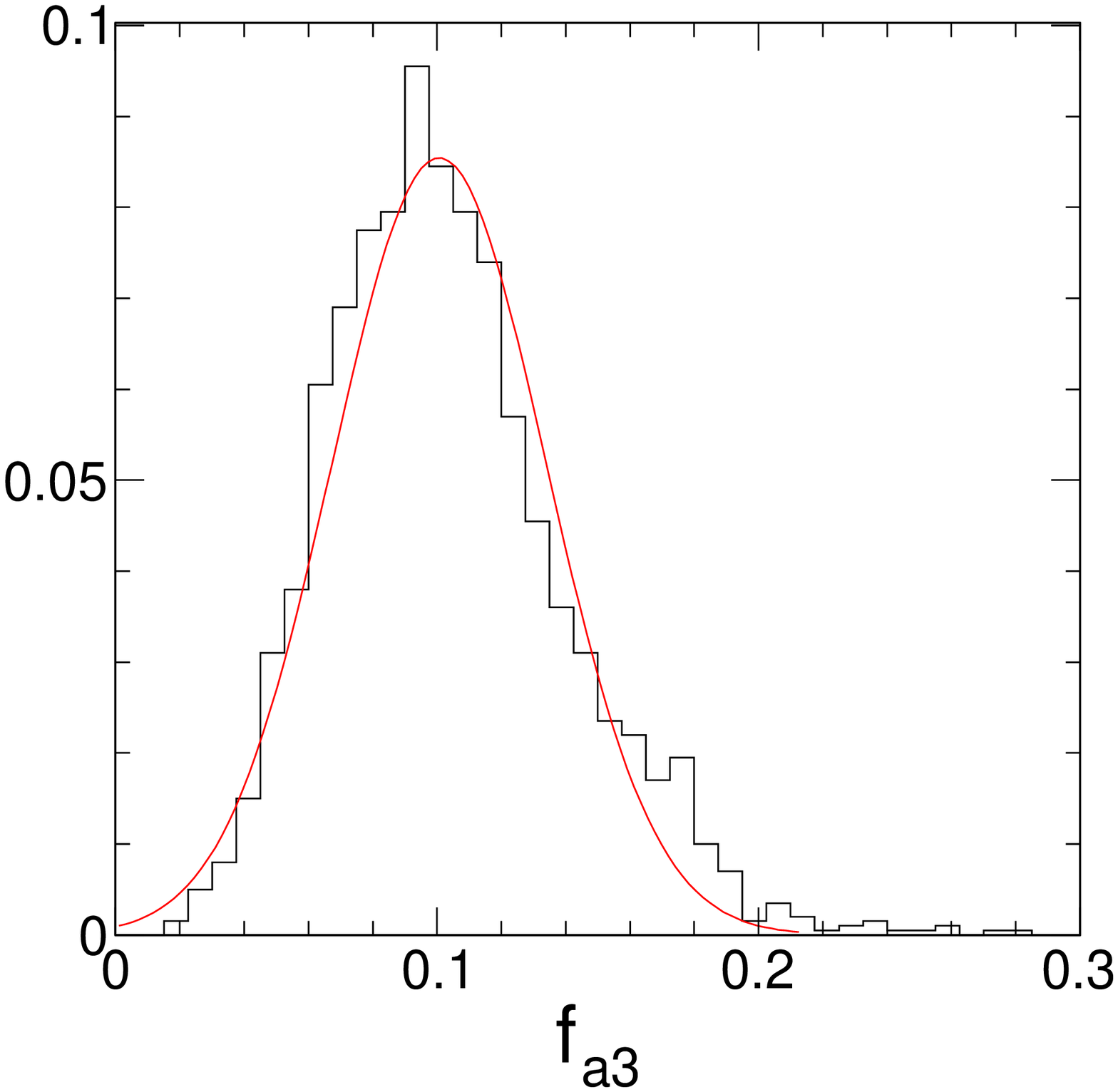}
}
\caption{
Distribution of fitted values of $f_{a3}$  in a large number of generated experiments in the channels
$pp\to ZH/WH\to (q\bar{q}^\prime)(ZZ^*)$ (left) and  $pp\to ZH\to (\ell\ell)(b\bar{b})$ (right)
with 300 fb$^{-1}$ (dotted) and 3000 fb$^{-1}$ (solid) of data collected at the  LHC.
}
\label{fig:vh_fa3}
\end{figure}

Results of one-parameter fits of $f_{a3}$ using each of the two processes discussed above 
 are shown in Fig.~\ref{fig:vh_fa3}
and presented in Table~\ref{table:sigmas}.
The conclusion is very similar to the VBF topology study. The $H\to ZZ$ channel is cleaner, but the $H\to b\bar{b}$ channel
provides higher statistics and as a result three times better precision for the same collected luminosity. 
The ultimate precision on $f_{a3}$
is in general comparable to that achieved in $H\to ZZ$ decay. However, due to large off-shell mass
of the $Z^*$ in production, this translates to a substantially better precision on $f_{a3}^{\rm dec}$ defined in decay,
$1.2\times 10^{-4}$ with 3000~fb$^{-1}$, similar to the expectation in the VBF channel.
We mitigate the effects of increased pileup in the 3000~fb$^{-1}$ scenario by increasing thresholds 
of jet $p_T>50$~GeV, which leads to about a factor of two degradation in precision in the $H\to ZZ$ channel.
We note that the $H\to b\bar{b}$ channel has tighter selection requirements and
could also benefit from jet substructure techniques~\cite{Butterworth:2008iy}.

We also note that while we considered only the $pp\to ZH\to (\ell\ell)(b\bar{b})$ channel in the $H\to b\bar{b}$ final state,
the technique is directly applicable to the $pp\to WH\to (\ell\nu)(b\bar{b})$ and $pp\to ZH\to (\nu\bar\nu)(b\bar{b})$ channels. 
In the  $pp\to WH\to (\ell\nu)(b\bar{b})$ case, the $\nu$ can be reconstructed as a missing transverse energy with the $W$
mass constraint and a two-fold ambiguity only remaining. Therefore, the full matrix element can be used. This technique 
can be used in the $Z\to\nu\bar\nu$ case where the $Z$ can be reconstructed as missing transverse energy as well, but some 
information is lost.



\section{Summary and Conclusions}
\label{sect:summary}

\begin{figure}[ht]
\centerline{
\setlength{\epsfxsize}{0.5\linewidth}\leavevmode\epsfbox{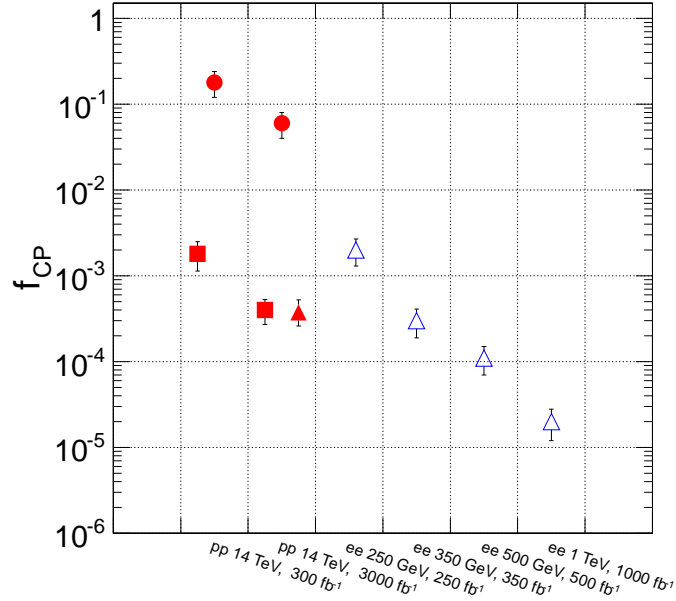}
}
\caption{
Summary of precision in $f_{C\!P}$  for $HVV$ couplings ($V=Z,W$) at the moment of  $3\sigma$ measurement.
Points indicate central values and error bars indicate $1\sigma$ deviations in the generated
experiments modeling different luminosity scenarios at proton (solid red) or $e^+e^-$ (open blue) colliders.
Measurements in three topologies $VH$ (triangles), WBF (squares), and decay $H\to VV$ (circles) are shown.
Different energy and luminosity scenarios are indicated on the $x$-axis.
}
\label{fig:cpscenarios}
\end{figure}

In summary, we have investigated the feasibility to measure anomalous couplings 
of the Higgs boson to electroweak gauge bosons and gluons, including $C\!P$-violating couplings. 
A coherent framework is presented to study these  anomalous couplings 
 in Higgs boson decays, vector boson fusion, or
associated production of a Higgs boson at either proton or lepton colliders. 
Both,  a Monte Carlo simulation program and a matrix element likelihood approach
are developed for these three types of processes. 
The expected sensitivity to the $f^{\rm dec}_{a3}$ parameter, defined as the ${C\!P}$-odd
cross section fraction in the decay to two vector bosons and which we will denote as $f_{C\!P}$ here,
is summarized in Table~\ref{table:cpscenarios} and Fig.~\ref{fig:cpscenarios}\footnote{
The measurement of $f_{C\!P}$ is independent of the coupling convention and therefore more convenient,
but it is equivalent to the measurement of $g_4/g_1$ coupling ratio. 
The translation between the two notations can be done using Eq.~(\ref{eq:fractions}) and comments below it. 
The translation between the $f_{a3}$ and $f^{\rm dec}_{a3}\equiv f_{C\!P}$ is not linear and may lead to asymmetric 
errors, from which we quote the uncertainty on the lower side. We omit the $VH$ point at the 300 fb$^{-1}$ LHC 
scenario because it does not quite reach the 3$\sigma$ threshold.
}.
At both the high-luminosity LHC and the first stage of the $e^+e^-$ collider, 
$ f_{C\!P}$ as small as $10^{-4}$ can be measured in the coupling to weak bosons ($W$ and $Z$). 
Higher precision seems to be achievable at a higher-energy
$e^+e^-$ collider, provided that $q^2$-dependence of effective couplings  does not yet lead 
to the suppression of non-renormalizable interactions. 

In the case of a parity-mixed $H$ state, 
the $f_{C\!P}$ value in either $H\to ZZ$ or $WW$ decay is expected to be small
since  the pseudoscalar coupling is loop-induced.  Therefore, values as small as $f_{C\!P}\sim 10^{-5}$
might be expected even in the case of sizable admixture of a pseudoscalar. As follows from 
Table~\ref{table:cpscenarios}, such small values cannot be measured  either at the LHC 
or at the initial-stage $e^+e^-$ collider, but expected precision is not out of
scale and interesting measurements could be achieved with higher luminosity and additional modes. 
Nonetheless, measuring $f_{C\!P}$ in couplings to massless vector bosons 
($gg,  \gamma\gamma, Z\gamma$) might be an interesting
alternative, since both scalar and pseudoscalar components are expected to be equally suppressed
by the loop effect, and $f_{C\!P}\sim 10^{-2}$ might be expected~\cite{Shu:2013uua}. 
We have tested the expected sensitivity to $f_{C\!P}$ in the $Hgg$ coupling at the  LHC. 
We found that kinematic features in the production of the Higgs boson in the association with jets  
are not strongly modified but interesting measurements could be made with sufficient statistics. 

Measuring $f_{C\!P}$ in the $H\to Z\gamma$ and $H \to \gamma\gamma$ modes at the  LHC is a challenge due to
their low branching fractions, and it is essentially impossible at an $e^+e^-$ collider. 
Measurements in the $H\to Z^*\gamma^*(\gamma^*\gamma^*)\to 4\ell$ process is also possible, 
but is challenging experimentally and requires high statistics. 
The $H\to\gamma\gamma$ final states does not allow measurement of ${C\!P}$ properties without the photon
polarization measurement. The latter could be measured in photon conversion in the detector, but this makes
the analysis very challenging and demands large statistics. Alternatively, there is a proposal for a photon
collider which could be built in association with a linear $e^+e^-$ collider and its strong feature is the ability
to collide polarized photons, with which ${C\!P}$ properties could be studied~\cite{Grzadkowski:1992sa}. 
Measuring polarization of the $Z$ in $H\to Z\gamma$ is not sufficient for ${C\!P}$ property measurements, 
unless there are complex phases in the couplings, see Appendix~\ref{sect:me}. 
Nonetheless, we provide the tools to study angular correlations in the $H\to Z\gamma$ process.

Finally, we comment on some further extensions of this analysis. First, similar measurements can be performed 
in $H\to WW^*$ decay mode. However, we have already shown~\cite{Bolognesi:2012mm}
that spin-zero coupling measurement is less precise in this channel compared to  $H\to ZZ^*$. 
Both decays could be studied at  the $e^+e^-$ collider, but the strongest feature of the $e^+e^-$ collider
is to measure these coupling in production, not in decay, due to larger statistics available and also due to
cross-section effects. Prospects for measuring anomalous couplings 
in the VBF process  $Z^*Z^*\to H$ at an  $e^+e^-$ collider are similar 
to what  we discussed at the  LHC. The number of events in this mode is in fact much larger than in the $Z^*\to ZH$ 
production mode with $Z\to\ell\ell$ at higher energies~\cite{sp6}, as shown in Table~\ref{table:XS},
but we do not expect enhanced sensitivity to $f_{C\!P}$ in this mode due to limited $q^2$ range for the virtual $Z^*$ bosons. 
We leave further studies in this mode to future work, while the tools will be very similar to those already employed in 
LHC studies shown here.

\begin{table}[t]
\begin{center}
\caption{
List of $f_{C\!P}$ values in $HVV$ couplings expected to be observed with $3\sigma$ significance
and the corresponding uncertainties $\delta f_{C\!P}$ for several collider scenarios,
with the exception of  $V^*\to V\!H$ mode at $pp$ 300 fb$^{-1}$ where the simulated measurement
does not quite reach 3$\sigma$.
Numerical estimates are given for the effective couplings  $Hgg$, $H\gamma\gamma$,  $HZ\gamma$,
$HZZ / HWW$, assuming custodial $Z/W$ symmetry and using $HZZ$ couplings as the reference. 
The $\checked$ mark indicates that a measurement is in principle possible but is not covered in this study.
}
\begin{tabular}{|ccc|cc|cc|cc|cc|c|c|c|}
\hline\hline
 &  &  &  \multicolumn{6}{c|}{ $HZZ / HWW$ } &   \multicolumn{2}{c|}{ $Hgg$} &  $HZ\gamma$  &    \multicolumn{2}{c|}{ $H\gamma\gamma$ } \\
\hline
collider & energy & ${\cal L}$ &  \multicolumn{2}{c|}{ $H\to VV^*$ } &  \multicolumn{2}{c|}{ $V^*\to V\!H$}  &  \multicolumn{2}{c|}{ $V^*V^*\to H$} 
                                &  \multicolumn{2}{c|}{  $gg\to H$}  &   $H\to Z\gamma$ & $\gamma\gamma\to H$ &  $H\to\gamma\gamma$  \\
       &  GeV   & fb$^{-1}$ &  $f_{C\!P}$ & $\delta f_{C\!P}$ &  $f_{C\!P}$ & $\delta f_{C\!P}$ &  $f_{C\!P}$ & $\delta f_{C\!P}$ &  $f_{C\!P}$ & $\delta f_{C\!P}$ &   &  &   \\
\hline 
$pp$ & 14\,000   & 300 & 0.18 & 0.06 & $ 6\times\!10^{-4}$ & $4\times\!10^{-4}$ &  $18\times\!10^{-4}$ & $7\times\!10^{-4}$ & -- & 0.50   & ~~ & ~~  & ~~   \\
$pp$ & 14\,000   & 3\,000 & 0.06 & 0.02 & $3.7\times\!10^{-4}$ & $1.2\times\!10^{-4}$ &  $4.1\times\!10^{-4}$ & $1.3\times\!10^{-4}$ & 0.50& 0.16   & \checked  & ~~  &  \checked \\
$e^+e^-$ & 250   & 250 &  \multicolumn{2}{c|}{ \checked}  & $21\times\!10^{-4}$ & $7\times\!10^{-4}$ &   \multicolumn{2}{c|}{ \checked}   &  &   & ~~ & ~~  & ~~   \\
$e^+e^-$ & 350   & 350 &  \multicolumn{2}{c|}{ \checked}  & $3.4\times\!10^{-4}$ & $1.1\times\!10^{-4}$ &   \multicolumn{2}{c|}{ \checked}   &  &   & ~~ & ~~  & ~~   \\
$e^+e^-$ & 500   & 500 &  \multicolumn{2}{c|}{ \checked}  & $11\times\!10^{-5}$ & $4\times\!10^{-5}$ &   \multicolumn{2}{c|}{ \checked}   &  &   & ~~ & ~~  & ~~   \\
$e^+e^-$ & 1\,000   & 1\,000 &  \multicolumn{2}{c|}{ \checked}  & $20\times\!10^{-6}$ & $8\times\!10^{-6}$ &   \multicolumn{2}{c|}{ \checked}   &  &   & ~~ & ~~  & ~~   \\
$\gamma\gamma$  & 125 & & \multicolumn{2}{c|}{ \checked} &&&&&&&&  \checked & \\
\hline\hline
\end{tabular}
\label{table:cpscenarios}
\end{center}
\end{table}

\bigskip

\noindent
{\bf Acknowledgments}:
We would like to acknowledge the long-term planning exercise for the U.S. high-energy physics community,
also known as ``Snowmass," from which this study emerged~\cite{Dawson:2013bba}.
We would like to thank Snowmass participants and CMS collaboration colleagues for feedback,
and in particular Michael Peskin and Tao Han for encouragement of the $e^+e^-$ studies and
Serguei Ganjour for discussion of the $\gamma\gamma$ channel on LHC.
We acknowledge contribution of our CMS collaboration colleagues to the MELA project development,
and in particular Meng Xiao for support of the MELA package. 
We are grateful to Jonathan Aguilar, Roberto Covarelli, Ben Kreis, Candice You, Xiaozhou Zhou
for help with the generator validation. 
We acknowledge significant contribution of Ulascan Sarica to development of statistical analysis tools. 
This research is partially supported by U.S. NSF under grants PHY-1100862 and PHY-1214000, 
and by U.S. DOE under grants DE-AC02-06CH11357 and DE-AC02-07CH11359.
We also acknowledge support from the LPC-CMS Fellows program operated through FNAL.
Calculations reported in this paper were performed on the Homewood High Performance Cluster 
of the Johns Hopkins University. 
%


\bigskip
\bigskip
\bigskip
\bigskip
\bigskip
\bigskip
\bigskip
\bigskip

\appendix 

\section{Event description with the matrix element likelihood approach (MELA)}
\label{sect:me}

The main  tool that we use in the analyses described in this paper is the likelihood method 
that employs expected probability distributions for various processes that can be used to 
measure anomalous Higgs boson couplings.  In this Appendix, we provide the necessary information for 
finding these probability distributions and give  a few examples of how they can be used. 

\subsection{The $H\to VV^*$ process}

We begin by describing the decay process $H\to VV \to 4f$, following 
notation of Refs.~\cite{Gao:2010qx, Bolognesi:2012mm}. This process is important  not only because 
it can be used directly to constrain anomalous couplings  but also because 
various crossings of $H \to VV$ amplitude  give amplitudes for 
associated Higgs boson production and vector boson fusion.
Complete description of the decay amplitude for $H \to VV^*$ requires two invariant masses and five angles, defined in Fig.~\ref{fig:decay}.   
We collectively denote these angles as $\vec\Omega=(\cos\theta^\ast,\Phi_1,\cos\theta_1,\cos\theta_2,\Phi)$.
The probability distribution that describes the decay of a Higgs boson to two gauge bosons $V$ is written as 
%
\begin{eqnarray}
 &&
  \frac{d\Gamma(m_1,m_2,\vec\Omega)}{dm_1\, dm_2\, d\vec\Omega} 
  \propto
    |\vec{p}_V(m_1,m_2)|
     \times\frac{m_1^3}{(m_1^2-m_{\sss V}^2)^2+m_{\sss V}^2\Gamma_{\sss V}^2}
     \times\frac{m_2^3}{(m_2^2-m_{\sss V}^2)^2+m_{\sss V}^2\Gamma_{\sss V}^2}
     \times \frac{d\Gamma(m_1,m_2,\vec\Omega)}{d\vec\Omega} \,,
\label{eq:differential-2}
\end{eqnarray}
%
\begin{figure}[t]
\centerline{
\setlength{\epsfxsize}{0.25\linewidth}\leavevmode\epsfbox{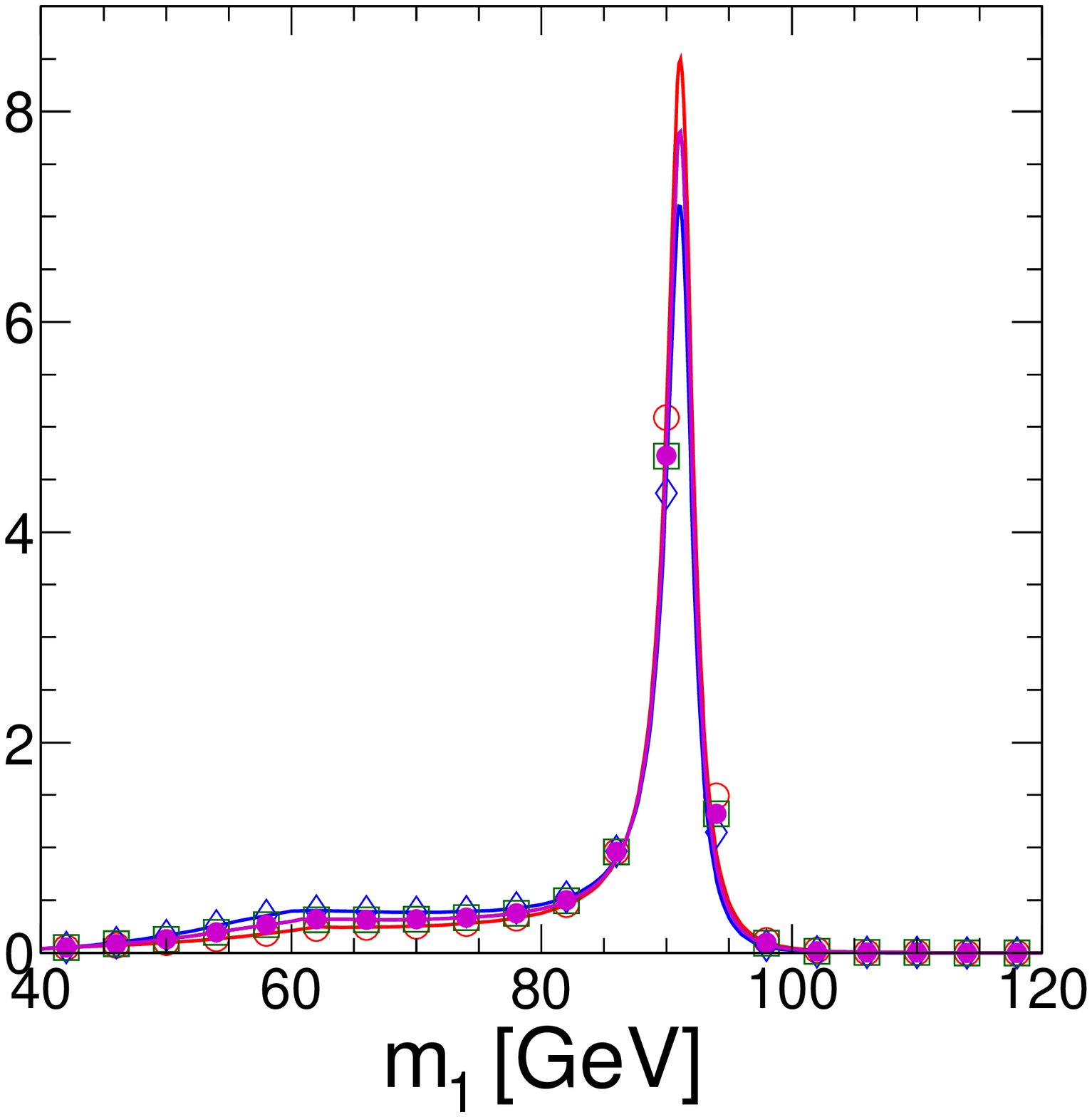}
\setlength{\epsfxsize}{0.25\linewidth}\leavevmode\epsfbox{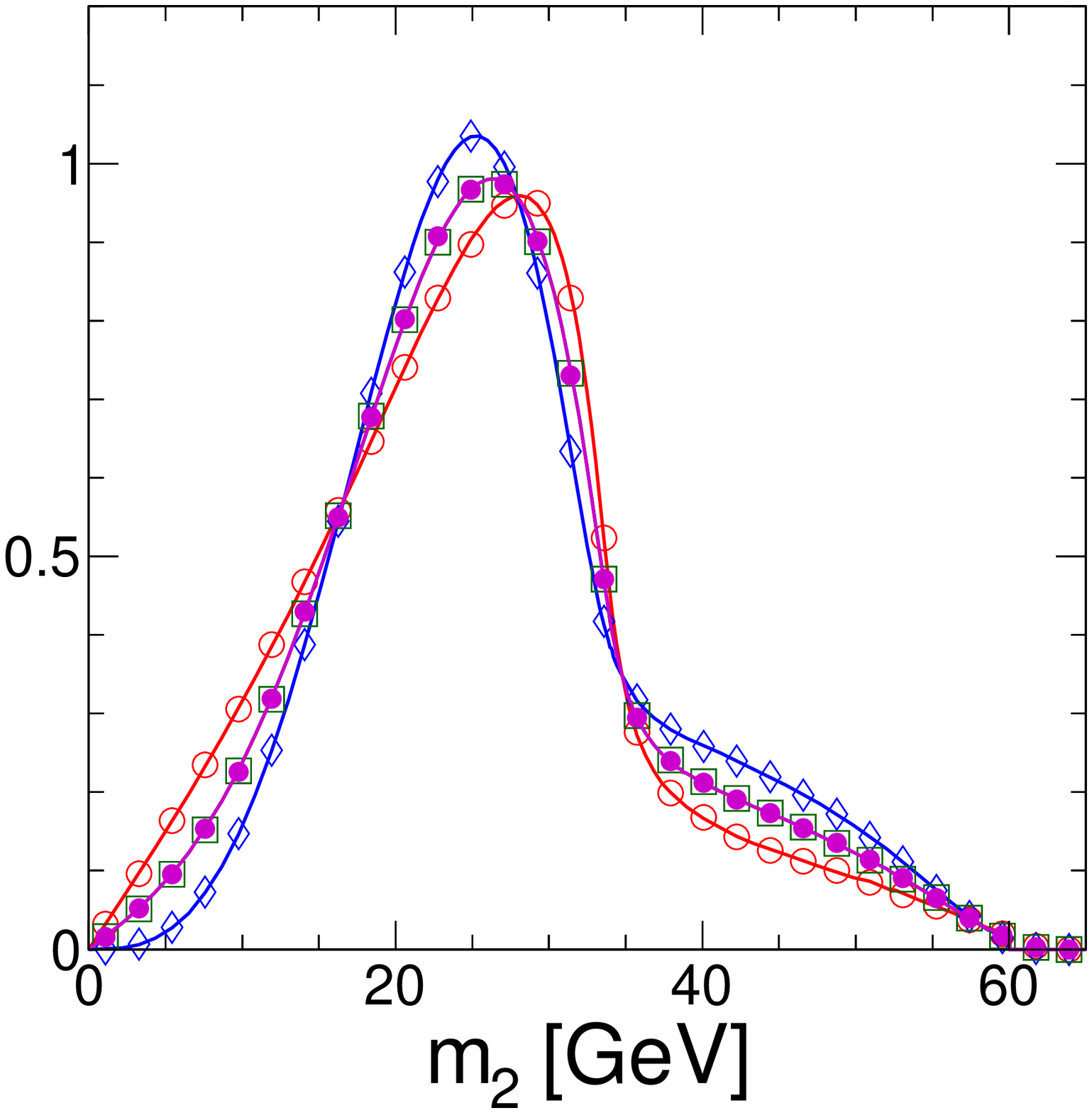}
\setlength{\epsfxsize}{0.25\linewidth}\leavevmode\epsfbox{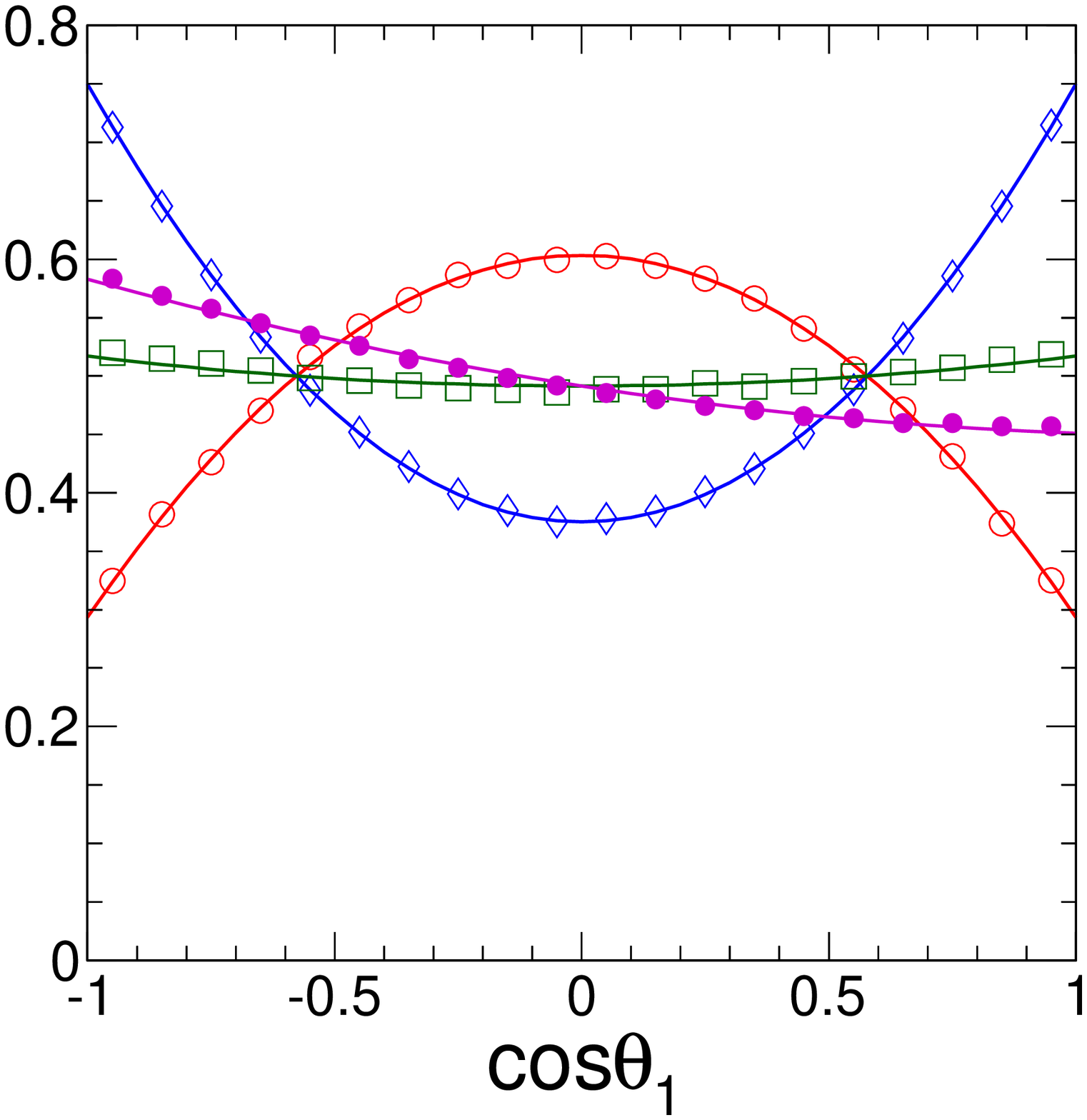}
\setlength{\epsfxsize}{0.25\linewidth}\leavevmode\epsfbox{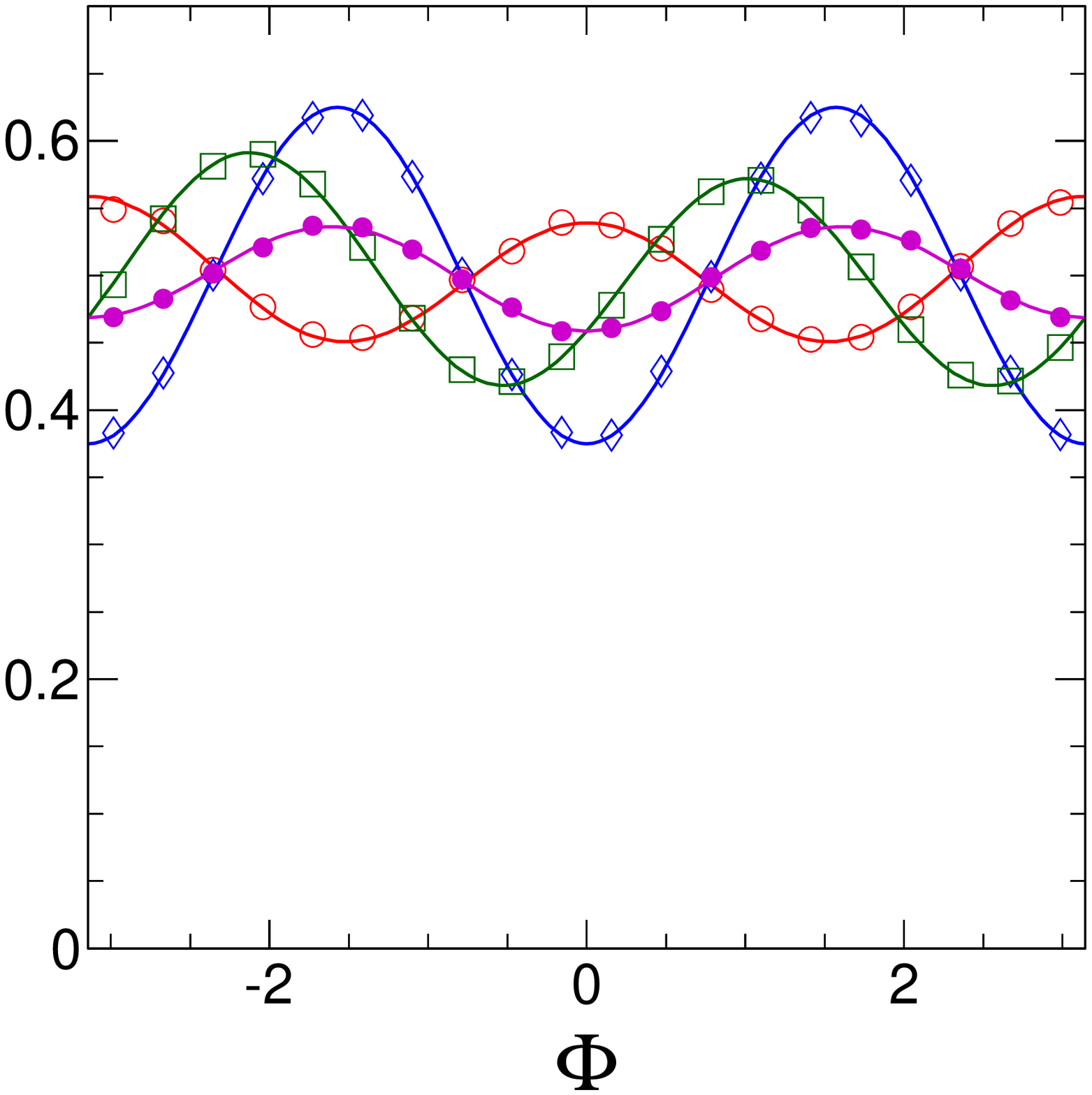}
}
\caption{
Distributions of the observables in the $H\to ZZ$ analysis, from left to right:
 $m_1$, $m_2$ (where $m_1>m_2$), $\cos\theta_1$ (same as $\cos\theta_2$), and $\Phi$.
Points show simulated events and lines show projections of analytical distributions.
Four scenarios are shown: SM ($0^+$, red open circles), pseudoscalar ($0^-$, blue diamonds), 
and two mixed states corresponding to $f_{a3}=0.5$ with $\phi_{a3}=0$ (green squares)
and $\pi/2$ (magenta points). 
For a spin-zero particle, distributions in  $\cos\theta^\ast$ and $\Phi_1$ are trivially flat, 
but this is not true  for  higher-spin states~\cite{Bolognesi:2012mm} or with detector effects.
}
\label{fig:lhc_angles}
\end{figure}
where the fully analytical expression for ${d\Gamma}/{d\vec\Omega}$ 
is given in Eq.~(A1) of Ref.~\cite{Bolognesi:2012mm},
and $\vec{p}_V$ is the $V$ boson momentum in $H$ rest frame. 
We show examples of kinematic distributions obtained for different types of tensor couplings 
in Fig.~\ref{fig:lhc_angles}.   Simulated events and projections of analytic distributions from Eq.~(\ref{eq:differential-2})
are compared there, illustrating an agreement between the two computations. 
Additional examples, including angular distributions for other spin hypotheses, can be found in Ref.~\cite{Bolognesi:2012mm}.
We note that lepton interference in the final states with identical leptons  changes the 
expected performance of the analysis by only a few percent. 
We therefore neglect this interference in the feasibility studies presented,
but provide the tools to take it into account~\cite{Bolognesi:2012mm,support}.
For example, lepton interference leads to variation of the fraction of the same-flavor four-lepton
events with respect to opposite-flavor events and this effect depends on the tensor structure of interactions.  
This interference is constructive in the Standard Model and destructive for a pseudoscalar decay.
Therefore, when $f_{a3}$ is defined, we use the $H\to ZZ^*\to 2e2\mu$ mode without lepton interference 
for the cross section calculations.

\begin{figure}[t]
\centerline{
\setlength{\epsfxsize}{0.25\linewidth}\leavevmode\epsfbox{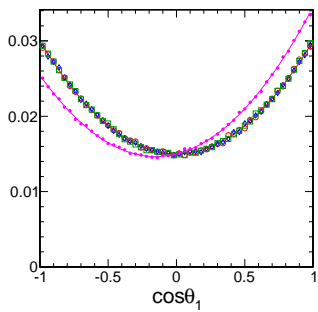}
}
\caption{
Distributions of the $\cos\theta_1$ observable in the $H\to Z\gamma$ analysis.
Four scenarios are shown: SM ($0^+$, red open circles), pseudoscalar ($0^-$, blue diamonds), 
and two mixed states corresponding to $f_{a3}=0.5$ with $\phi_{a3}=0$ (green squares)
and $\pi/2$ (magenta points). 
For a spin-zero particle, distributions in  $\cos\theta^\ast$ and $\Phi_1$ are trivially flat.
}
\label{fig:decay-zgamma}
\end{figure}

We illustrate the decay process $H\to  Z\gamma \to 2f\gamma$ in Fig.~\ref{fig:decay-zgamma}.
For a spin-zero particle, only one angular distribution is non-trivial, $\cos\theta_1$.
The distribution reads $(1+\cos^2\theta_1)$, unless a complex phase $\phi_{a3}=arg(g_4/g_2)$
appears in the couplings. The angular distribution can be easily derived from formulas in
Refs.~\cite{Gao:2010qx, Bolognesi:2012mm}; for the case $f_{a3}=0.5$ with $\phi_{a3}=\pi/2$
the angular distribution reads  $(1+2A_f\cos\theta_1+\cos^2\theta_1)$, see 
Fig.~\ref{fig:decay-zgamma}.  Note that in this case 
the forward-backward  asymmetry is maximal.
 Here $A_{f}=2\bar{g}_V^f\bar{g}_A^f/(\bar{g}_V^{f2}+\bar{g}_A^{f2})$ 
is the parameter characterizing the decay $Z\to f\bar{f}$~\cite{pdg} and it is approximately 0.15 for 
$Z\to\ell^+\ell^-$. Since non-trivial asymmetry appears in the $H\to  Z\gamma \to 2f\gamma$ decay in the
special case of complex $g_4/g_2$ coupling ratio only, we do not consider this mode further for the 
measurement of anomalous $HVV$ couplings, but we point out that such a study is in principle possible.

\subsection{The $e^+e^-\to ZH$ process}

\begin{figure}[t]
\centerline{
\setlength{\epsfxsize}{0.33\linewidth}\leavevmode\epsfbox{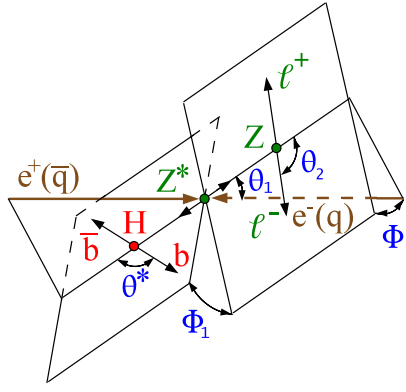}
}
\caption{
Higgs  production and decay at the  $e^+e^-$ or $pp$ collider  with
$e^+e^-(q\bar{q})\to Z^*\to ZH\to\ell^+\ell^-b\bar{b}$ as shown in the parton collision frame.
}
\label{fig:decay-vh}
\end{figure}

We obtain the matrix element for the $e^+e^- \to Z^* \to ZH$  process by crossing the amplitudes for $H \to ZZ^*$ described above. 
Since the intermediate $Z^*$ boson has fixed invariant mass\footnote{The invariant mass 
 obviously coincides with the energy $\sqrt{s}$ of an $e^+e^-$ collider.} 
 and all final state particles are on shell, the probability distribution  depends on five angles $\vec\Omega$,
defined in the middle panel of Fig.~\ref{fig:decay}. It might be easier to understand the decay kinematics 
in Fig.~\ref{fig:decay-vh}, but we would like to stress that the two are equivalent and 
Fig.~\ref{fig:decay} allows direct analogy with the already established process of a Higgs boson decay.

\begin{figure}[t]
\centerline{
\setlength{\epsfxsize}{0.45\linewidth}\leavevmode\epsfbox{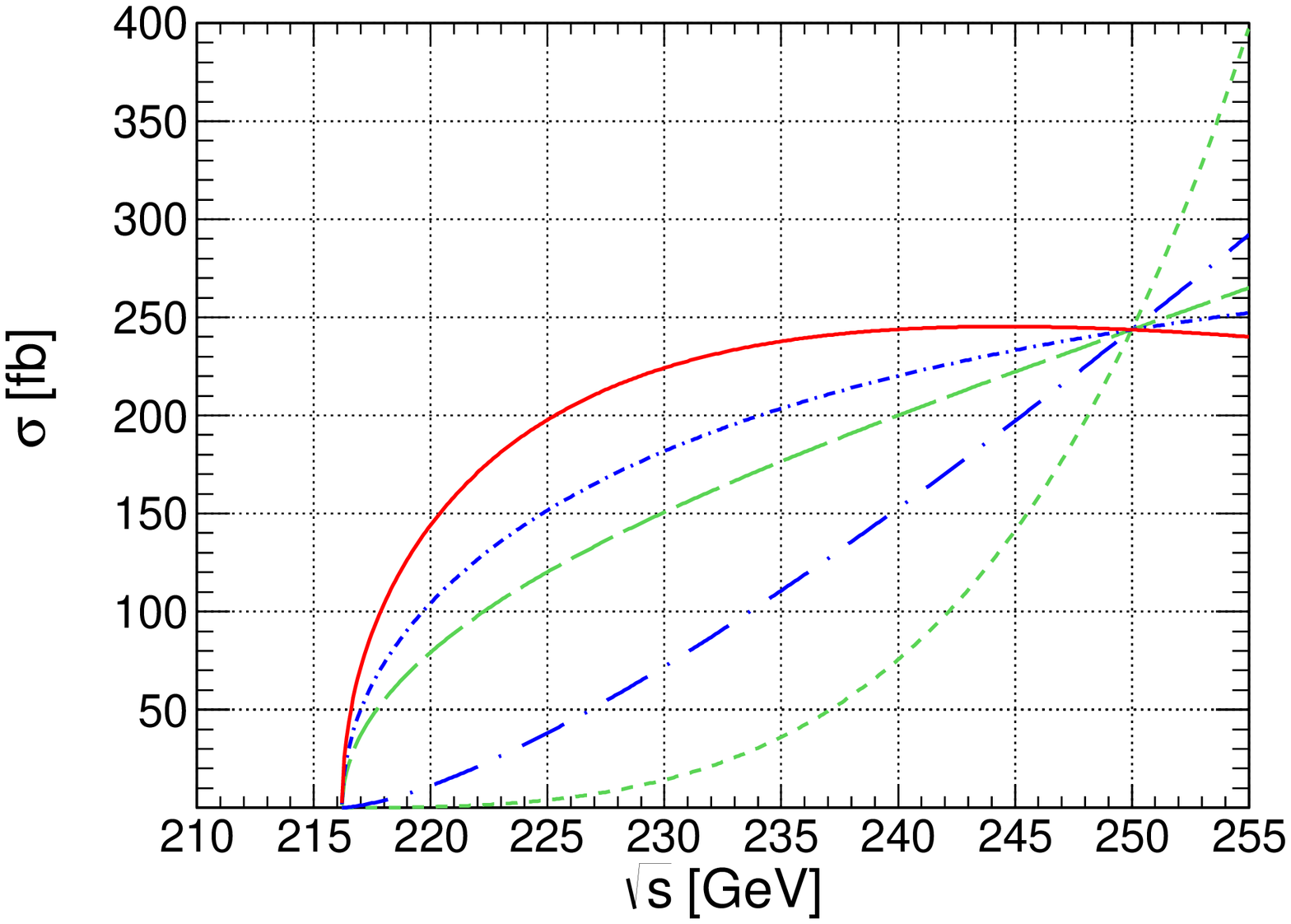}
\setlength{\epsfxsize}{0.45\linewidth}\leavevmode\epsfbox{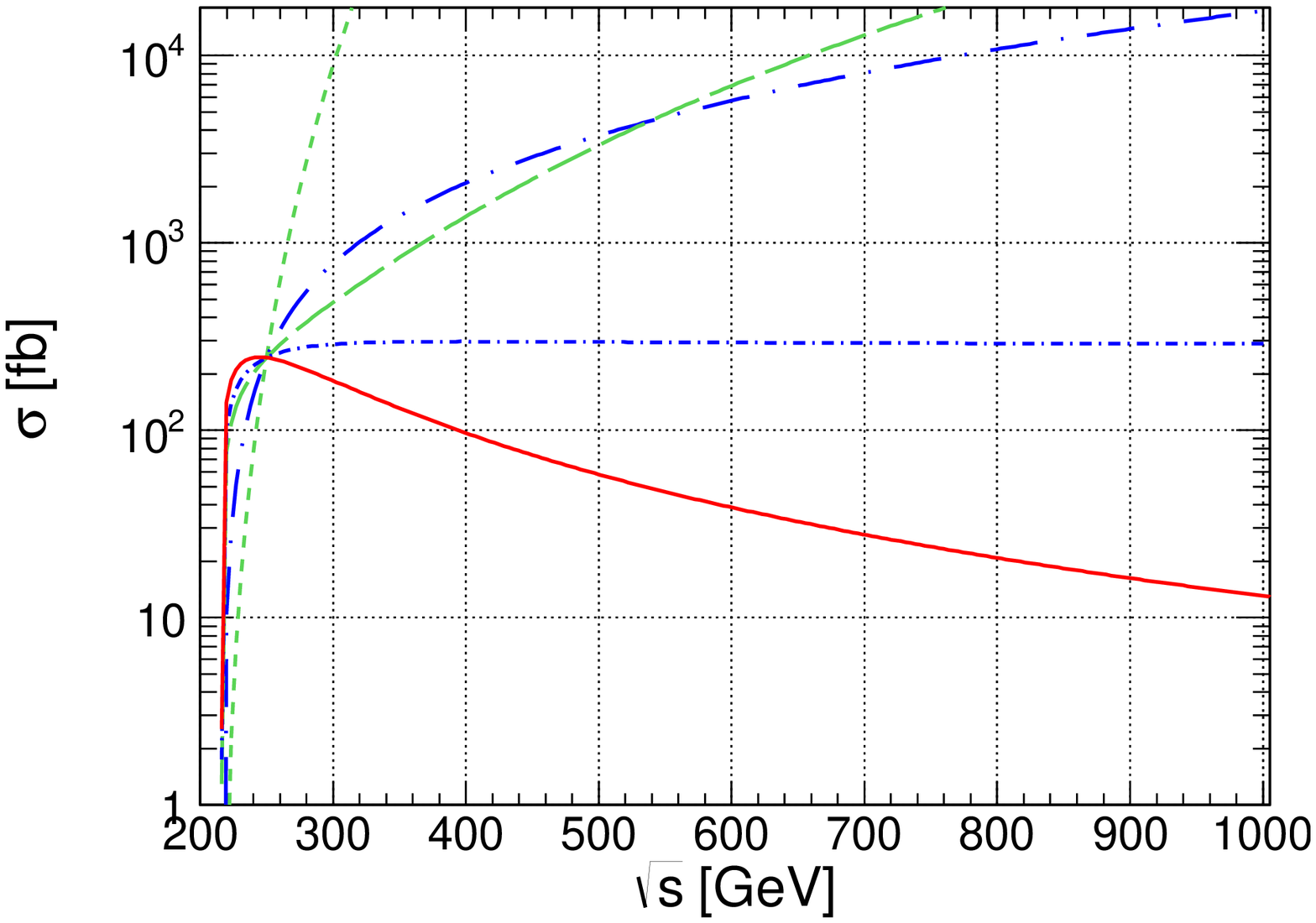}
}
\caption{
Cross section of  $e^+e^-\to Z^*\to ZX$ process  as a function of $\sqrt{s}$ for several representative models:
SM Higgs boson ($0^+$, solid red), 
vector ($1^-$, dot-long-dashed blue), 
axial vector ($1^+$, dot-short-dashed blue), 
Kaluza-Klein graviton with minimal couplings ($2^+_m$, long-dashed green), 
spin-2 with higher-dimension operators ($2^+_h$, short-dashed green).
All cross sections are normalized to SM value at $\sqrt{s}=250$ GeV.
}
\label{fig:eeXS-spin2}
\end{figure}

To compute the differential cross section for $e^+e^- \to ZH \to \mu^+\mu^- H$, 
we modify  ${d\Gamma}/{d\vec\Omega}$ in Eq.~(A1) of Ref.~\cite{Bolognesi:2012mm} to account for changes in kinematics. 
In particular,  $s^\prime=q_1q_2$ in Eq.~(13) of 
Ref.~\cite{Bolognesi:2012mm}\footnote{We add prime to $s^\prime$ to avoid confusion with $\sqrt{s}=m_1$ in this case.} is defined 
for two outgoing momenta of $Z$-bosons.  
If instead we use the four-momentum $P_1$ of the initial $e^+e^-$ state, we must write $q_1=-P_1$ 
and, as a result, $s^\prime = -P_1 q_2 = -(m_{\sss H}^2 - m_{1}^2 - m_2^2)/2 \,,$
where $m_{1}^2=P_1^2$ and $m_2^2=m_Z^2$.
This leads to  the following differential angular distributions for a spin-zero particle production
\begin{eqnarray}
&&\frac{d \Gamma_{J=0}(s, \vec\Omega)}{  \; d\vec\Omega } 
             \propto  4\, |A_{00}|^2\,\sin^2\theta_1 \sin^2\theta_2 
\nonumber \\ 
&& ~~~~~ ~~~~~ ~~~ + |A_{+0}|^2 \left(1-2 R_{1} \cos\theta_1+\cos^2\theta_1 \right) \left(1+2 A_{f_2} \cos\theta_2+\cos^2\theta_2 \right) \nonumber \\ 
&& ~~~~~ ~~~~~ ~~~ + |A_{-0}|^2 \left(1+2 R_{1} \cos\theta_1+\cos^2\theta_1\right) \left(1-2 A_{f_2} \cos\theta_2+\cos^2\theta_2\right) \nonumber \\ 
&& ~~~~~ ~~~~~ ~~~ - 4|A_{00}| |A_{+0}| (R_{1} - \cos\theta_1) \sin\theta_1 (A_{f_2} + \cos\theta_2) \sin\theta_2 \cos(\Phi + \phi_{+0}) \nonumber \\ 
&& ~~~~~ ~~~~~ ~~~ - 4|A_{00}| |A_{-0}| (R_{1} + \cos\theta_1) \sin\theta_1 (A_{f_2} - \cos\theta_2) \sin\theta_2 \cos(\Phi - \phi_{-0}) 
 \nonumber \\ 
&& ~~~~~ ~~~~~ ~~~ + 2|A_{+0}| |A_{-0}| \sin^2\theta_1 \sin^2\theta_2 \cos(2\Phi - \phi_{-0} + \phi_{+0}) \,.
\label{eq:ilc_vhangular}
\end{eqnarray}
In Eq.~(\ref{eq:ilc_vhangular}), $R_1=(A_{f_1}+P^-)/(1+A_{f_1} P^-)$, where 
$A_{f_i}=2\bar{g}_V^f\bar{g}_A^f/(\bar{g}_V^{f2}+\bar{g}_A^{f2})$ is the parameter
characterizing the decay $Z_i\to f_i\bar{f}_i$~\cite{pdg} with 
$A_{f_1}\simeq 0.15$ for the $Zee$ coupling,  $A_{f_2}$ is for the coupling to fermions in the $Z$ decay, 
and $P^-$ is the effective polarization of the electron beam defined in such a way that  
$P^-=0$ corresponds to  the unpolarized beam.
Amplitudes $|A_{\lambda_1\lambda_2}|$ and their phases $\phi_{\lambda_1\lambda_2}$ are obtained by crossing 
the corresponding expressions in Eqs.~(9)--(15) of Ref.~\cite{Bolognesi:2012mm}.
Examples of kinematic  distributions in the $e^+e^- \to ZH$ process can be found  in Fig.~\ref{fig:ilc_angles}; they show 
good agreement between analytical parameterization and numerical computations and exhibit features 
similar to those seen in decay in Fig.~\ref{fig:lhc_angles}. 
Extension to higher spins follows the same logic and can be easily written using expressions in Ref.~\cite{Bolognesi:2012mm},
such as Eqs.~(A1), (17), (21).
Applications to spin-zero, -one, and -two particle production can be found in Figs.~\ref{fig:eeXS} and~\ref{fig:eeXS-spin2}.


\subsection{The $q\overline{q}^\prime\to VH$ process on LHC}

To describe associated $ZH$ and $WH$ production in proton collisions  we modify Eq.~(\ref{eq:ilc_vhangular})
to account for the fact that we now have quarks and antiquarks colliding and that the energy and 
luminosity distribution of these partonic collisions is described by products of parton distribution functions.
The probability distribution for $pp \to ZH$ and $pp \to WH$ processes is described by 
\begin{eqnarray}
 && 
 \frac{d \Gamma({\hat{s}}, Y, \vec\Omega)}{ \; d{\hat{s}}\, dY\, d\vec\Omega }  \propto
\sum_{q,\bar{q}^\prime}
{\cal P}_{q\overline{q}^\prime}({\hat{s}}, \vec\Omega)
\times P({\hat{s}})
\times F_{q\overline{q}^\prime}(\hat{s}, Y) \,,
\label{fig:lhc_vhangular}
\end{eqnarray}
where the sum runs over the five ${q\bar{q}}$ flavors in the $Z^*\to ZH$ production and 
over 12 ${q\bar{q}^\prime}$ flavors in the $W^*\to WH$ process, 
$\hat{s}=m_{VH}^2$,
${\cal P}_{q\overline{q}^\prime}({\hat{s}}, \vec\Omega)$ is the amplitude squared from Eq.~(\ref{eq:ilc_vhangular}),
$P(\hat{s})$ is the kinematic factor~\cite{Barger:1993wt}, 
and $F_{q\overline{q}^\prime}(\hat{s}, Y)$ is the partonic luminosity function 
\begin{eqnarray}
F_{q\overline{q}^\prime}(\hat{s}, Y)= f_q(x_+,\hat s) f_{\bar{q}^\prime}(x_-,\hat s) + (x_+ \leftrightarrow x_-) \,,
\label{eq:dilution-amplitude-parton}
\end{eqnarray}
where $x_\pm = \sqrt{\hat s / s}\,e^{\pm Y}$.
All angular variables are defined in the partonic center-of-mass frame. 

\begin{figure}[t]
\centerline{
\setlength{\epsfxsize}{0.25\linewidth}\leavevmode\epsfbox{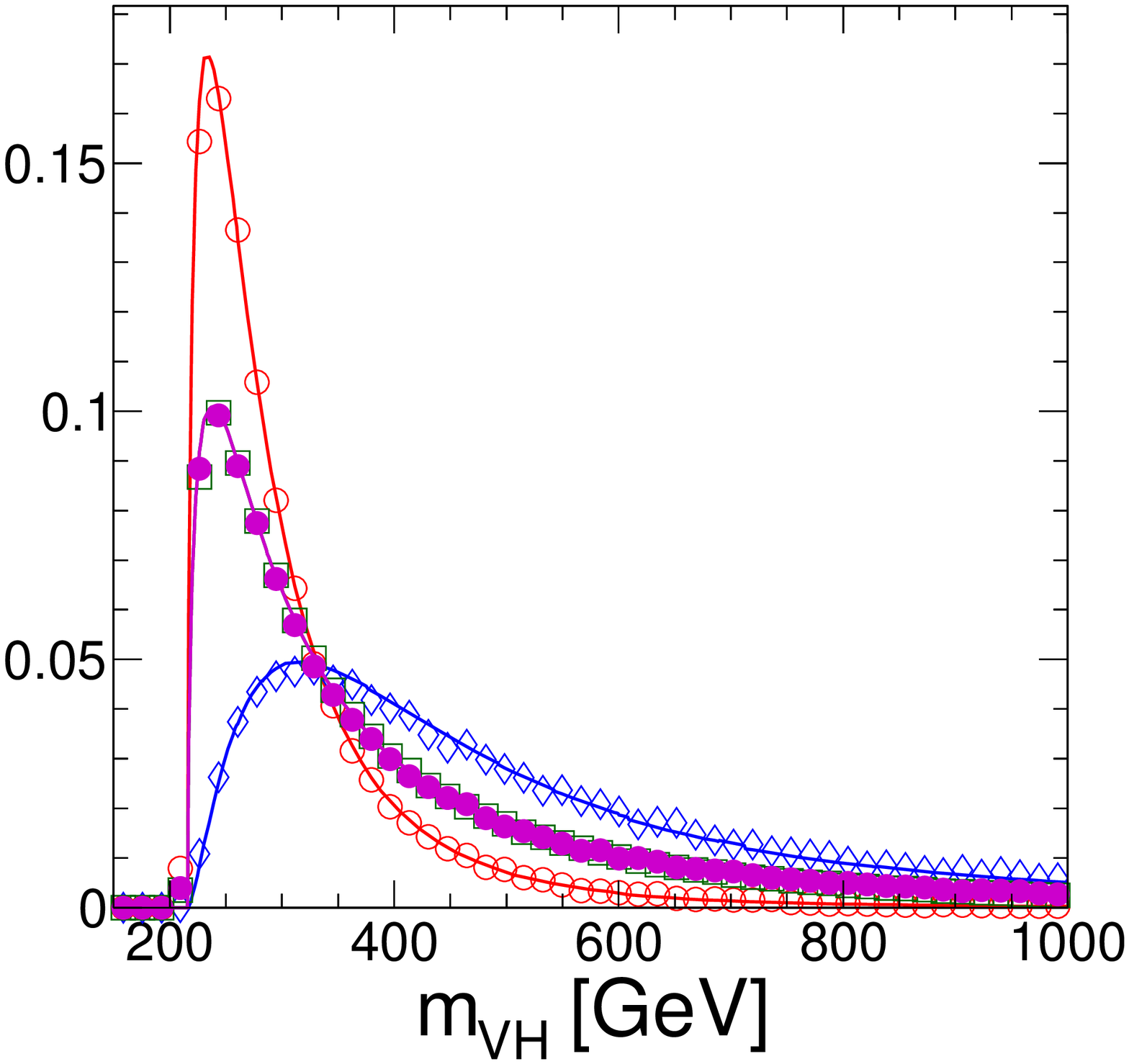}
\setlength{\epsfxsize}{0.25\linewidth}\leavevmode\epsfbox{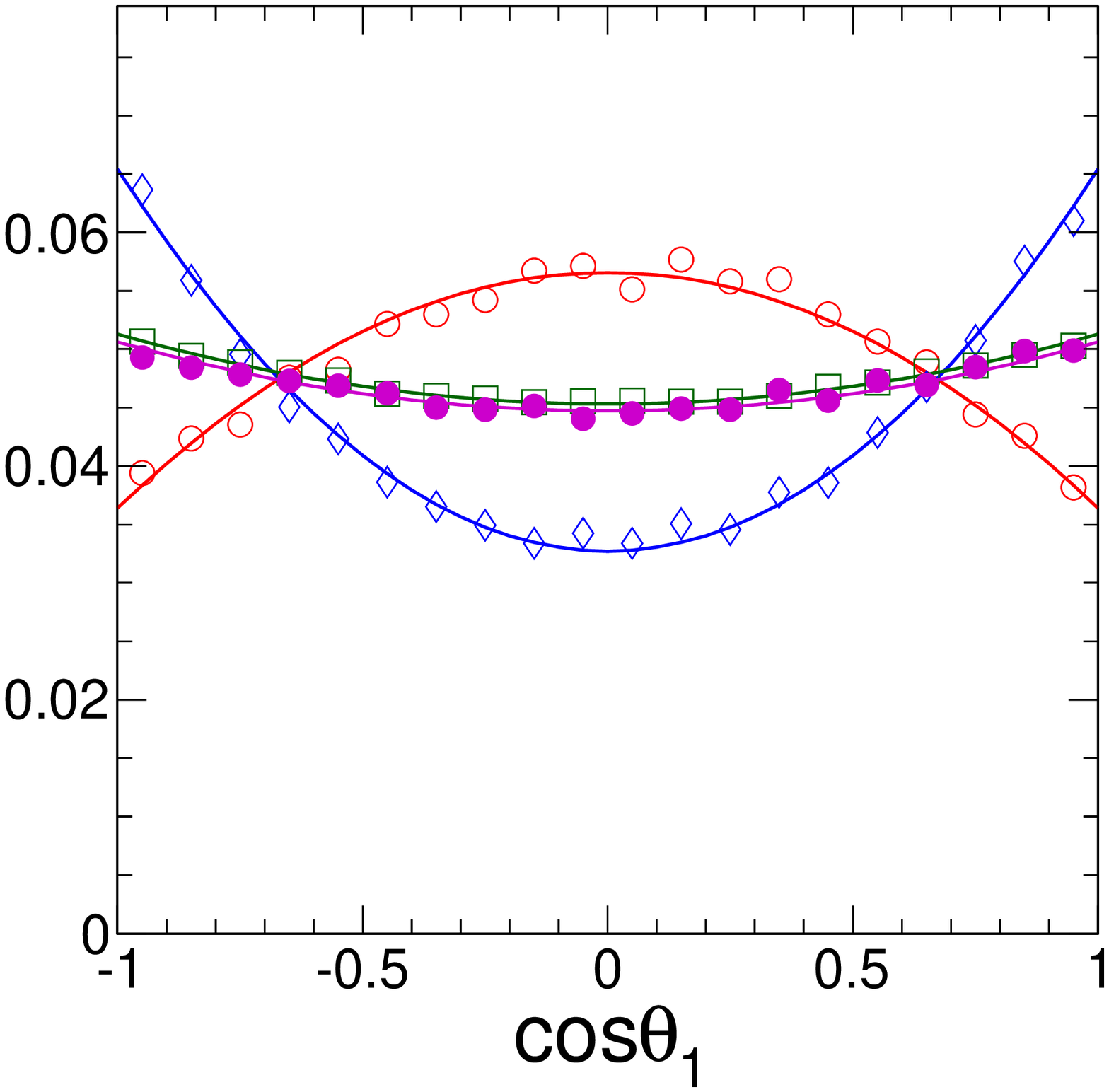}
\setlength{\epsfxsize}{0.25\linewidth}\leavevmode\epsfbox{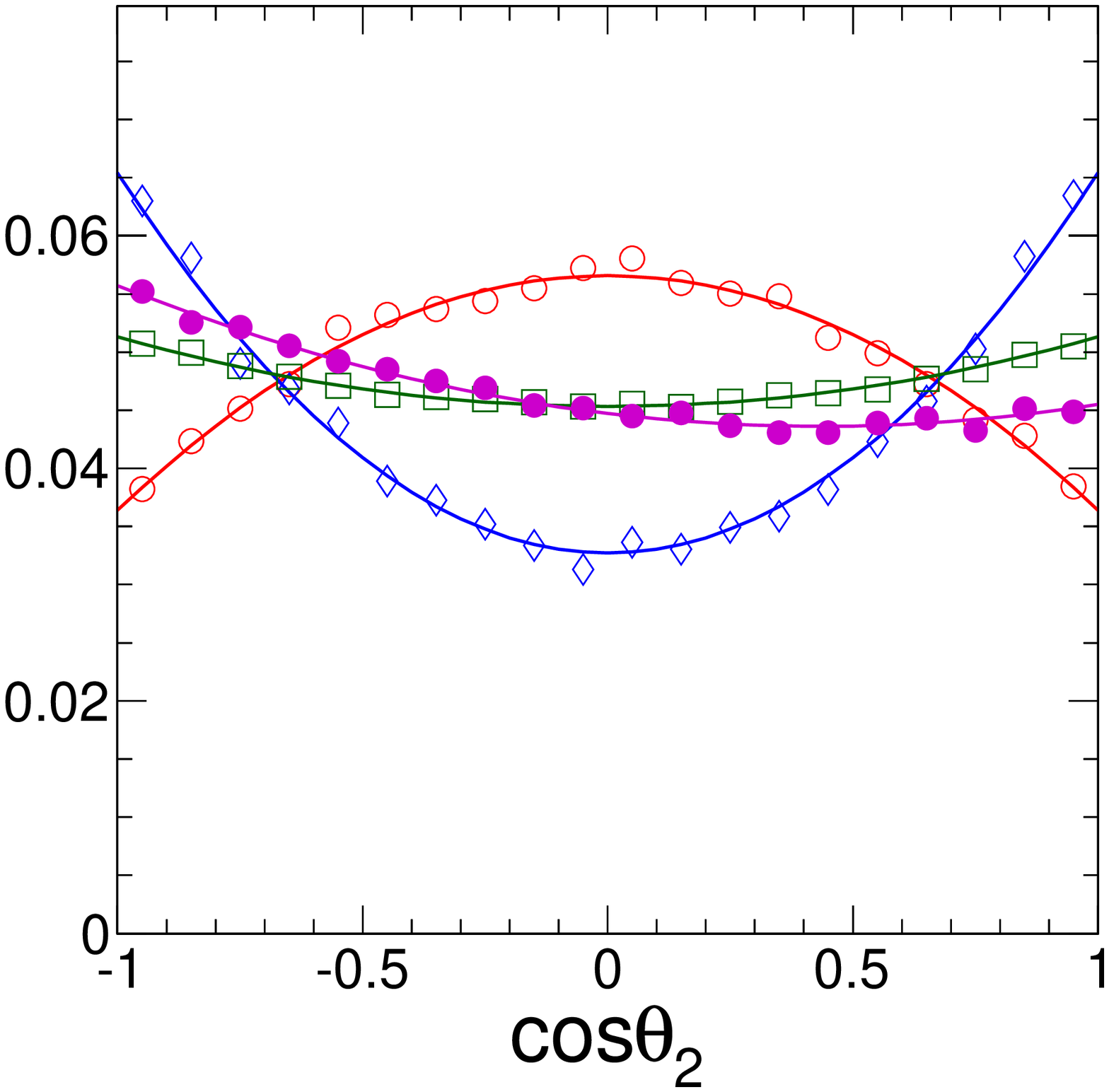}
\setlength{\epsfxsize}{0.25\linewidth}\leavevmode\epsfbox{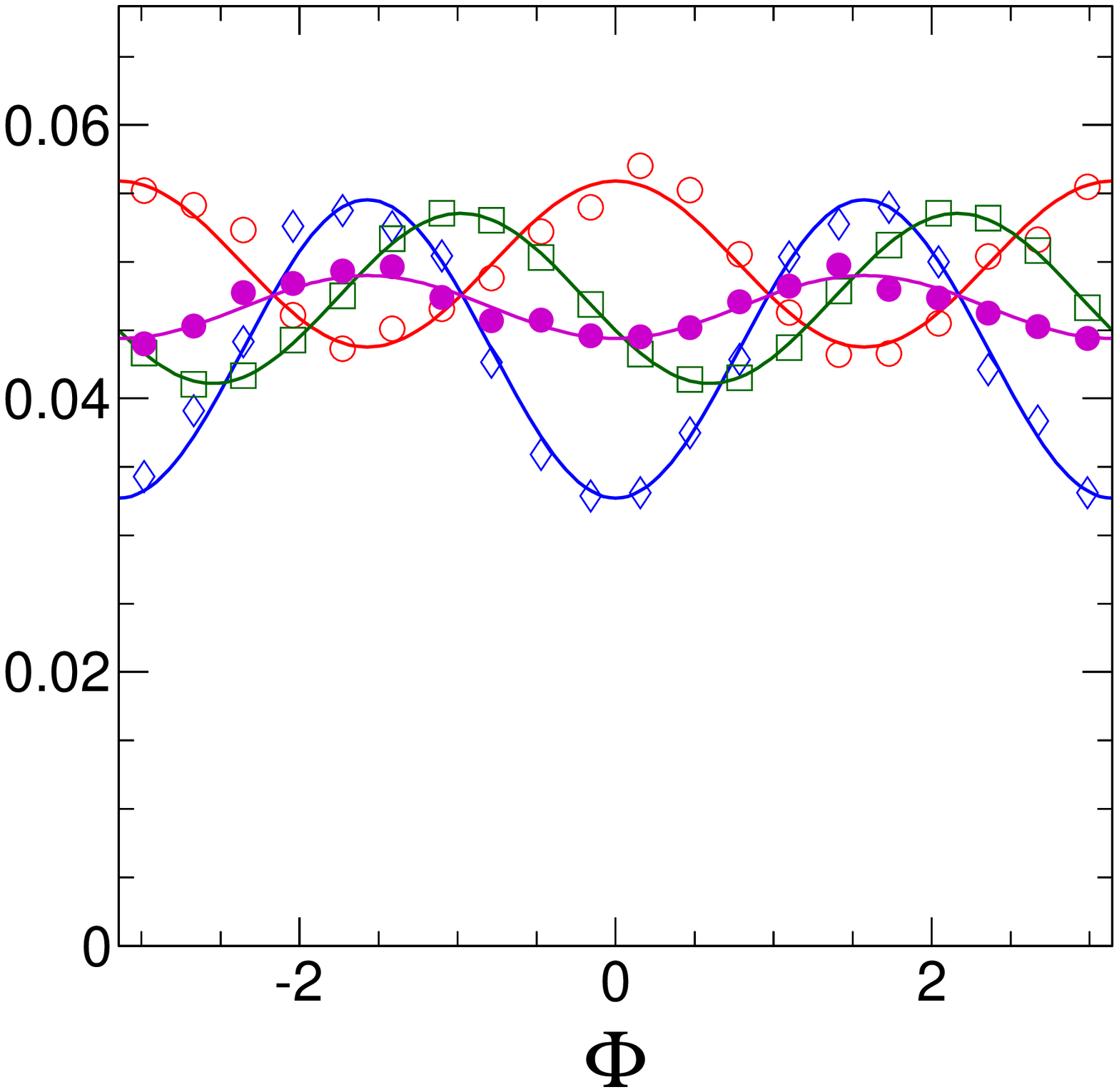}
}
\caption{
Distributions of the observables in the $pp\to ZH$ analysis, 
from left to right: $m_{VH}$, $\cos\theta_1$, $\cos\theta_2$, $\Phi$.
Points show simulated events and lines show projections of analytical distributions.
Four scenarios are shown: SM ($0^+$, red open circles), pseudoscalar ($0^-$, blue diamonds), 
and two mixed states corresponding to $f_{a3}=0.5$ with $\phi_{a3}=0$ (green squares)
and $\pi/2$ (magenta points). 
}
\label{fig:lhc_vhangles}
\end{figure}

Sample kinematic distributions are shown in Fig.~\ref{fig:lhc_vhangles}. There is a good 
agreement between  numerical simulations and analytic probability distributions. 
We note that continuous distribution of the invariant mass $m_{VH}=\sqrt{\hat{s}}$ 
scans the range of a few hundred GeV which  is in the ballpark of center-of-mass energies 
proposed for $e^+e^-$ colliders. 

\subsection{Higgs production in association with two jets}

\begin{figure}[t]
\centerline{
\setlength{\epsfxsize}{0.25\linewidth}\leavevmode\epsfbox{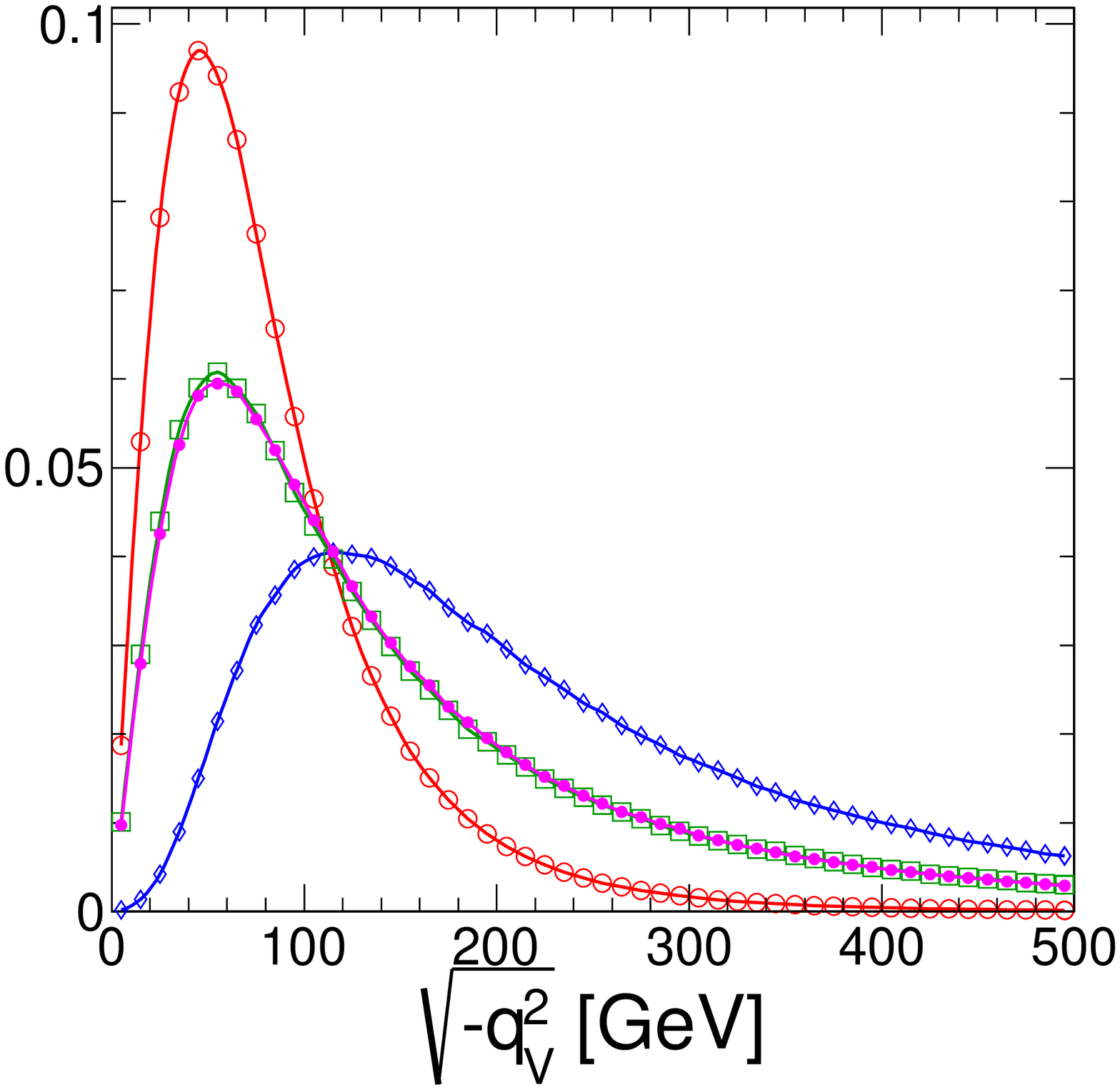}
\setlength{\epsfxsize}{0.25\linewidth}\leavevmode\epsfbox{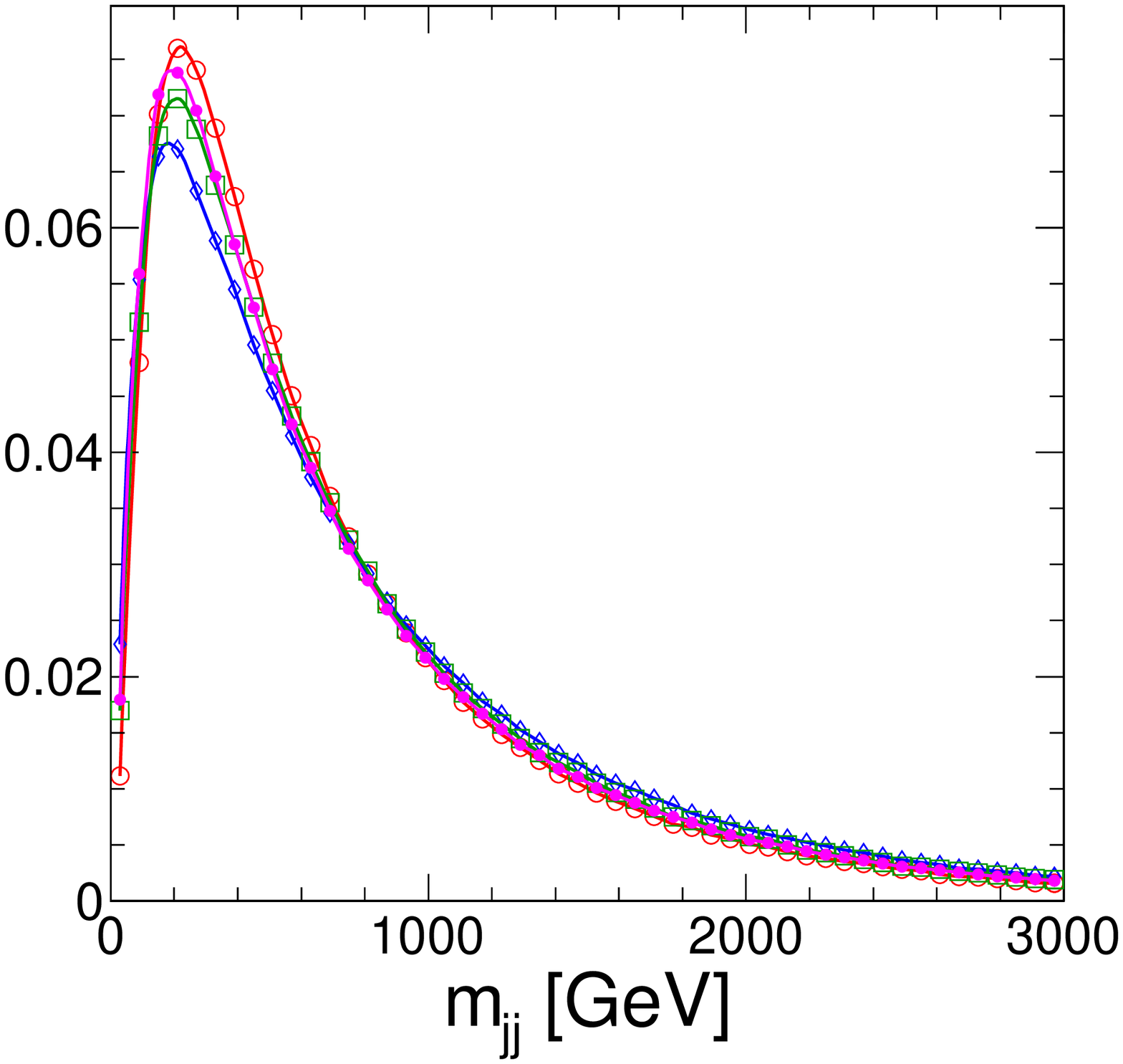}
\setlength{\epsfxsize}{0.25\linewidth}\leavevmode\epsfbox{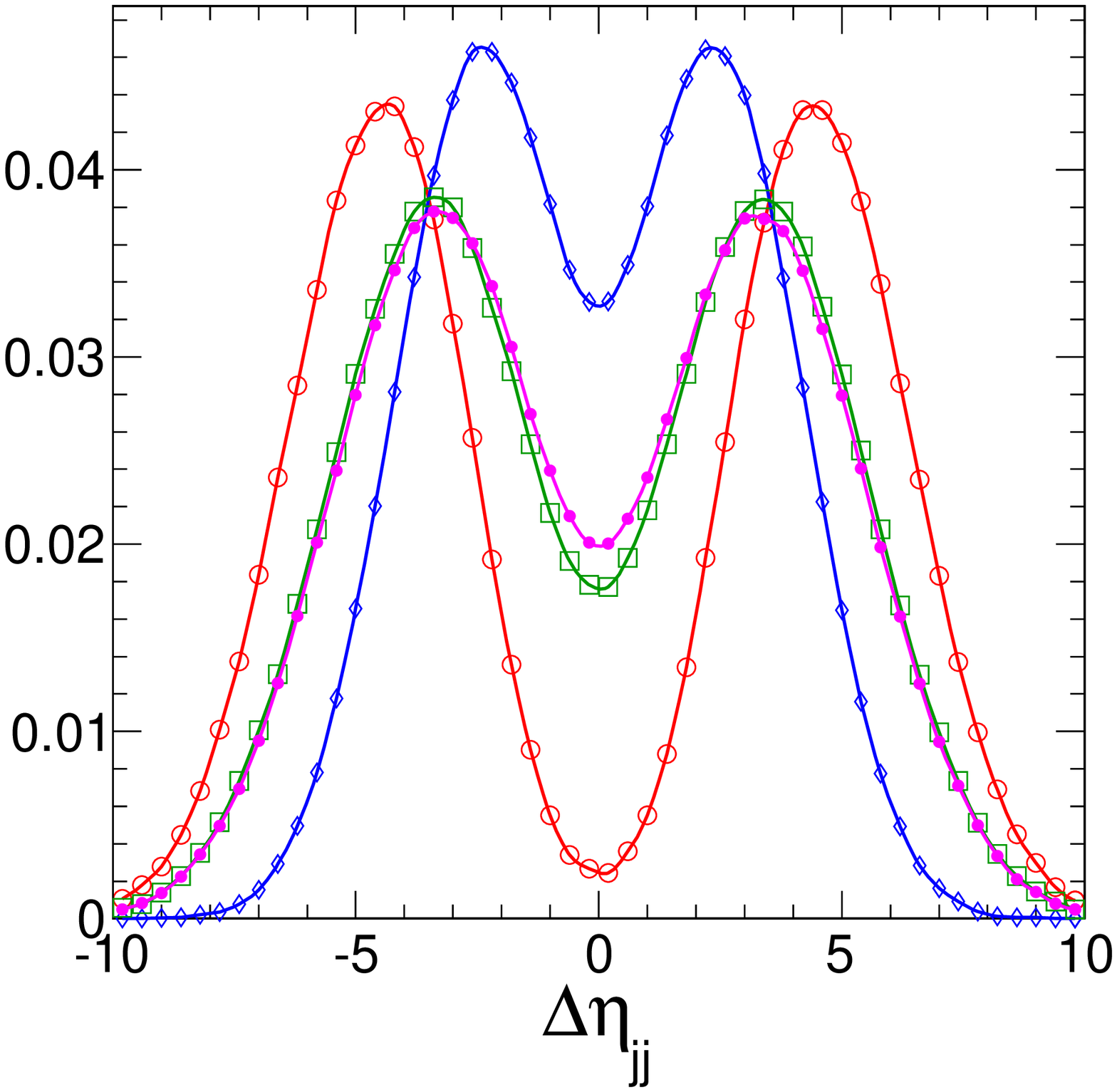}
\setlength{\epsfxsize}{0.25\linewidth}\leavevmode\epsfbox{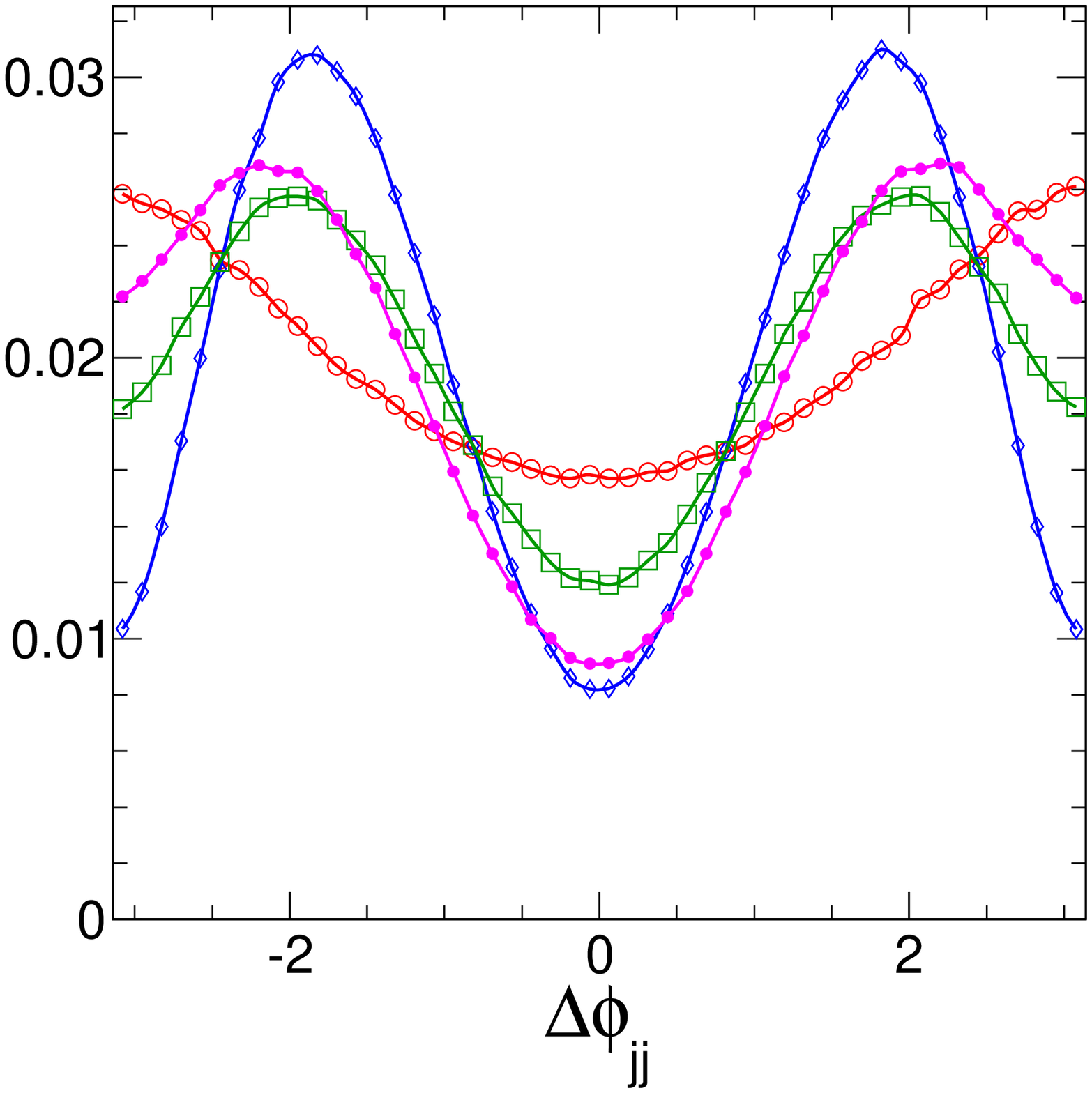}
}
\caption{
Distributions of di-jet observables $\sqrt{-q^2_{\sss V}}$,
$m_{jj}$, $\Delta\eta_{jj}$, $\Delta\phi_{jj}$ in the 
weak vector boson fusion production of a 125~GeV Higgs boson at LHC with 14 TeV energy.
Points connected by lines show simulated events.
Four scenarios are shown: SM ($0^+$, red open circles), pseudoscalar ($0^-$, blue diamonds), 
and two mixed states corresponding to $f_{a3}=0.5$ with $\phi_{a3}=0$ (green squares)
and $\pi/2$ (magenta points). 
}
\label{fig:vbf_angles}
%
\centerline{
\setlength{\epsfxsize}{0.25\linewidth}\leavevmode\epsfbox{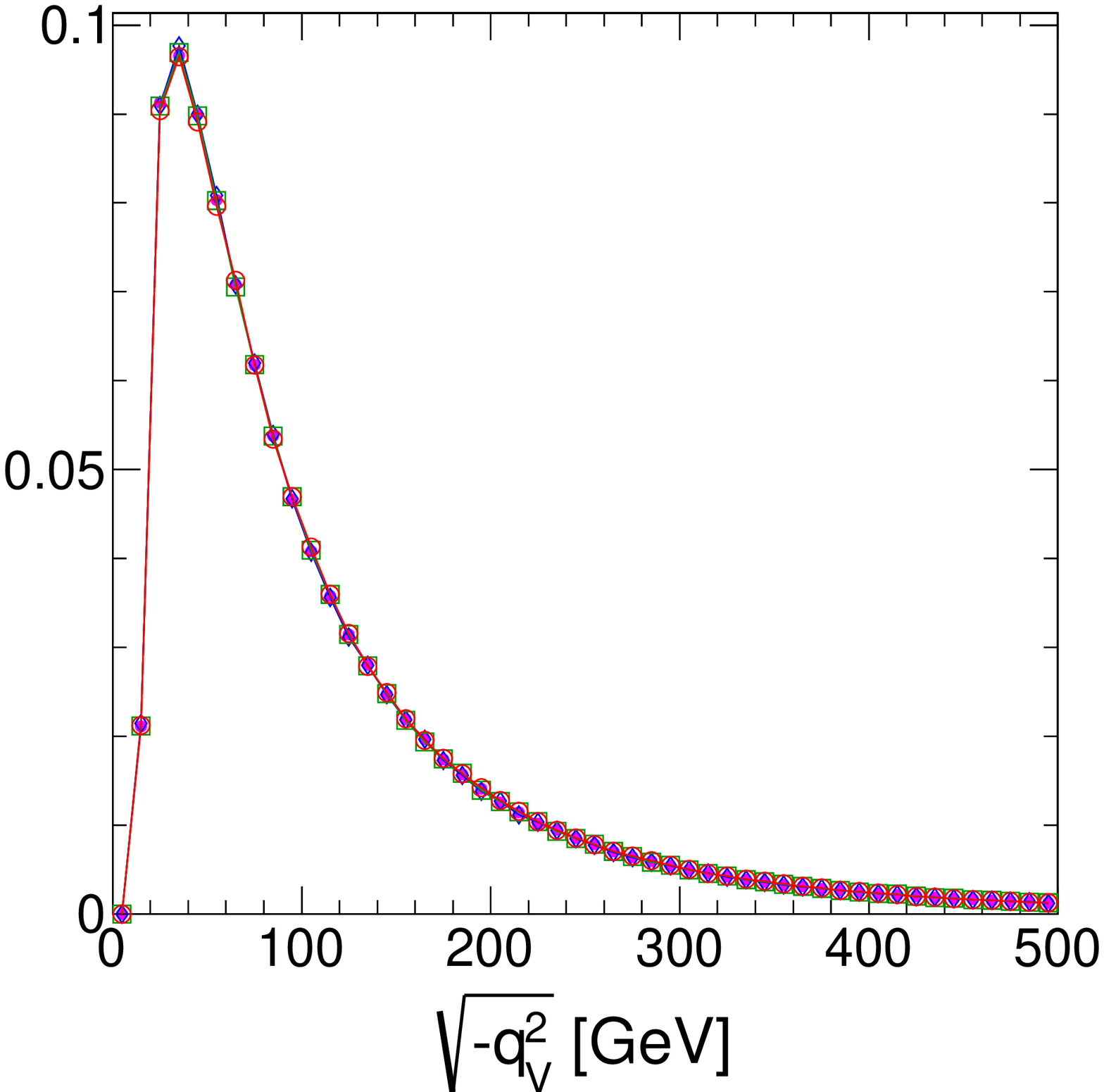}
\setlength{\epsfxsize}{0.25\linewidth}\leavevmode\epsfbox{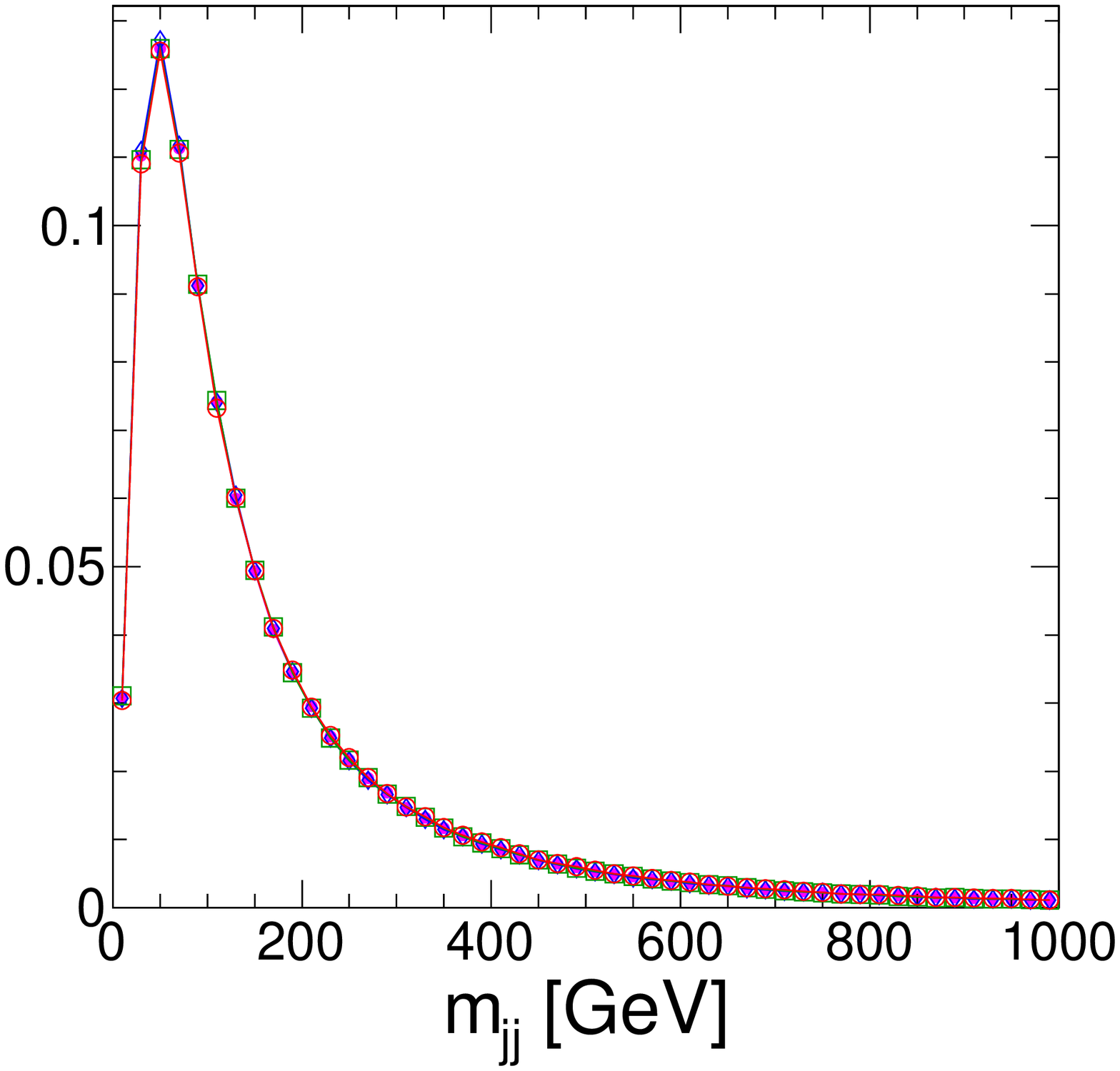}
\setlength{\epsfxsize}{0.25\linewidth}\leavevmode\epsfbox{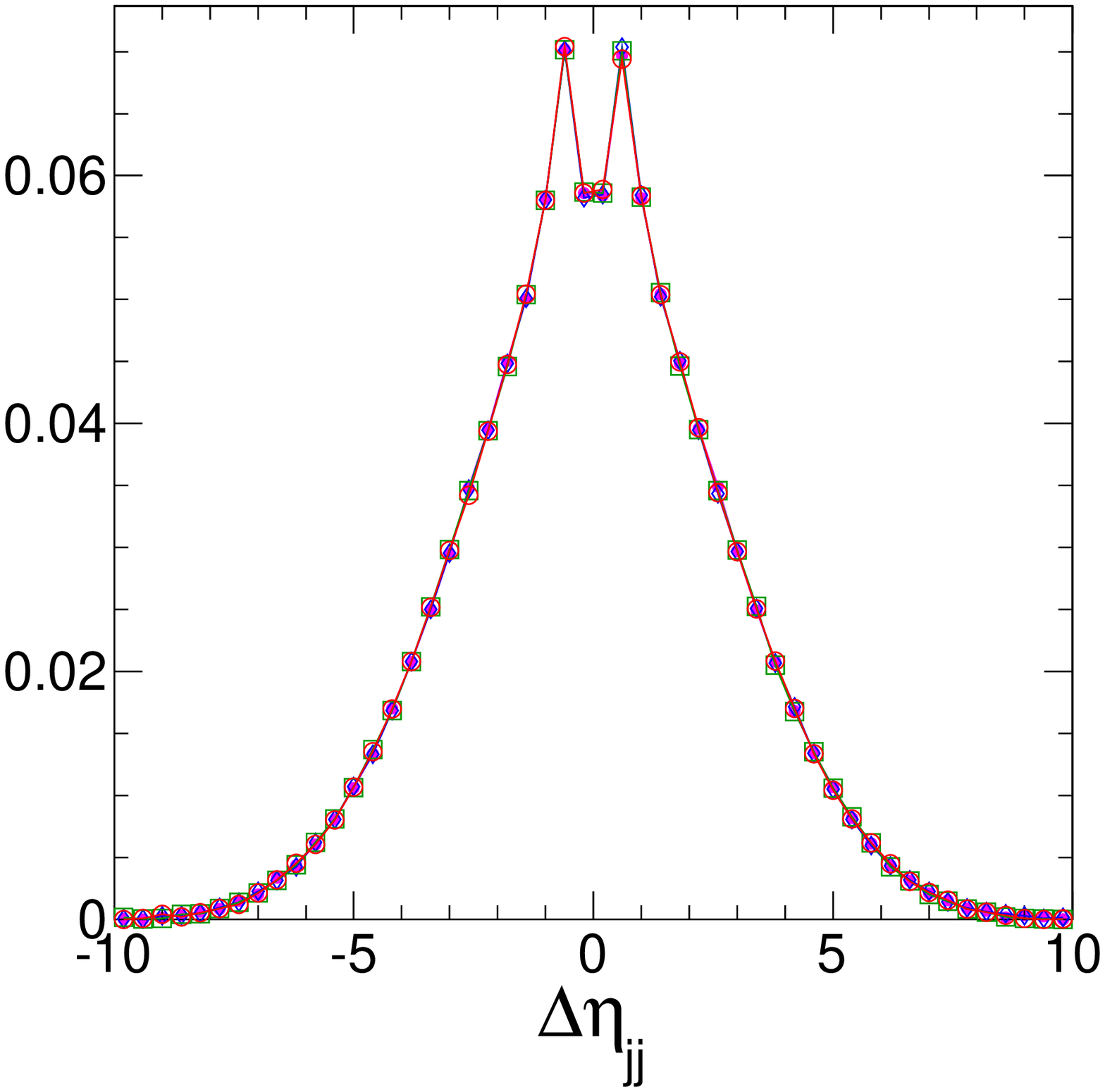}
\setlength{\epsfxsize}{0.25\linewidth}\leavevmode\epsfbox{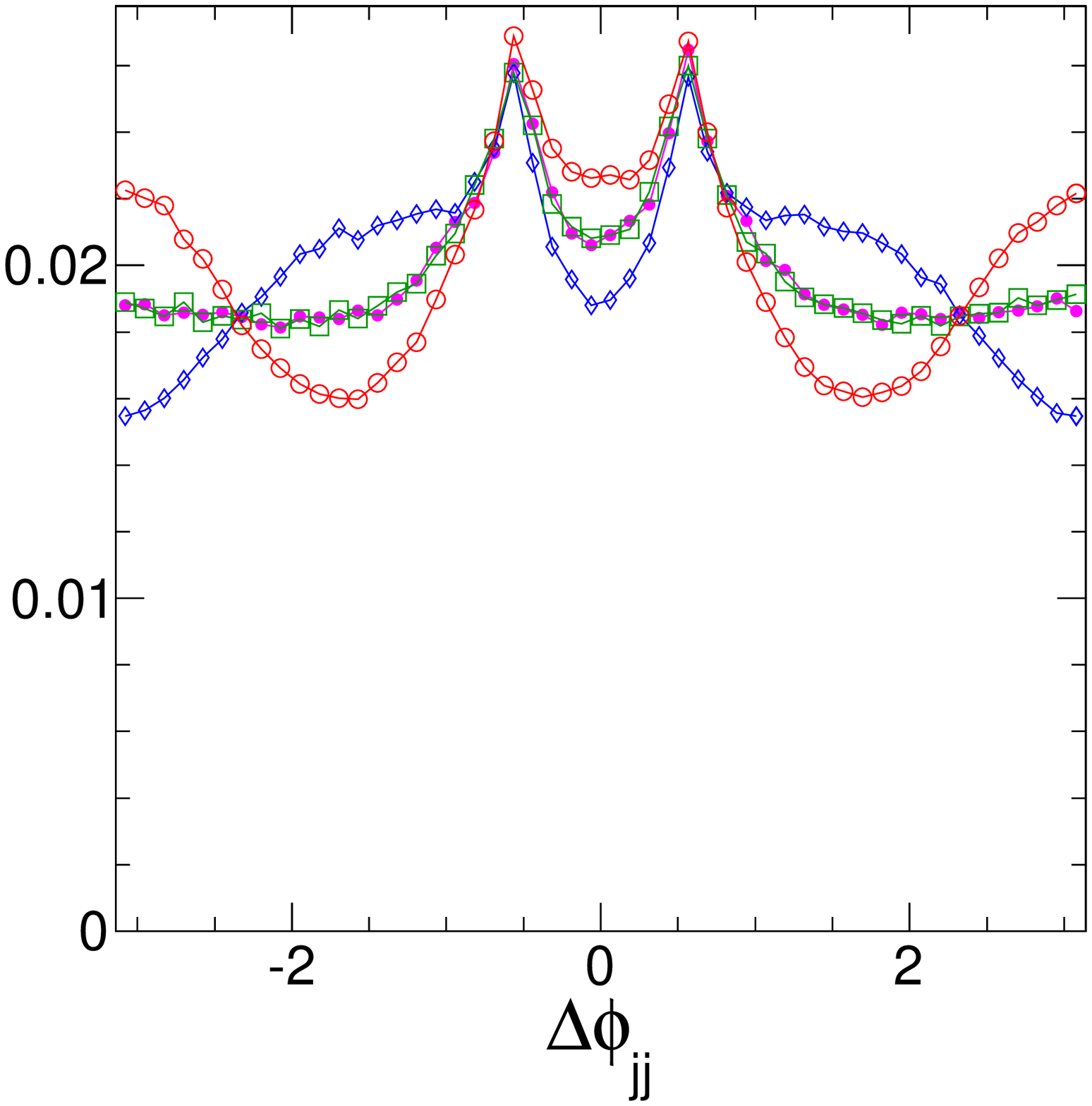}
}
\caption{
Distributions of di-jet observables $\sqrt{-q^2_{\sss V}}$,
$m_{jj}$, $\Delta\eta_{jj}$, $\Delta\phi_{jj}$ in the 
strong vector boson fusion production of a 125~GeV Higgs boson at LHC with 14 TeV energy.
Points connected by lines show simulated events.
Selection requirements are applied on jets $p_T>15$ GeV and $\Delta R_{jj}>0.5$.
Four scenarios are shown: SM ($0^+$, red open circles), pseudoscalar ($0^-$, blue diamonds), 
and two mixed states corresponding to $f_{a3}=0.5$ with $\phi_{a3}=0$ (green squares)
and $\pi/2$ (magenta points). 
}
\label{fig:h2j_angles}
\end{figure}

For studies of the Higgs boson production in association with two jets for both weak boson fusion and 
gluon fusion see e.g. Ref.~\cite{DelDuca:2006hk}.  
Analytic parameterization of the probability distribution in this case is more involved 
because the two vector bosons have negative virtualities $q_i^2<0$, 
and because parton distribution functions  of a proton need to be incorporated. Although a 
partial analytic description of probability distributions is available, see e.g. Ref.~\cite{Hagiwara:2009wt}, 
in this analysis we employ the matrix elements for $pp \to H+2j$  as implemented in the {\sf JHU} generator. 
The matrix elements likelihood approach describes the full kinematics of the two jets
and the Higgs bosons candidate as a single function without information about the decay of the Higgs.
On the other hand, all correlations between the Higgs momentum and momenta of the two jets are included. 
In Fig.~\ref{fig:vbf_angles} we show representative distributions of di-jet observables 
$m_{jj}$, $\Delta\eta_{jj}$, $\Delta\phi_{jj}$, and $\sqrt{|q_V^2|}$ of the two vector bosons
calculated from momenta of the jets and Higgs candidate, for the scalar, pseudoscalar, and mixed states
produced in weak boson fusion.
The same distributions are shown in Fig.~\ref{fig:h2j_angles}  for Higgs boson production in gluon fusion. 
Several observables, in particular $\Delta\phi_{jj}$, exhibit differences between the
scalar and pseudoscalar couplings. 
The enhanced production of events with anomalous couplings at higher values of $|q_i^2|$
in WBF is similar to the $VH$ process; this effect is significantly weaker 
in the gluon fusion.

\subsection{Background}

Parameterization of background matrix elements is important for signal-to-background separation. 
Indeed, this was a crucial part of  the Higgs boson discovery by the CMS collaboration~\cite{discovery-cms} 
with the MELA technique which identifies  kinematic differences  between dilepton pairs 
produced in the decay of the Higgs boson via $H\to ZZ^*\to 4\ell$ and in $q \bar q$ annihilation, 
$q\overline{q}\to ZZ^*/Z\gamma^*$,  to distinguish them from each other. 
We use MCFM generator~\cite{Campbell:2011bn}  matrix elements for both $q\overline{q}\to ZZ^*/Z\gamma^*/Z\gamma$ and $gg \to ZZ^*$  
processes to describe relevant backgrounds~\cite{support}.
We also provide interference of $gg\to ZZ^*$~\cite{Campbell:2011bn}  and $gg\to H^*\to ZZ$~\cite{support}
for optimal analysis above the $ZZ$ threshold, such as a study suggested in Refs.~\cite{Kauer:2012hd, 1307.4935}.
We note that analytic parameterization of the $q\bar{q}\to ZZ^*/Z\gamma^*$ background is also available~\cite{Chen:2012jy}
and we also use it for the background parameterization.
A similar approach to $q\bar{q}\to ZZ^*/Z\gamma^*$ background is also discussed in~Ref.~\cite{Avery:2012um}.


\subsection{Analytic parameterization of parton distribution functions}

\begin{figure}[t]
\centerline{
\setlength{\epsfxsize}{0.5\linewidth}\leavevmode\epsfbox{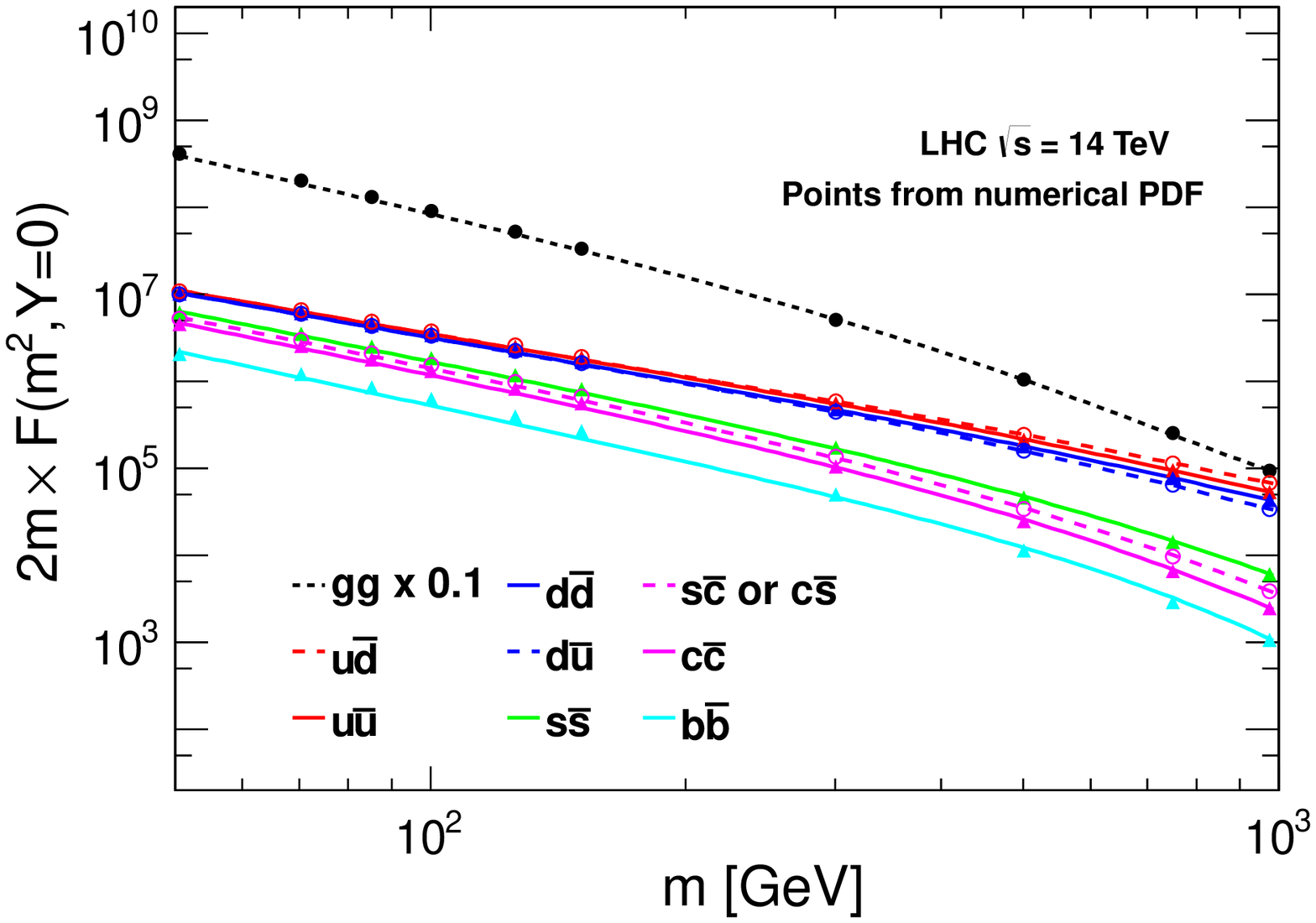}
\setlength{\epsfxsize}{0.5\linewidth}\leavevmode\epsfbox{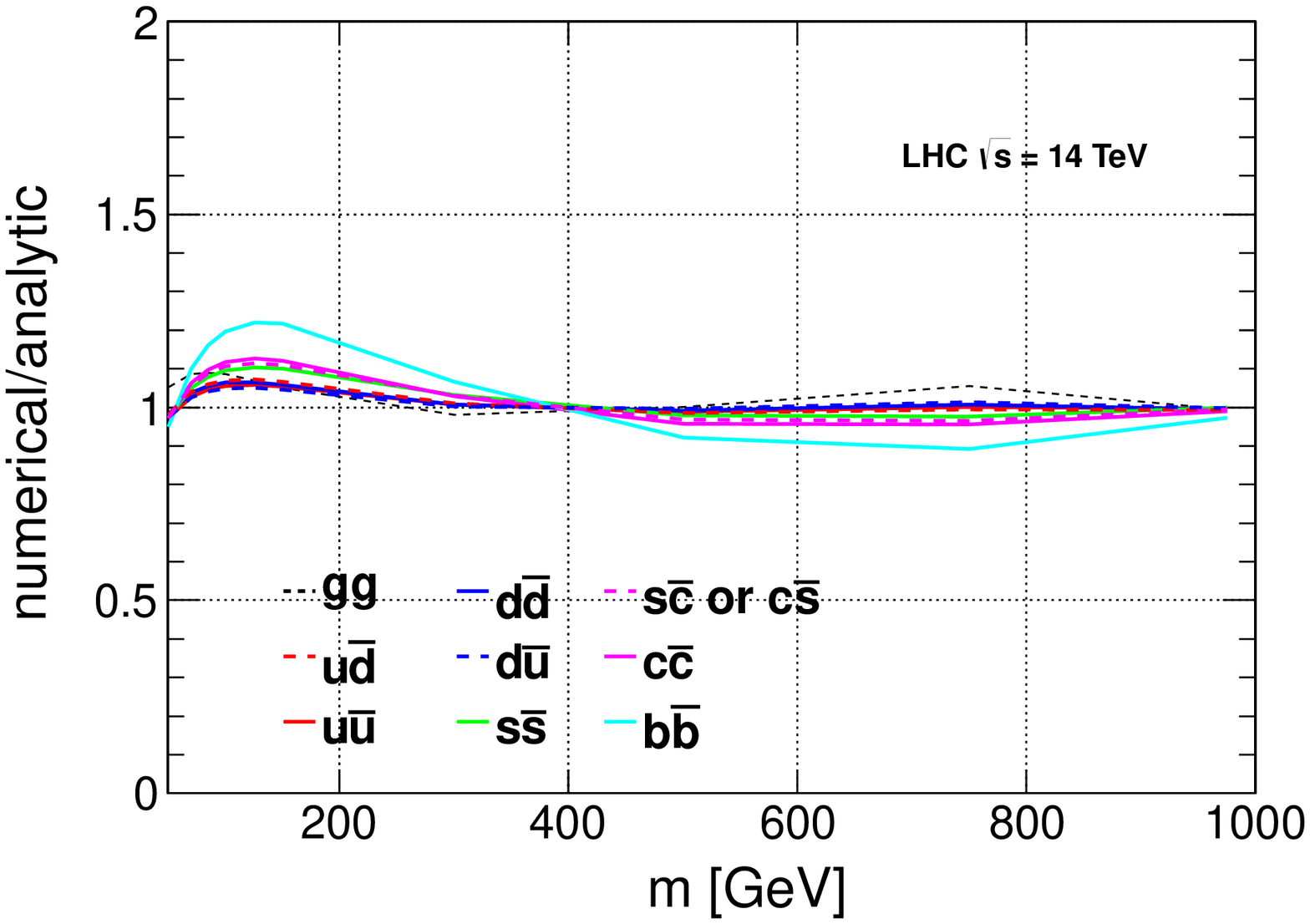}
}
\caption{
Distribution of  $2mF_{q\overline{q}^\prime}(m^2,Y=0)$ defined in Eq.~(\ref{eq:dilution-amplitude-parton}) 
for proton-proton collision energies of 14 TeV as a function of the parton invariant mass $m$. 
Curves show analytical approximation and points show exact numerical calculation using CTEQ6L1 PDFs,
${q\overline{q}^\prime}$ combinations suppressed by Cabibbo-Kobayashi-Maskawa (CKM) mechanism are not shown.
On the right plot, ratio between the numerical and analytical parameterizations is shown.
An equivalent factor for gluon-fusion production is shown for comparison and is scaled by a factor of 0.1.
}
\label{fig:lhc_partonlumi}
\end{figure}

Calculation of both signal and background processes at a hadron collider involves parton distribution 
functions (PDFs). These functions are usually calculated numerically by solving Altarelli-Parisi equations using dedicated numerical programs. 
It may be desirable, in some cases,  to have an analytic parameterization  of the parton distribution functions. For example, such parameterization
may allow faster computations or even analytic integrations of the products of PDFs and partonic cross sections.  
Parton distribution functions $f_q(x,\hat{s})$ are extracted from \textsc{cteq6} PDF set~\cite{cteq1, cteq2}
and are parameterized analytically using polynomial and exponential functions in the relevant range of $x$ with coefficients 
that are also functions of $\hat{s}$~\cite{thesis-nhan, Chatrchyan:2011ya}.
The resulting set of analytically-parameterized CTEQ6L1 PDFs can be found in  Ref.~\cite{support}.
The partonic luminosity functions from Eq.~(\ref{eq:dilution-amplitude-parton}) are shown in Fig.~\ref{fig:lhc_partonlumi}.

\bigskip

\section{Statistical Approaches}
\label{sect:statistics}

The ultimate goal of the analysis described in this paper is the measurement of all anomalous couplings 
of the Higgs bosons to the gauge bosons.  This can be accomplished by performing a multi-dimensional 
fit to match observed kinematic distributions in various processes  to theory predictions. 
Theoretical input to the fit involves real parameters such as for example
$\vec{\zeta}=\{f_{a2}, \phi_{a2}, f_{a3} , \phi_{a3},...\}$ in Eq.~(\ref{eq:fractions}) 
which, once known, can be used to derive the couplings.  To set up a fit process, we  follow Ref.~\cite{Gao:2010qx} and  
introduce  the likelihood function for $N$ candidate events
%
\begin{equation}
{{\cal L}} =  \exp\left( - n_{\rm sig}-n_{\rm bkg}  \right) 
\prod_i^{N} \left( n_{\rm sig} \times{\cal P}_{\rm sig}(\vec{x}_{i};~\vec{\zeta})  
+n_{\rm bkg} \times{\cal P}_{\rm bkg}(\vec{x}_{i})  
\right)\,,
\label{eq:likelihood}
\end{equation}
where $n_{\rm sig}$ is the number of signal events, $n_{{\rm bkg}}$ is the number of background events, and 
${\cal P}(\vec{x}_{i};\vec{\zeta})$ is the probability density function for signal or background.
Each candidate event  $i$ is characterized by a set of eight observables, 
for example $\vec{x}_{i}=\{m_1, m_2, \vec\Omega\}_i$ as defined in Fig.~\ref{fig:decay}.
The number of observables and free parameters can be extended or reduced, depending on the desired fit. 

The advantage of this approach is that the likelihood ${{\cal L}}$ in Eq.~(\ref{eq:likelihood}) 
can be maximized for a large set of parameters in the most optimal way without losing information. The disadvantage 
is the difficulty to describe the detector response and background parameterization in a multi-dimensional space.
In addition, convergence of the fit for a limited number of events may be an issue as well. 

Nonetheless, successful implementation 
can be achieved with certain approximations, for example by allowing for a single anomalous coupling constant at a time. 
Consider a case where $f_{a3}$ and $\phi_{a3}$ are  non-vanishing and write the signal probability as
%
\begin{equation}
{\cal P}_{\rm sig}(\vec{x}_{i}; f_{a3}, \phi_{a3}) =  
(1-f_{a3}) \, {\cal P}_{0^+}(\vec{x}_{i}) + f_{a3} \, {\cal P}_{0^-}(\vec{x}_{i}) + \sqrt{f_{a3}(1-f_{a3})} \, {\cal P}_{\rm int}(\vec{x}_{i}; \phi_{a3})
\,,
\label{eq:fractions-P}
\end{equation}
where ${\cal P}_{\rm int}$ describes interference of  $0^+$ ($g_1$) and $0^-$ ($g_4$) terms.

In this simplified approach, we consider a single observable 
$\vec{x}_{i}=\{D_{0^-}\}_i$ and one free parameter $\vec{\zeta}=\{f_{a3} \}$.
A kinematic discriminant is constructed from the ratio of probabilities for the SM signal and alternative signal $0^-$ hypothesis
\begin{eqnarray}
\label{eq:melaSig}
{D_{0^-}}=\frac{{\cal P}_{0^+}}{{\cal P}_{0^+}+{\cal P}_{0^-}}
=\left[1+\frac{{\cal P}_{0^-} (m_1, m_2, \vec\Omega) }
{{\cal P}_{0^+} (m_1, m_2, \vec\Omega) } \right]^{-1}
\,.
\end{eqnarray}

\begin{figure}[t]
\centerline{
\setlength{\epsfxsize}{0.33\linewidth}\leavevmode\epsfbox{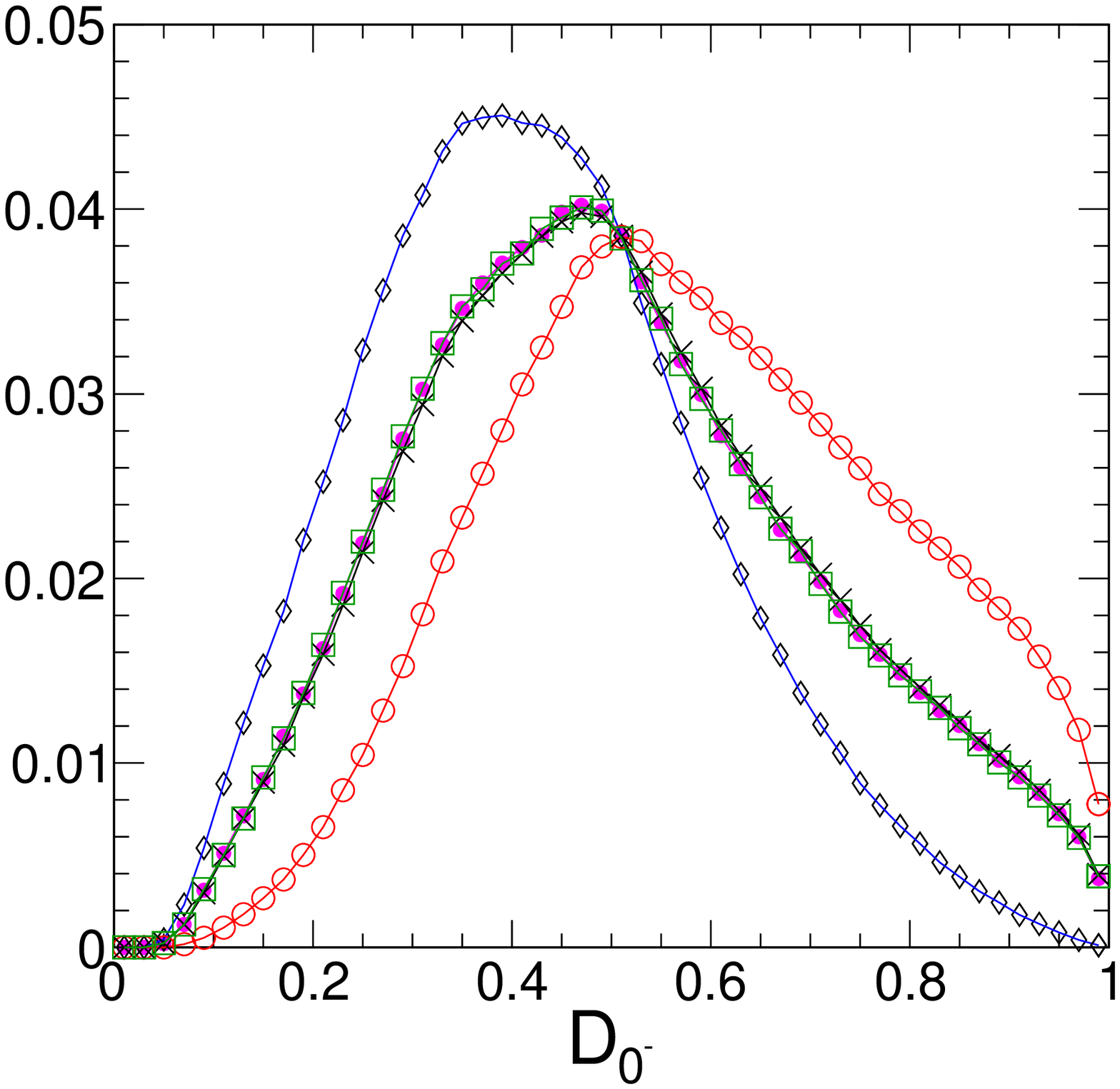}
\setlength{\epsfxsize}{0.33\linewidth}\leavevmode\epsfbox{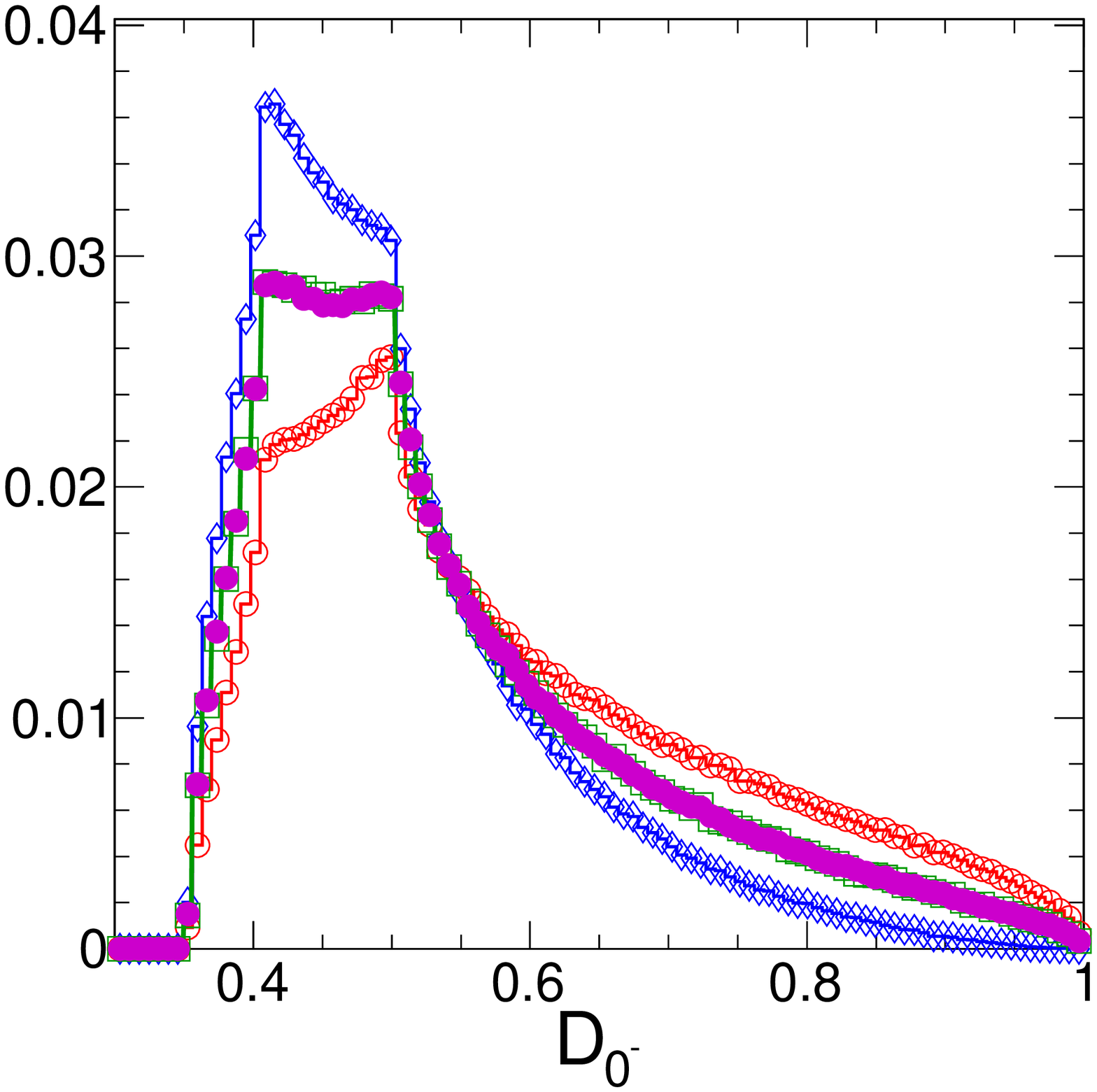}
\setlength{\epsfxsize}{0.33\linewidth}\leavevmode\epsfbox{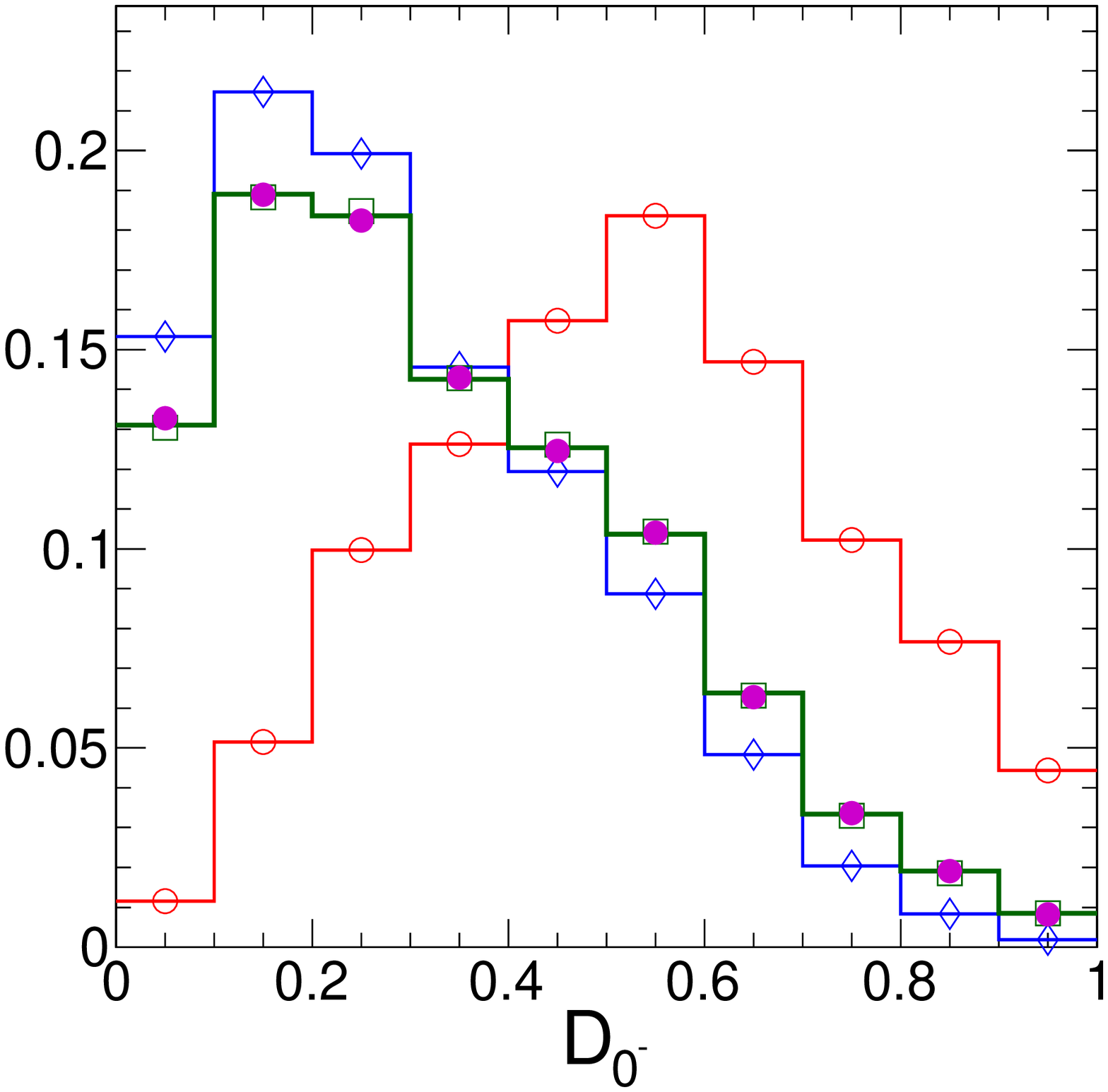}
}
\caption{
Distribution of $D_{0^-}$ for generated events in three topologies:
$H\to ZZ\to 4\ell$ (left),
$e^+e^-\to Z^*\to ZH\to (\ell\ell)(b\bar{b})$ (middle) at $\sqrt{s}=250$ GeV, and
$q\bar{q}\to Z^*\to ZH\to (\ell\ell)(b\bar{b})$ (right) at a 14 TeV $pp$ collider.
Four processes  are shown: SM scalar ($0^+$, red open circles), pseudoscalar ($0^-$, blue diamonds), 
and two mixed states corresponding to $f_{a3}=0.5$ with $\phi_{a3}=0$ (green squares)
and $\pi/2$ (magenta points). 
Also shown is the $f_{a3}=0.5$ sample with no interference between the scalar and pseudoscalar
terms simulated (black crosses).
Events are shown after selection requirements, 
which have different efficiencies for $0^+$ and $0^-$ samples in the $q\bar{q}\to Z^*\to ZH$ channel. 
}
\label{fig:D0mn}
\end{figure}

\begin{figure}[t]
\centerline{
\setlength{\epsfxsize}{0.33\linewidth}\leavevmode\epsfbox{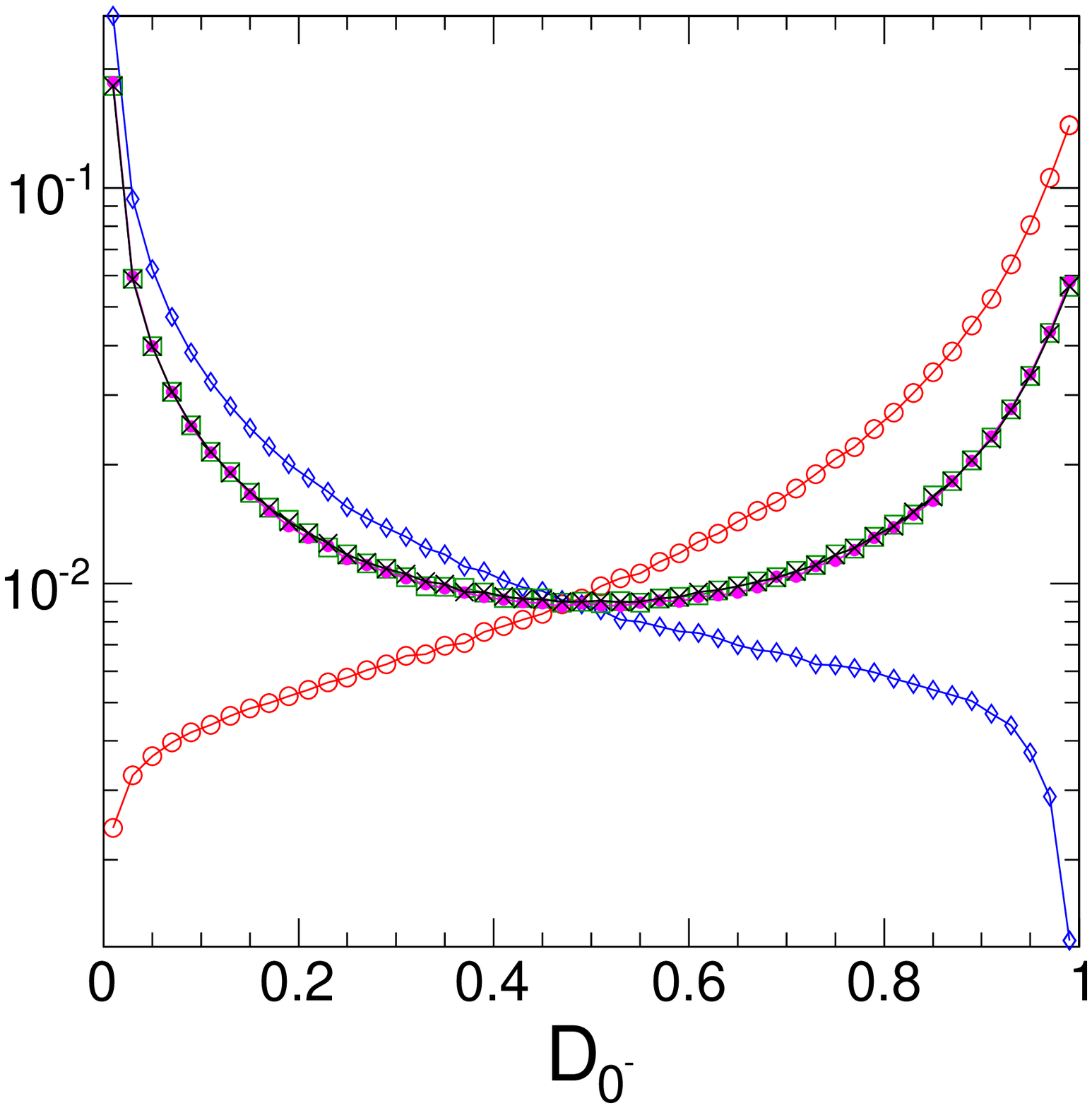}
\setlength{\epsfxsize}{0.33\linewidth}\leavevmode\epsfbox{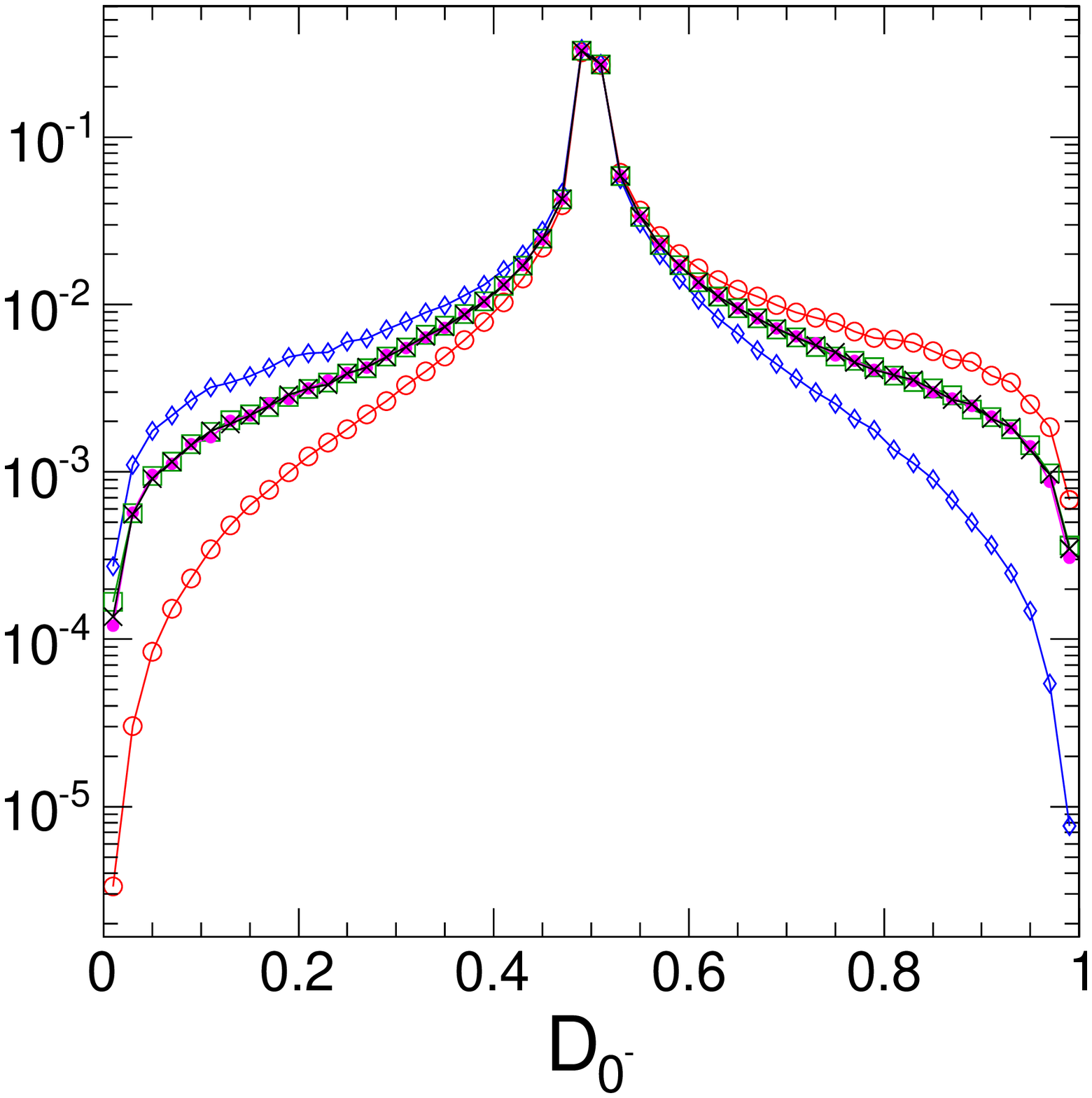}
\setlength{\epsfxsize}{0.33\linewidth}\leavevmode\epsfbox{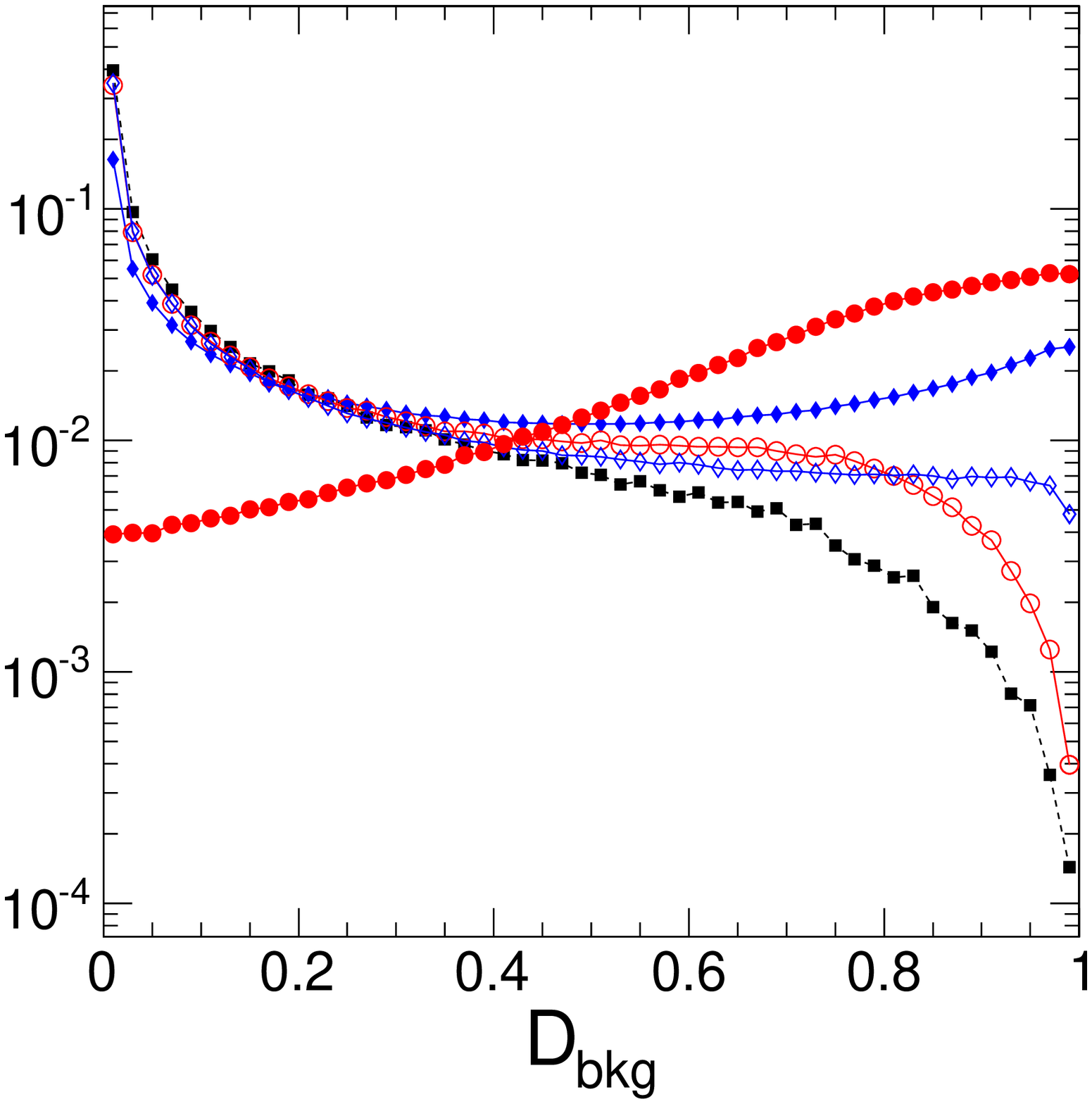}
}
\caption{
Distribution of $D_{0^-}$ for generated events in two VBF-like topologies:
$V^*V^*\to H$ + 2 jets (left) and $gg \to H$ + 2 jets (middle), with $H\to ZZ^*$ as an example.
Four processes are shown: SM ($0^+$, red open circles), pseudoscalar ($0^-$, blue diamonds), 
and two mixed states corresponding to $f_{a3}=0.5$ with $\phi_{a3}=0$ (green squares)
and $\pi/2$ (magenta points). 
Also shown is the $f_{a3}=0.5$ sample with no interference between the scalar and pseudoscalar
terms simulated (black crosses).
Right: Distribution of ${D_{\rm bkg}}$ for VBF topology considering 
$V^*V^*\to H$ + 2 jets as signal and $gg \to H$ + 2 jets as background.
The following processes are shown: 
WBF $0^+$ (red solid circles),
WBF $0^-$ (blue solid diamonds),
SBF $0^+$ (red open circles),
SBF $0^-$ (blue open diamonds),
$ZZ$ + 2 jets background (black squares).
Events are shown after selection requirements.
}
\label{fig:D0mn-vbf}
\end{figure}

We now make a technical comment that allows us to simplify fitting for $f_{a3}$  when distribution of 
$D_{0^-}$ is employed.  Consider a $C\!P$-mixed case. 
The matrix element squared, which is used to generate events for the $D_{0^-}$ distribution,
contains the square of $C\!P$-even part, the square of $C\!P$-odd part, and the 
interference of the two, as shown in Eq.~(\ref{eq:fractions-P}).
We observe that the interference part does not contribute to 
the distribution of $D_{0^-}$ variable; 
the illustration for five production and decay processes considered
in this paper can be found in  Figs.~\ref{fig:D0mn} and \ref{fig:D0mn-vbf}.
This allows us to set up a simple procedure 
by generating $f_{a3}$-independent $C\!P$-even and $C\!P$-odd events {\it once} and then combining 
them in appropriate proportion.  This feature is unique for $f_{a3}$ measurements. 
As long as only a limit is set on  $f_{a3}$, such an analysis may be sufficient.
Note that this approach is equivalent to averaging over all possible phases of the
amplitude, $\phi_{a3}$, which is generally unknown until measured. 

\begin{figure}[t]
\centerline{
\setlength{\epsfxsize}{0.33\linewidth}\leavevmode\epsfbox{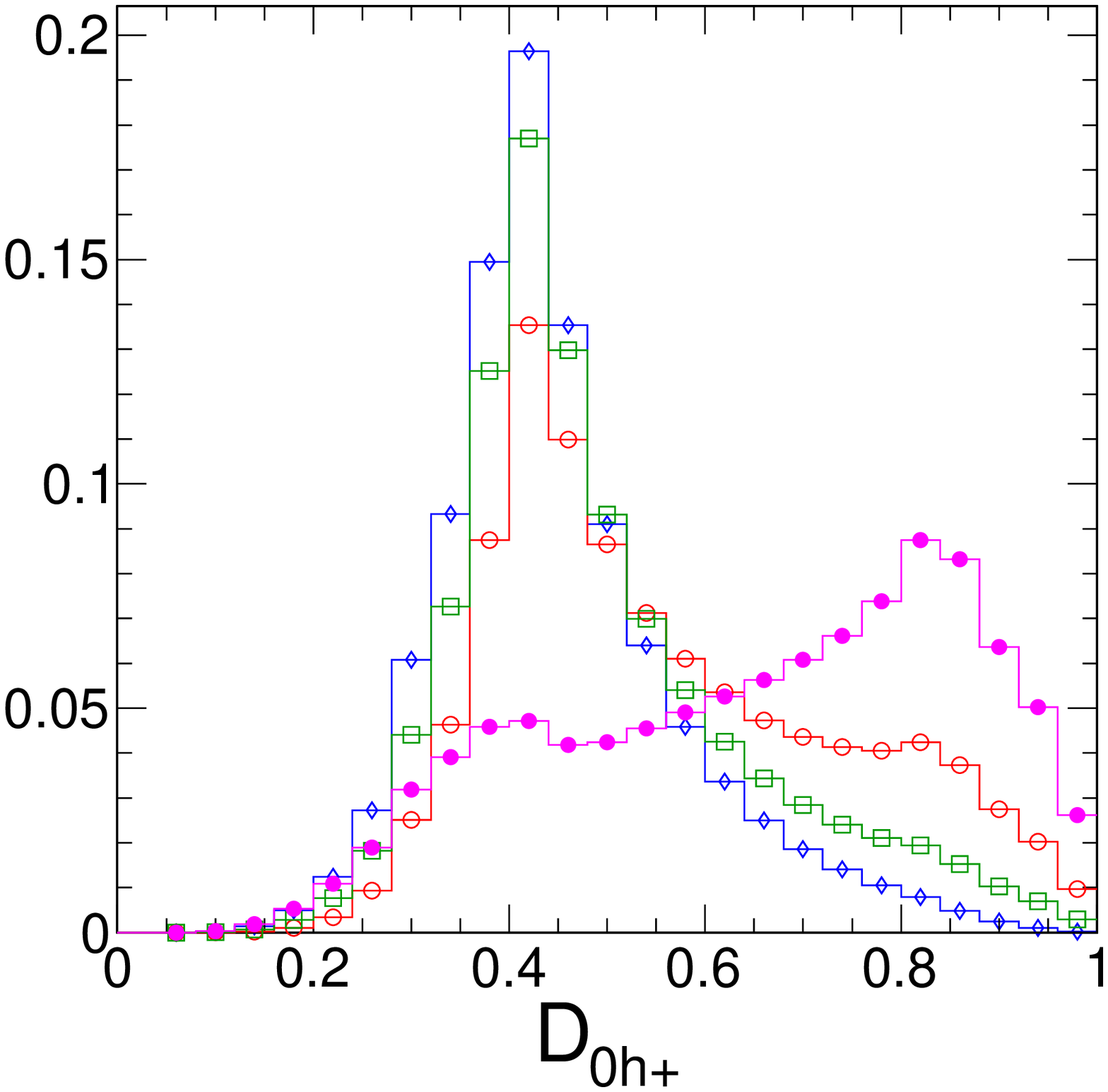}
\setlength{\epsfxsize}{0.33\linewidth}\leavevmode\epsfbox{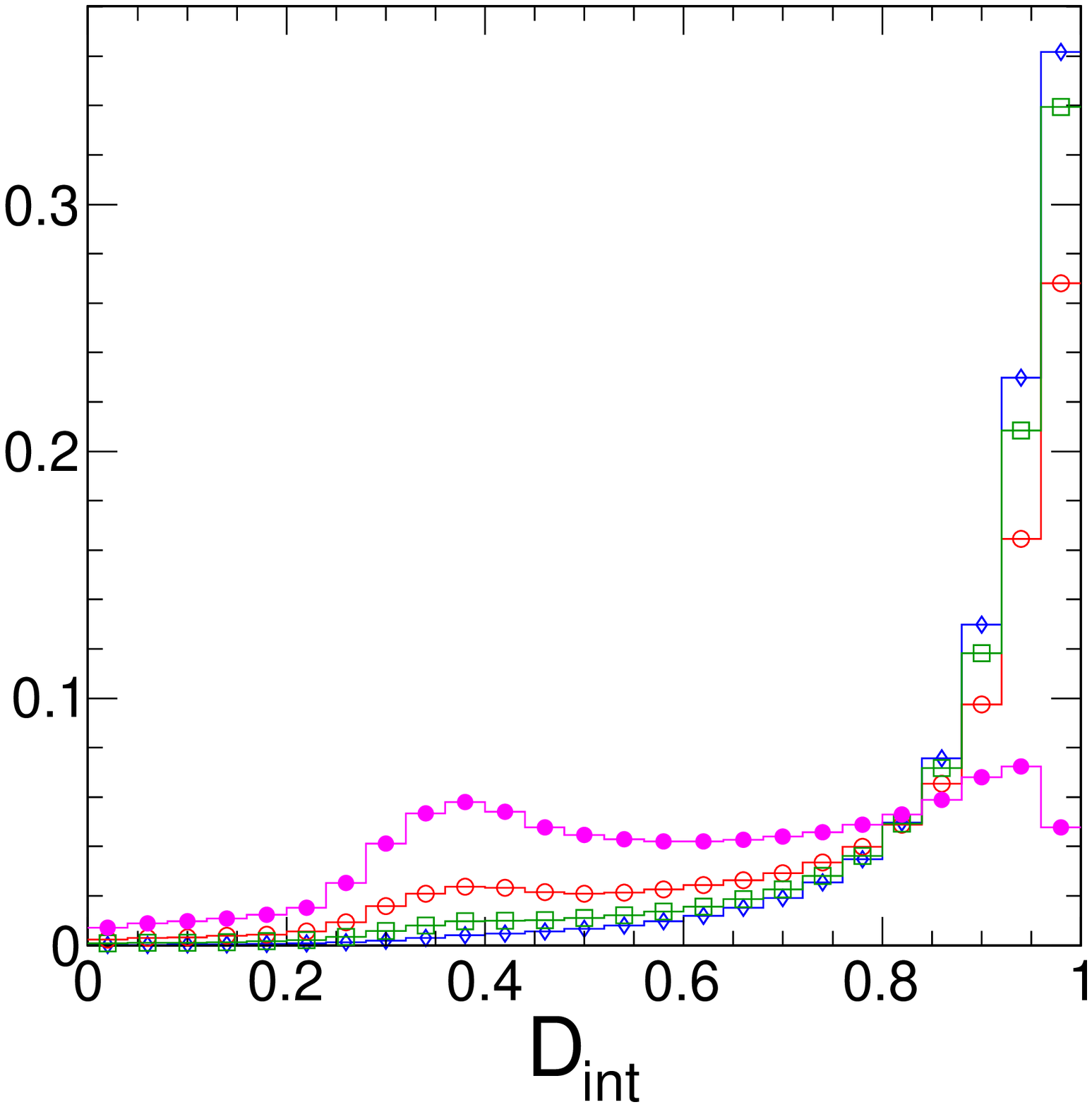}
\setlength{\epsfxsize}{0.33\linewidth}\leavevmode\epsfbox{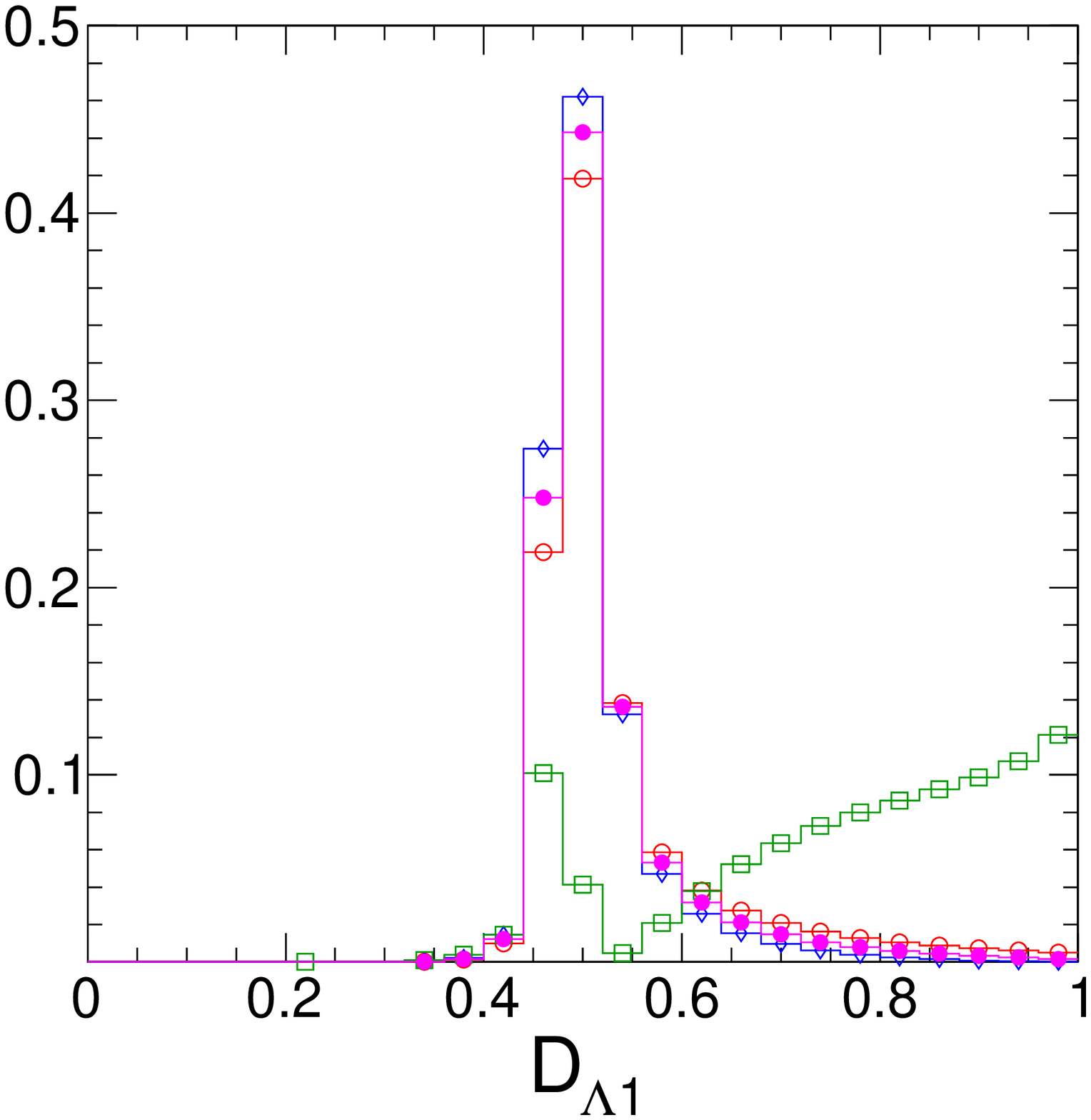}
}
\caption{
Distribution of $D_{0h^+}$ (left), $D_{\rm int}$ (middle), and  $D_{\Lambda1}$ (right) for generated events in the $H\to ZZ\to 4\ell$ process.
Four scenarios are shown. Left and middle plots: SM scalar ($0^+$, red open circles), BSM scalar ($0_h^+$, blue diamonds), 
and two mixed states corresponding to $f_{a2}=0.5$ with $\phi_{a2}=0$ (green squares) and $\pi$ (magenta points). 
Right plot: SM scalar ($0^+$, red open circles), BSM scalar ($f_{\Lambda1}=1$, blue diamonds), 
and two mixed states corresponding to $f_{\Lambda1}=0.5$ with $\phi_{\Lambda1}=0$ (green squares) and $\pi$ (magenta points). 
}
\label{fig:D0hpl}
\end{figure}

It is possible to extend the above approach and create a discriminant, $D_{C\!P}$, which is sensitive to interference
of the $0^+$ ($g_1$) and $0^-$ ($g_4$) terms
\begin{eqnarray}
\label{eq:melaSig-interf}
{D_{C\!P}}=\frac{{\cal P}_{\rm int}(m_1, m_2, \vec\Omega; \phi_{a3}) }{{\cal P}_{0^+}(m_1, m_2, \vec\Omega) +{\cal P}_{0^-}(m_1, m_2, \vec\Omega) }
\,.
\end{eqnarray}
This analysis includes two observables $\vec{x}_{i}=\{D_{0^-}, D_{C\!P}\}_i$ 
and one parameter $\vec{\zeta}=\{f_{a3} \}$ for a given value of $\phi_{a3}$.
Such an  approach can be also applied to other cases, such as a measurement of the 
parameter $f_{a2}$, where the ${\cal P}_{\rm int}$ cannot be omitted. 
The corresponding discriminants are called  $D_{0h^+}$ and $D_{\rm int}$, instead of  
$D_{0^-}$ and $D_{C\!P}$; their distributions are shown in Fig.~\ref{fig:D0hpl}. 
The strong interference effect is visible in Fig.~\ref{fig:D0hpl} and the full treatment 
Eq.~(\ref{eq:fractions-P}) is needed.
Finally a discriminant $D_{\Lambda1}$ is also shown in Fig.~\ref{fig:D0hpl} which is designed to 
separate the $g_1(q_1^2,q_2^2)=-g_1^{\prime\prime}\times(q_1^2+q_2^2)/\Lambda_1^2$ 
anomalous coupling term from the Standard Model coupling.
 
Equation~(\ref{eq:fractions-P}) can be easily extended to an arbitrary number of contributing amplitudes.
For example, an arbitrary complex phase $\phi_{a3}$ can be easily incorporated noting that 
${\cal P}_{\rm int}(\vec{x}_{i}; \phi_{a3})={\cal P}_{\rm int}(\vec{x}_{i}; \phi_{a3}=0)\times\cos\phi_{a3}+
{\cal P}_{\rm int}(\vec{x}_{i}; \phi_{a3}=\pi/2)\times\sin\phi_{a3}$. This and three discriminants 
$D_{0^-}$, $D_{C\!P}$ and $D_{C\!P}^\perp$ computed in Eq.~(\ref{eq:melaSig-interf}) for $\phi_{a3}=0$ and $\pi/2$, 
provide full information for the measurement of $f_{a3}$ and $\phi_{a3}$ simultaneously. 
Equivalently, three terms in Eq.~(\ref{eq:fractions-P}) would be extended
to six terms when interference of three amplitudes is considered, such as simultaneous measurement of
$f_{a2}$ and $f_{a3}$ with real couplings. The number of terms is increased to nine when arbitrary phases 
$\phi_{a2}$ and $\phi_{a3}$ are considered. The number of relevant discriminants is also increased with
one discriminant for each term in the probability distribution, except for the Standard Model coupling.
However, some discriminants carry most of the relevant information and a reduced set may be sufficient, 
as we illustrate in Sec.~\ref{sect:HZZ_LHC}. Nonetheless, the present goal is to test the presence of one 
anomalous coupling at a time, and the approach with one or two optimal discriminants is sufficient for many 
of such measurements.

We also comment on the technical implementation of Eq.~(\ref{eq:fractions-P}) and its extensions to a
larger number of interfering amplitudes. Each probability distribution as a function of observables (such as
one or two discriminants) can be easily obtained with MC simulation including all detector effects for
signal and full parameterization for background. It is sufficient to generate only as many signal MC samples as 
there are terms in the equation, three in the case of Eq.~(\ref{eq:fractions-P}). Interference parameterization can be
easily extracted from the combination of the mixed and pure samples following the same Eq.~(\ref{eq:fractions-P}).
For signal, it is also possible to generate just one MC sample covering the phase-space of observables,
and then re-weight the MC parameterization using the matrix element ratios discussed in Appendix~\ref{sect:me}. 
 
Background  treatment  requires special consideration. The set $\vec{x}_{i}$ can be extended to
include observables discriminating against background, such as reconstructed Higgs boson invariant mass.
For studies presented here, we adopt a simplified approach where instead of including Higgs boson
invariant mass we fix the number of background events $n_{\rm bkg}$ to expected yields. 
Nonetheless, in some cases an effective background suppression can be achieved with a matrix
element approach as well. In such a case we employ a discriminant optimal for background suppression
\begin{eqnarray}
\label{eq:melaBkg}
{D_{\rm bkg}}=\left[1+\frac{{\cal P}_{\rm bkg} (m_1, m_2, \vec\Omega) }
{{\cal P}_{0^+} (m_1, m_2, \vec\Omega) } \right]^{-1}
\,
\end{eqnarray}
and extend the set of observables to include $\vec{x}_{i}=\{D_{\rm bkg}, D_{0^-},...\}_i$. We note that this is needed only
in the approach employing discriminants. In the case of multidimensional fits, complete kinematic information is already 
contained in the set of observables. 

To illustrate the use of $D_{\rm bkg}$, we show the separation of the gluon-fusion ``background'' and 
the weak-boson ``signal'' in $H+2j$ events in Fig.~\ref{fig:D0mn-vbf}.  
A similar approach can be used  in the analysis of $VH$ production with the decay 
$V\to2$ jets, where the gluon fusion process $H+2$ jets is treated as a background. 
Alternatively, extraction of  $f_{a3}$ in gluon fusion should treat the WBF process as a background. 

\begin{figure}[t]
\centerline{
\setlength{\epsfxsize}{0.25\linewidth}\leavevmode\epsfbox{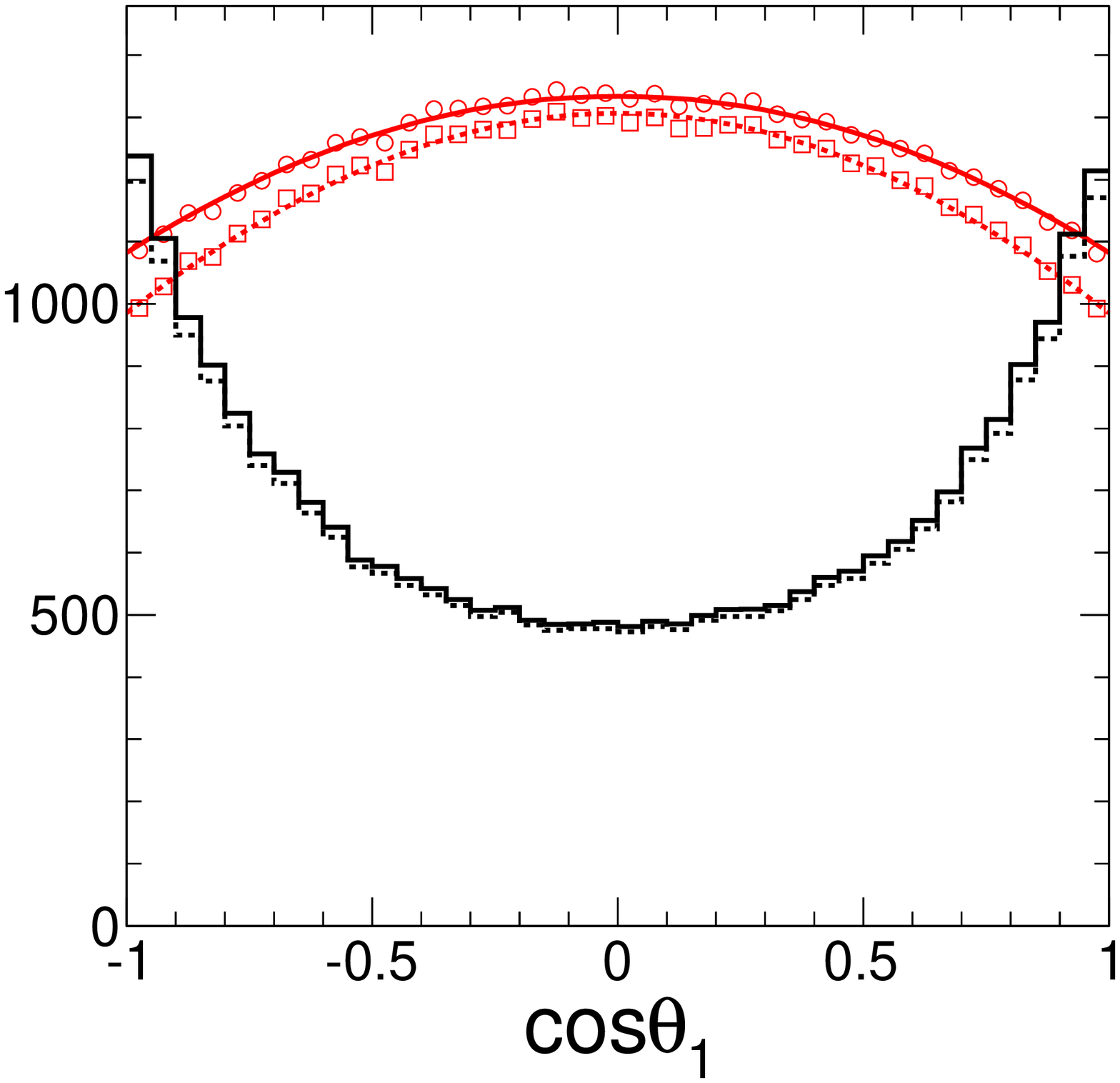}
\setlength{\epsfxsize}{0.25\linewidth}\leavevmode\epsfbox{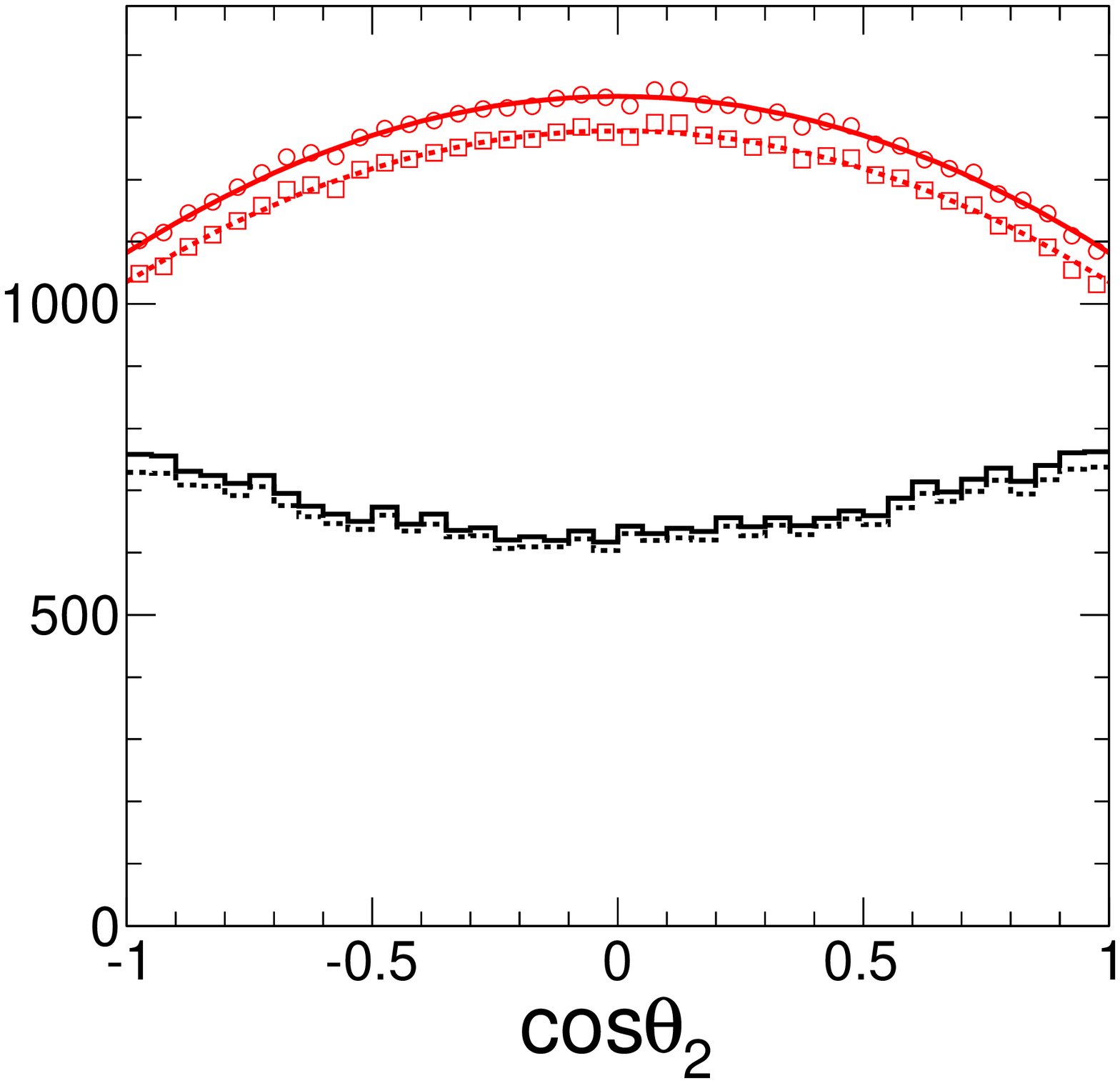}
\setlength{\epsfxsize}{0.25\linewidth}\leavevmode\epsfbox{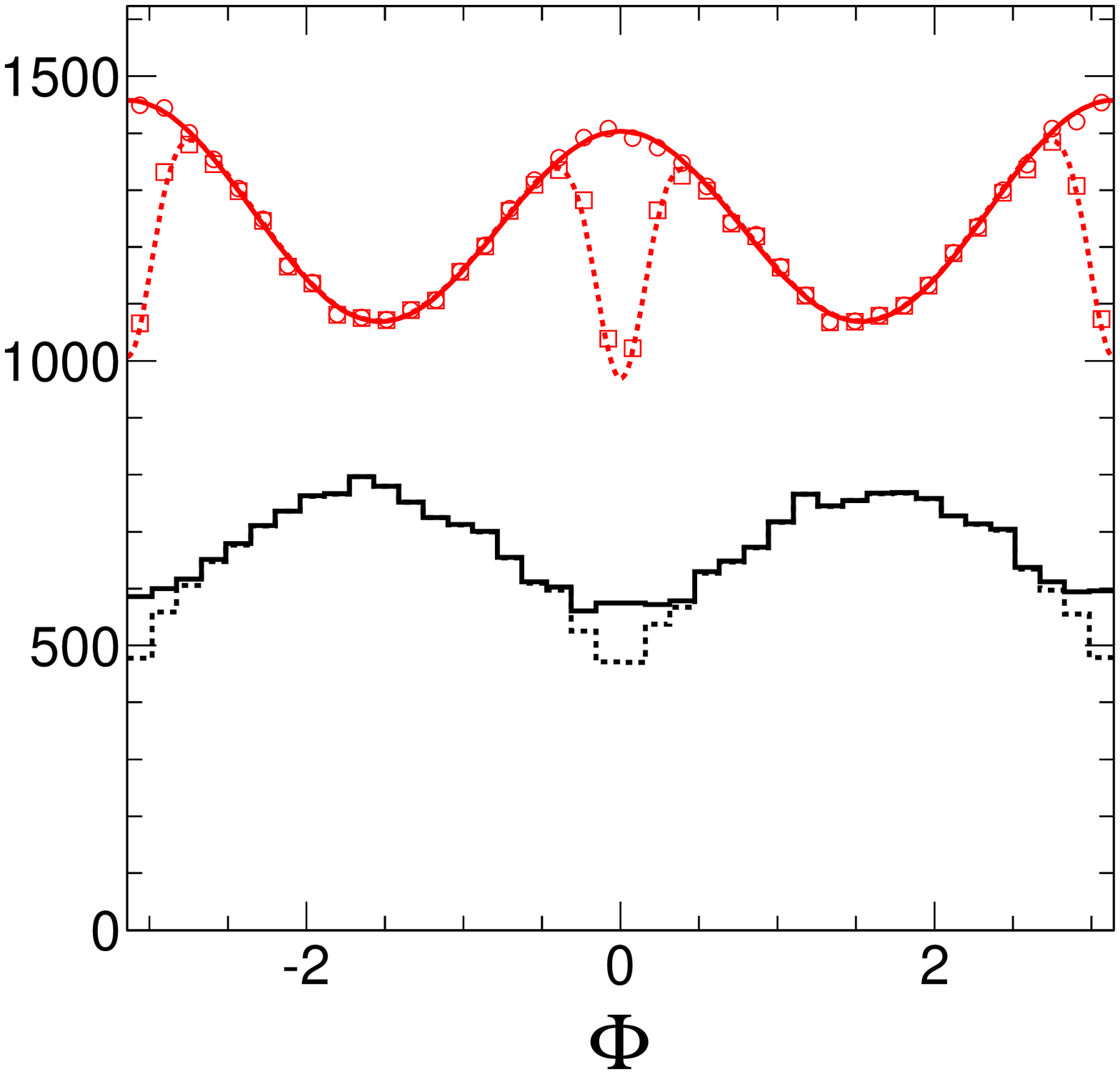}
}
\caption{
Distributions of the observables in the $e^+e^-\to ZH$ analysis at $\sqrt{s}=250$ GeV, 
from left to right: $\cos\theta_1$, $\cos\theta_2$, and $\Phi$.
Points (red) show simulated events for the SM Higgs boson with curves showing projections of analytical distributions.
Histograms (black) show background distributions. 
Distributions before (solid) and after (dashed) detector acceptance effects are shown.
}
\label{fig:ilc_detector}
\end{figure}

The above approaches with kinematic discriminants simplify parameterization of detector effects and backgrounds. 
The idea behind those approaches is to store most relevant information in as few observables as possible,
simplifying the analysis and focussing on the most interesting measurements. A complementary approach is to
try to describe both detector effects and backgrounds in a multi-dimensional space of observables.
This approach allows the full multi-parameter implementation in Eq.~(\ref{eq:likelihood}).
For the final states with leptons, the resolution effects are typically small and can be ignored for
most of the observables. When such effects become important, detector transfer functions between the
ideal and the reconstructed observables can in principle be incorporated into the probability distributions.  
The non-uniform reconstruction efficiency can be modeled with the acceptance function ${\cal G}$ 
which enters the ${\cal P}_{\rm sig}$ parameterization and is given by the step-function
\be
{\cal G}(m_1,m_2,\vec{\Omega}) = \prod_{\ell} 
\theta(|\eta_{\rm max}| - |\eta_{\ell}(m_1,m_2,\vec{\Omega})|)\,,
\label{eq:ac1}
\ee
where $\eta_{\ell}=\ln\cot(\theta_{\ell}/2)$ is the pseudorapidity of a lepton
and $|\eta_{\rm max}|$ is the maximal pseudorapidity in reconstruction.
We also assume that the detection efficiency does not change
within the detector acceptance, otherwise ${\cal G}$ is multiplied by the non-uniform function.
We illustrate the effect on observables in the $e^+e^-\to ZH\to \ell\ell H$ analysis at $\sqrt{s}=250$~GeV
in Fig.~\ref{fig:ilc_detector}, where the acceptance function from Eq.~(\ref{eq:ac1}) is implemented
analytically.

\begin{figure}[t]
\centerline{
\setlength{\epsfxsize}{0.25\linewidth}\leavevmode\epsfbox{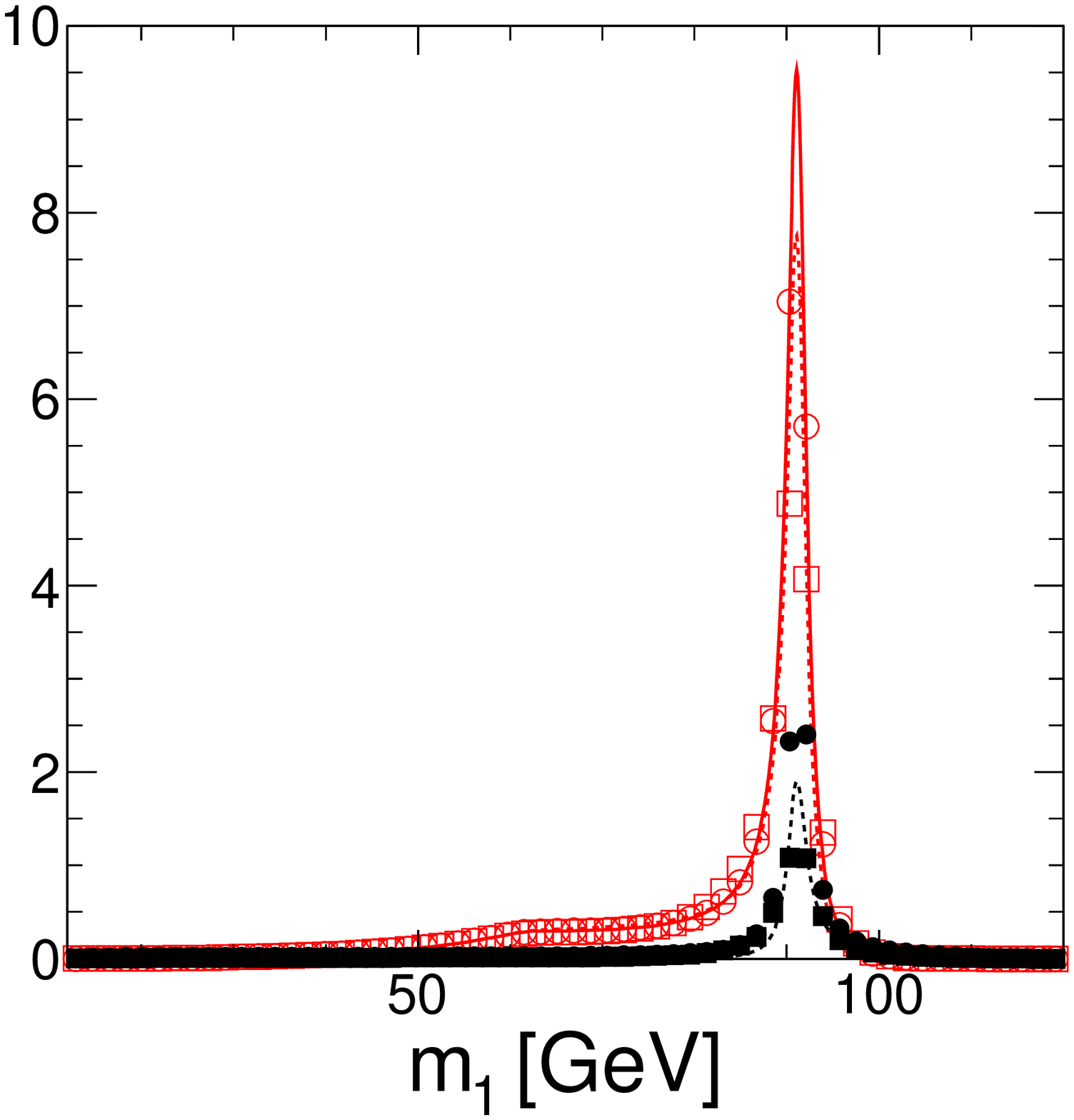}
\setlength{\epsfxsize}{0.25\linewidth}\leavevmode\epsfbox{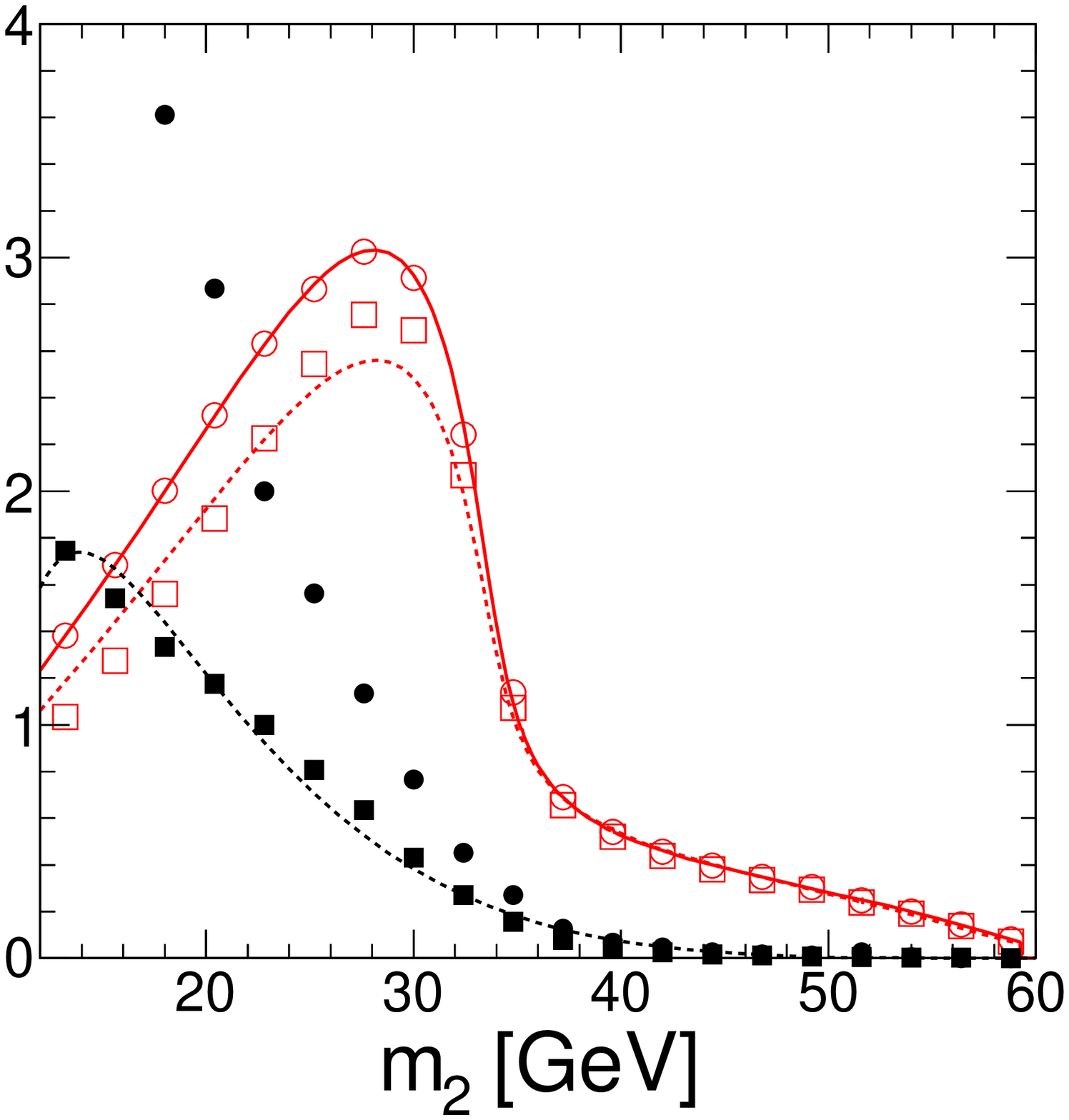}
\setlength{\epsfxsize}{0.25\linewidth}\leavevmode\epsfbox{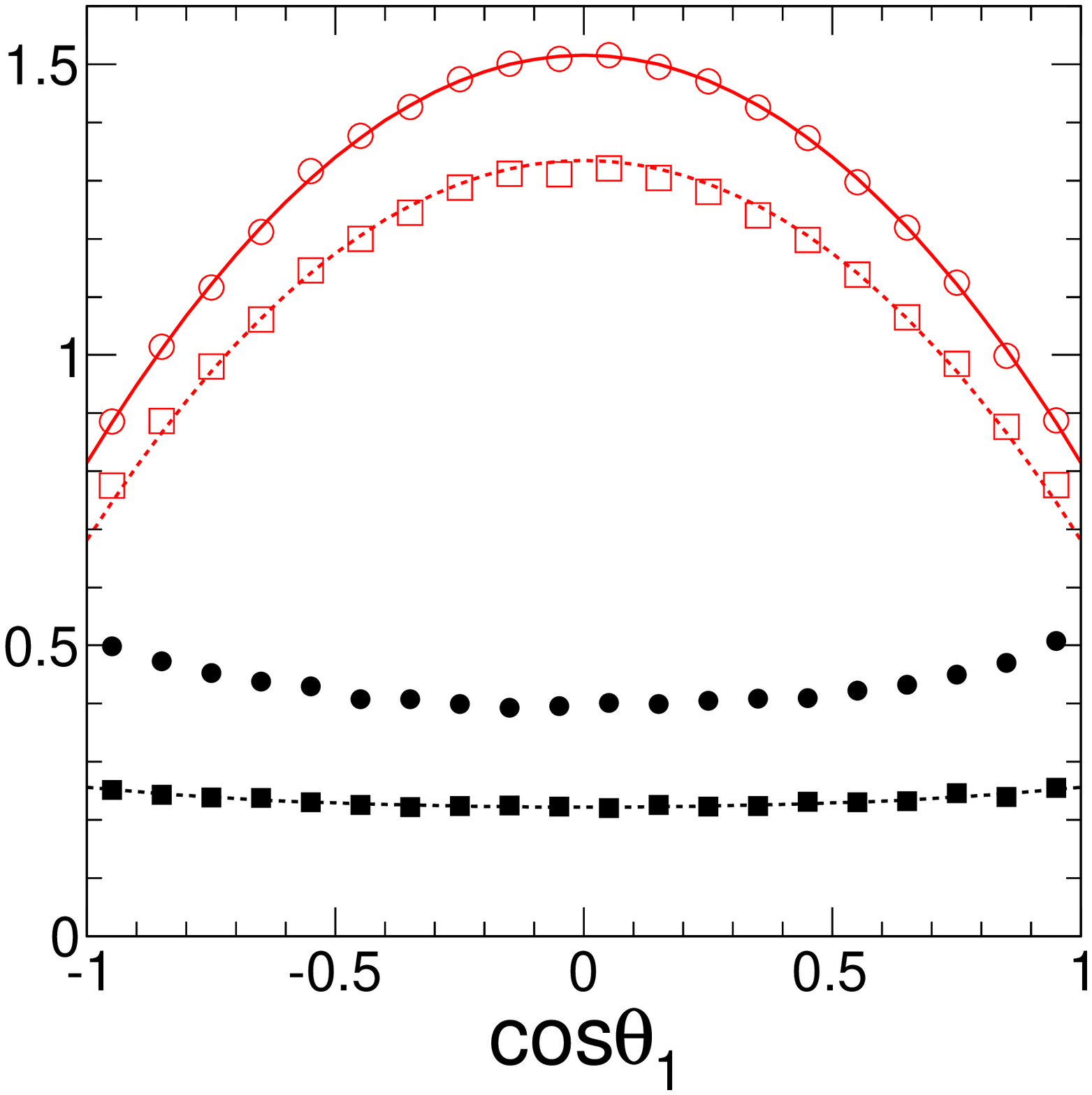}
\setlength{\epsfxsize}{0.25\linewidth}\leavevmode\epsfbox{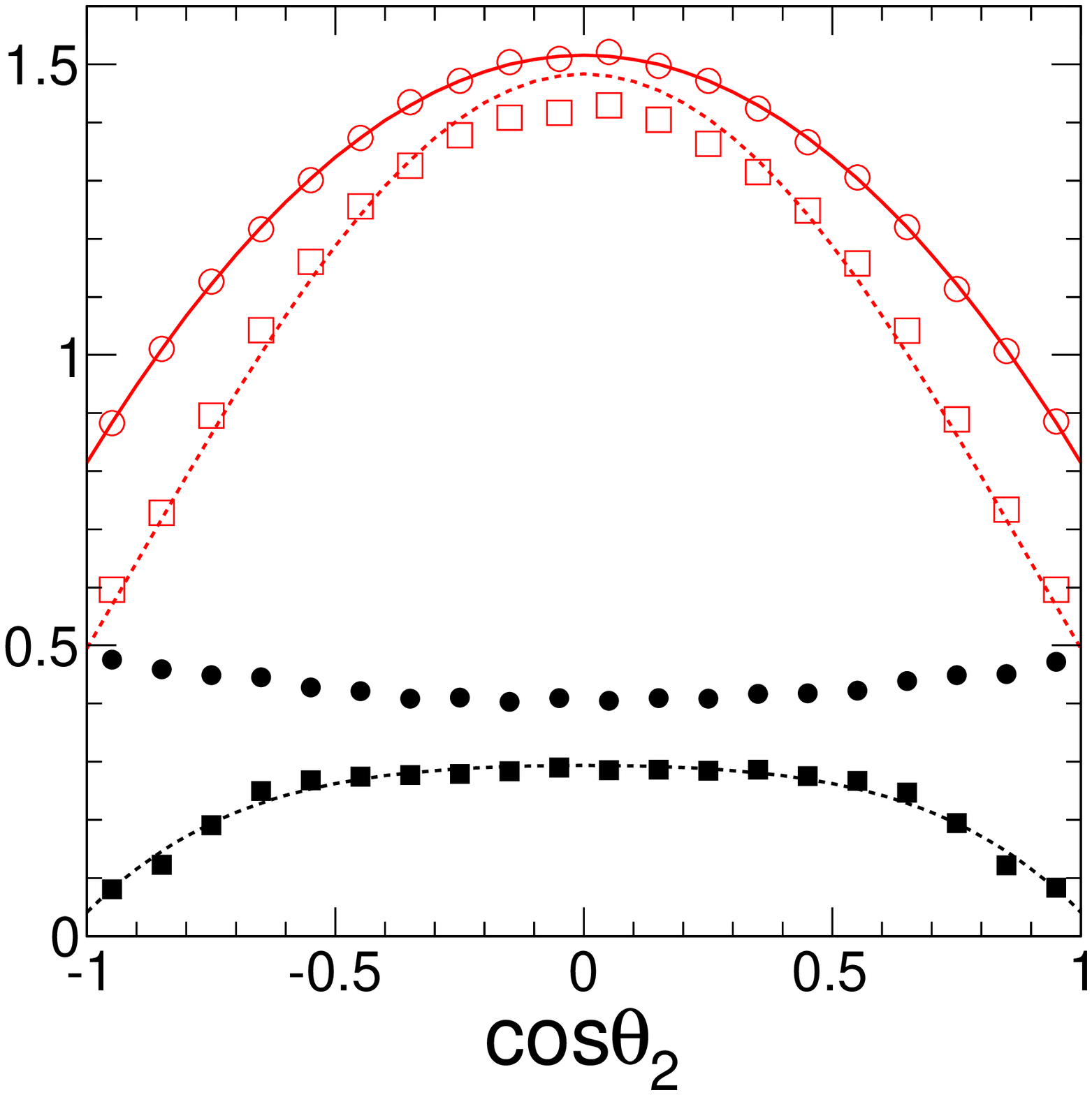}
}
\centerline{
\setlength{\epsfxsize}{0.25\linewidth}\leavevmode\epsfbox{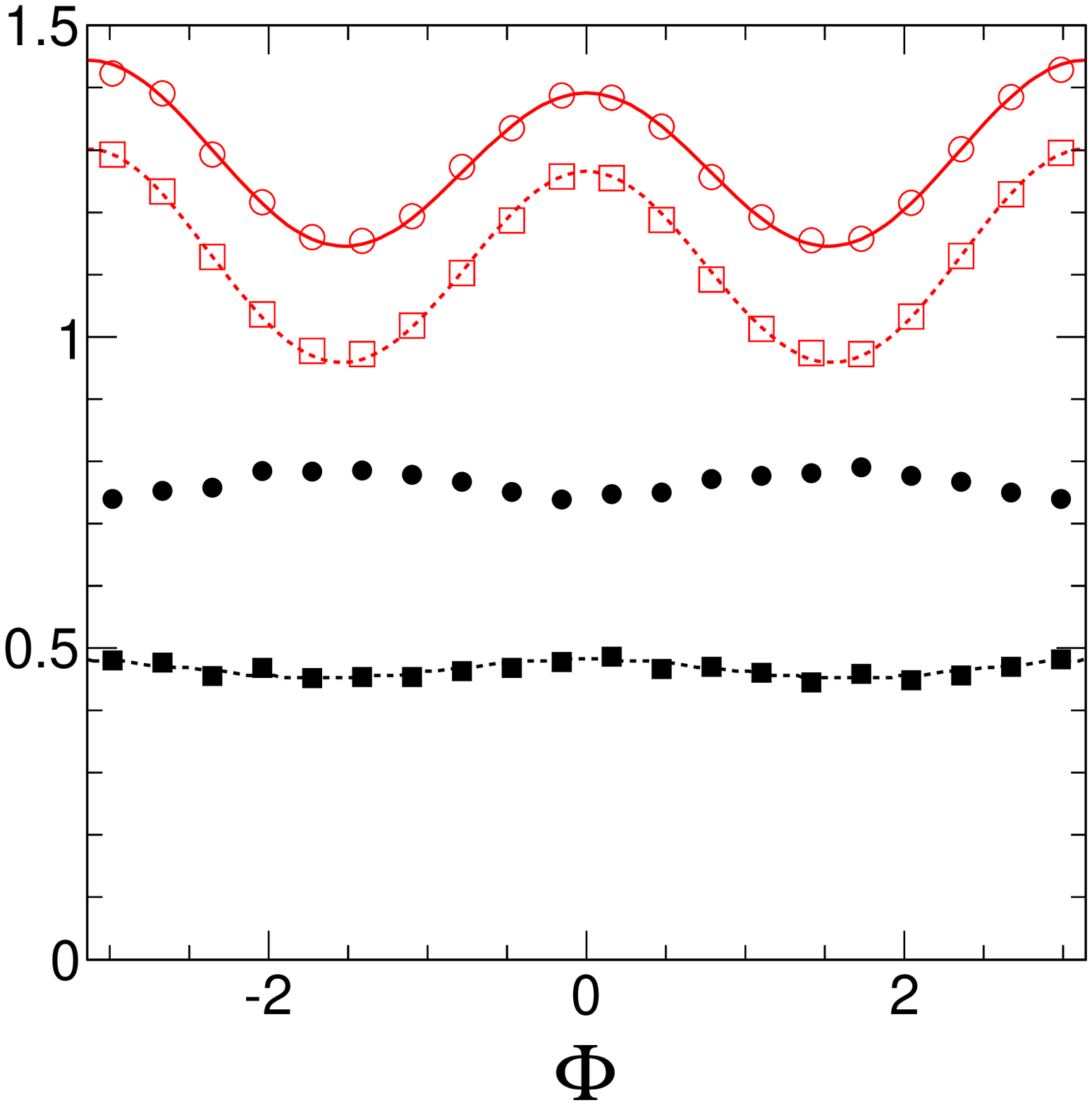}
\setlength{\epsfxsize}{0.25\linewidth}\leavevmode\epsfbox{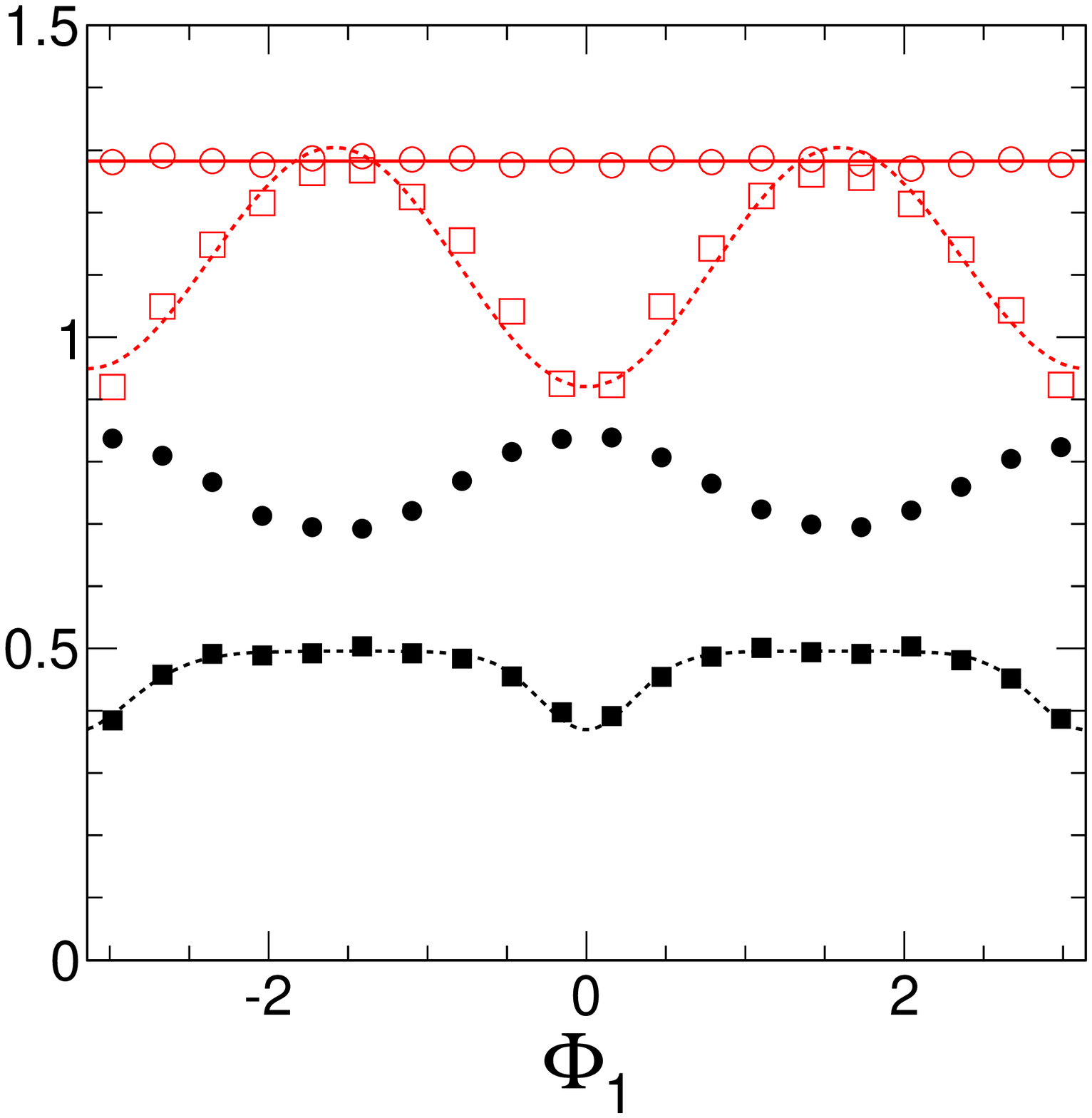}
\setlength{\epsfxsize}{0.25\linewidth}\leavevmode\epsfbox{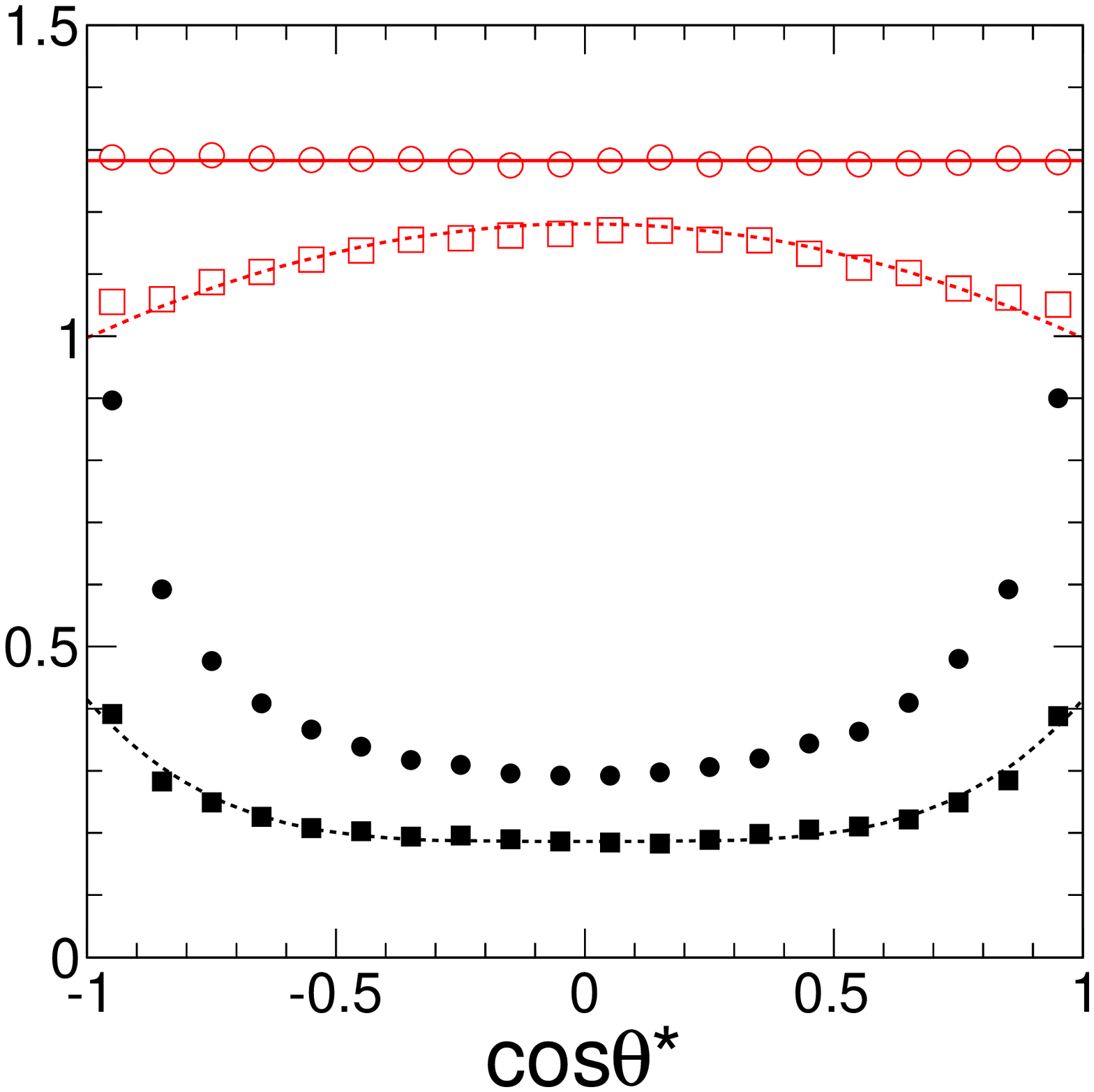}
}
\caption{
 Distributions of various observables 
($m_1$, $m_2$, $\cos\theta_1$, $\cos\theta_2$,  $\Phi$, $\Phi_1$, and $\cos\theta^*$) 
in the $H\to ZZ^*\to 4\ell$ analysis at the  LHC.
Open red points show simulated events for the SM Higgs boson with curves showing projections of analytical distributions.
Solid black points show background distributions with curves showing projections of analytical parameterization.
Distributions before (circles) and after (squares) detector acceptance effects are shown.
}
\label{fig:lhc_detector}
\end{figure}

Parameterization of background distributions, ${\cal P}_{\rm bkg}$, with multiple observables is also 
possible analytically. For example, in the $H\to ZZ^*\to 4f$ analysis, 
parameterization of the $q\bar{q}\to ZZ^*/Z\gamma^*$ process is available in Ref.~\cite{Chen:2012jy}, 
and detector effects can be included in a manner similar to what we described for the signal. 
Alternatively, a multi-dimensional template histogram can be used in place of such parameterization, 
potentially with proper smoothing of the distributions if there are sufficient statistics of simulated events. 
We have investigated both of these approaches for signal and background parameterization and 
found both of them feasible. However, some of the technical limitations include normalization 
of probabilities in multidimensional space, which may slow down the data analysis considerably.
Therefore, for multi-parameter fits presented in this paper, we employ
a simplified approach when both acceptance functions, ${\cal G}$, and background distributions,
${\cal P}_{\rm bkg}$, are approximated with analytical functions describing generated distributions 
in either one or two dimensions, see for example Fig.~\ref{fig:lhc_detector}.
The results of such studies are verified to give correct expectations for 
measurement precision by comparing to the expectations without detector effects or background
(optimistic), and with the full treatment of detector effects and background using discriminant approach 
as in Eq.~(\ref{eq:melaSig}), which serve as two bounds of expected performance. 
In most cases, all three results provide similar expectations and we quote results from the 
multi-parameter fit. When analytical parameterization is not readily available, we quote results from the 
discriminant approach.


So far we have discussed the case of spin-zero boson, but  the tools and ideas presented in this paper 
can be extended 
to any spin-parity study, such as multi-parameter fits of a spin-two hypothesis.  In such a case, non-trivial
$\cos\theta^*$ and $\Phi_1$ distributions appear and depend on the production mechanism.
It is desirable to extend the matrix element approach in such 
a way that it does not depend on the production model of a particle with non-zero spin
but considers only its decay.
This feature can be easily achieved by considering the unpolarized $X$-boson production
by either averaging over the spin degrees of freedom of the produced $X$-boson or, equivalently,
integrating over the two production angles $\cos\theta^*$ and $\Phi_1$, defined in Fig.~\ref{fig:decay},
in the probability distribution ${\cal P}$~\cite{Gao:2010qx,Bolognesi:2012mm}. 
This leads to  the following expression for the spin-averaged matrix element squared for 
the decay of a new boson $X$
\begin{eqnarray}
\label{eq:melaSigProd}
\int \mathrm d\Phi_1  \mathrm d\cos\theta^{*}  \,\,{\cal P} (m_1, m_2, \vec\Omega) 
\,.
\end{eqnarray}
This method applies to any possible hypothesis with non-zero spin and small residual effects
arising from detector acceptance can be addressed in experimental analysis.
We provide tools that allow one to pursue this approach using  both analytic and numerical 
computations of the probability distribution ${\cal P}$~\cite{support}.




\begin{thebibliography}{99}

\bibitem{discovery-atlas} 
ATLAS collaboration, G.~Aad {\it et al.},
  Phys.\ Lett.\ B {\bf 716}, 1 (2012)
  [arXiv:1207.7214 [hep-ex]].
  
\bibitem{discovery-cms} 
CMS collaboration, S.~Chatrchyan {\it et al.}, 
  Phys.\ Lett.\ B {\bf 716}, 30 (2012)
  [arXiv:1207.7235 [hep-ex]];
  JHEP {\bf 1306}, 081 (2013)
  [arXiv:1303.4571 [hep-ex]].
  
  
\bibitem{evidence-tev} 
CDF and D0 Collaborations, T.~Aaltonen {\it et al.},
  Phys.\ Rev.\ Lett.\  {\bf 109}, 071804 (2012)
  [arXiv:1207.6436 [hep-ex]].
  
\bibitem{landau}
L. D. Landau, Dokl. Akad. Nauk SSSR {\bf 60}, 207 (1948);
C. N. Yang, Phys. Rev. {\bf 77}, 242 (1950).

\bibitem{properties-cms} 
CMS collaboration, S.~Chatrchyan {\it et al.}, Phys. Rev. Lett. {\bf 110}, 081803 (2013)
[arXiv:1212.6639 [hep-ex]];
  arXiv:1312.1129 [hep-ex];
  arXiv:1312.5353 [hep-ex];
%
preprints CMS-PAS-HIG-12-041, 
CMS-PAS-HIG-13-001, 
CMS-PAS-HIG-13-002, 
CMS-PAS-HIG-13-003, 
CMS-PAS-HIG-13-005, 
CMS-PAS-HIG-13-012, 
CMS-PAS-HIG-13-016. 

\bibitem{properties-atlas} 
ATLAS collaboration, G.~Aad {\it et al.}, 
  Phys.\ Lett.\ B {\bf 726}, 88 (2013) [arXiv:1307.1427 [hep-ex]];
  Phys.\ Lett.\ B {\bf 726}, 120 (2013) [arXiv:1307.1432 [hep-ex]];
preprints ATL-CONF-2013-013, ATL-CONF-2013-029, ATL-CONF-2013-031. 


\bibitem{Gao:2010qx} 
  Y.~Gao, A.~V.~Gritsan, Z.~Guo, K.~Melnikov, M.~Schulze and N.~V.~Tran,
  Phys.\ Rev.\ D {\bf 81}, 075022 (2010)
  [arXiv:1001.3396 [hep-ph]].
  
\bibitem{Bolognesi:2012mm} 
  S.~Bolognesi, Y.~Gao, A.~V.~Gritsan, K.~Melnikov, M.~Schulze, N.~V.~Tran and A.~Whitbeck,
  Phys.\ Rev.\ D {\bf 86}, 095031 (2012)
  [arXiv:1208.4018 [hep-ph]].

\bibitem{Accomando:2006ga} 
  E.~Accomando, A.~G.~Akeroyd, E.~Akhmetzyanova, J.~Albert {\it et al.},
  hep-ph/0608079.
  
\bibitem{Heinemeyer:2013tqa} 
LHC Higgs Cross Section Working Group Collaboration, S. Heinemeyer {\it et al.},
arXiv:1307.1347 [hep-ph].

\bibitem{Barger:1993wt} 
  V.~D.~Barger, K.~-m.~Cheung, A.~Djouadi, B.~A.~Kniehl and P.~M.~Zerwas,
  Phys.\ Rev.\ D {\bf 49}, 79 (1994)
  [hep-ph/9306270].
  
\bibitem{sp6} 
  T.~Han and J.~Jiang,
  Phys.\ Rev.\ D {\bf 63}, 096007 (2001)
  [hep-ph/0011271].
  
\bibitem{sp4}  
  T.~Plehn, D.~L.~Rainwater and D.~Zeppenfeld,
  Phys.\ Rev.\ Lett.\  {\bf 88}, 051801 (2002)
  [hep-ph/0105325].

\bibitem{sp1} 
  S.~Y.~Choi, D.~J.~Miller, M.~M.~Muhlleitner and P.~M.~Zerwas,
  Phys.\ Lett.\ B {\bf 553}, 61 (2003) [hep-ph/0210077];
%
  R.~M.~Godbole, D.~J.~Miller, and M.~M.~Muhlleitner,
  JHEP {\bf 0712}, 031 (2007)
  [arXiv:0708.0458 [hep-ph]].
  
\bibitem{sp5} 
  V.~Hankele, G.~Klamke, D.~Zeppenfeld and T.~Figy,
  Phys.\ Rev.\ D {\bf 74}, 095001 (2006)
  [hep-ph/0609075].
  
\bibitem{DelDuca:2006hk} 
  V.~Del Duca, G.~Klamke, D.~Zeppenfeld, M.~L.~Mangano, M.~Moretti, F.~Piccinini, R.~Pittau and A.~D.~Polosa,
  JHEP {\bf 0610}, 016 (2006)
  [hep-ph/0608158].
  
\bibitem{Mahlon:2006zc} 
  G.~Mahlon and S.~J.~Parke,
  Phys.\ Rev.\ D {\bf 74}, 073001 (2006)
  [hep-ph/0606052].

\bibitem{sp10} 
  S.~Dutta, K.~Hagiwara and Y.~Matsumoto,
  Phys.\ Rev.\ D {\bf 78}, 115016 (2008)
  [arXiv:0808.0477 [hep-ph]].

\bibitem{Hagiwara:2009wt} 
  K.~Hagiwara, Q.~Li and K.~Mawatari,
  JHEP {\bf 0907}, 101 (2009)
  [arXiv:0905.4314 [hep-ph]].

\bibitem{sp7}  
  S.~D.~Rindani and P.~Sharma,
  Phys.\ Rev.\ D {\bf 79}, 075007 (2009)
  [arXiv:0901.2821 [hep-ph]].

\bibitem{sp8}  
  S.~S.~Biswal, D.~Choudhury, R.~M.~Godbole and Mamta,
  Phys.\ Rev.\ D {\bf 79}, 035012 (2009)
  [arXiv:0809.0202 [hep-ph]].

\bibitem{sp9} 
  S.~S.~Biswal and R.~M.~Godbole,
  Phys.\ Lett.\ B {\bf 680}, 81 (2009)
  [arXiv:0906.5471 [hep-ph]].
  
\bibitem{sp2} 
  A.~De Rujula, J.~Lykken, M.~Pierini, C.~Rogan and M.~Spiropulu,
  Phys.\ Rev.\ D {\bf 82}, 013003 (2010)
  [arXiv:1001.5300 [hep-ph]].

\bibitem{sp14} 
  N.~D.~Christensen, T.~Han and Y.~Li,
  Phys.\ Lett.\ B {\bf 693}, 28 (2010)
  [arXiv:1005.5393 [hep-ph]].

\bibitem{sp16} 
  N.~Desai, B.~Mukhopadhyaya and D.~K.~Ghosh,
  Phys.\ Rev.\ D {\bf 83}, 113004 (2011)
  [arXiv:1104.3327 [hep-ph]].

\bibitem{Gainer:2011aa} 
  J.~S.~Gainer, W.~-Y.~Keung, I.~Low and P.~Schwaller,
  Phys.\ Rev.\ D {\bf 86}, 033010 (2012)
  [arXiv:1112.1405 [hep-ph]].

\bibitem{sp3} 
  D.~Stolarski and R.~Vega-Morales,
  Phys.\ Rev.\ D {\bf 86}, 117504 (2012)
  [arXiv:1208.4840 [hep-ph]].

\bibitem{sp12}	
  J.~Ellis, D.~S.~Hwang, V.~Sanz and T.~You,
  JHEP {\bf 1211}, 134 (2012)
  [arXiv:1208.6002 [hep-ph]].
  
\bibitem{sp13}  
  C.~Englert, M.~Spannowsky and M.~Takeuchi,
  JHEP {\bf 1206}, 108 (2012)
  [arXiv:1203.5788 [hep-ph]].
  
\bibitem{Boughezal:2012tz} 
  R.~Boughezal, T.~J.~LeCompte and F.~Petriello,
  arXiv:1208.4311 [hep-ph].
  
\bibitem{Chen:2012jy} 
  Y.~Chen, N.~Tran and R.~Vega-Morales,
  JHEP {\bf 1301}, 182 (2013)
  [arXiv:1211.1959 [hep-ph]];
  J.~S.~Gainer, K.~Kumar, I.~Low and R.~Vega-Morales,
  JHEP {\bf 1111}, 027 (2011)
  [arXiv:1108.2274 [hep-ph]].
  
\bibitem{Artoisenet:2013puc} 
  P.~Artoisenet, P.~de Aquino, F.~Demartin, R.~Frederix, S.~Frixione, F.~Maltoni, M.~K.~Mandal and P.~Mathews {\it et al.},
JHEP {\bf 1311}, 043 (2013) [arXiv:1306.6464 [hep-ph]].

\bibitem{sp15} C.~Englert {\it  et al.}, 
JHEP {\bf 1301}, 148 (2013) [arXiv:1212.0843 [hep-ph]].
  
\bibitem{sp11} 
  A.~Djouadi, R.~M.~Godbole, B.~Mellado and K.~Mohan,
  Phys.\ Lett.\ B {\bf 723}, 307 (2013)
  [arXiv:1301.4965 [hep-ph]].
  
\bibitem{Gainer:2013rxa} 
  J.~S.~Gainer, J.~Lykken, K.~T.~Matchev, S.~Mrenna and M.~Park,
  Phys.\ Rev.\ Lett.\  {\bf 111}, 041801 (2013)
  [arXiv:1304.4936 [hep-ph]].
 
\bibitem{Sun:2013yra} 
  Y.~Sun, X.~-F.~Wang and D.~-N.~Gao,
  arXiv:1309.4171 [hep-ph].
 

\bibitem{support} 
The Monte-Carlo generator, the manual, and supporting material can be downloaded from 
http://www.pha.jhu.edu/spin/

\bibitem{Alwall:2006yp}
  J.~Alwall {\it et al.},
  Comput.\ Phys.\ Commun.\  {\bf 176} (2007) 300.

\bibitem{powheg} 
  P.~Nason,
  JHEP {\bf 0411}, 040 (2004) [hep-ph/0409146];
  %
  S.~Frixione, P.~Nason and C.~Oleari,
  JHEP {\bf 0711}, 070 (2007) [arXiv:0709.2092 [hep-ph]];
%
  S.~Alioli, P.~Nason, C.~Oleari and E.~Re,
  JHEP {\bf 1006}, 043 (2010)
  [arXiv:1002.2581 [hep-ph]].
  
\bibitem{Alwall:2011uj} 
  J.~Alwall, M.~Herquet, F.~Maltoni, O.~Mattelaer and T.~Stelzer,
  JHEP {\bf 1106}, 128 (2011)
  [arXiv:1106.0522 [hep-ph]].

\bibitem{Arnold:2008rz} 
  K.~Arnold, M.~Bahr, G.~Bozzi, F.~Campanario, C.~Englert, T.~Figy, N.~Greiner and C.~Hackstein {\it et al.},
  Comput.\ Phys.\ Commun.\  {\bf 180}, 1661 (2009)
  [arXiv:0811.4559 [hep-ph]].

\bibitem{Arnold:2011wj} 
  K.~Arnold, J.~Bellm, G.~Bozzi, M.~Brieg, F.~Campanario, C.~Englert, B.~Feigl and J.~Frank {\it et al.},
  arXiv:1107.4038 [hep-ph].

\bibitem{Arnold:2012xn} 
  K.~Arnold, J.~Bellm, G.~Bozzi, F.~Campanario, C.~Englert, B.~Feigl, J.~Frank and T.~Figy {\it et al.},
  arXiv:1207.4975 [hep-ph].

\bibitem{pythia} 
  T.~Sjostrand, S.~Mrenna and P.~Z.~Skands,
  Comput.\ Phys.\ Commun.\  {\bf 178}, 852 (2008)
  [arXiv:0710.3820 [hep-ph]].
  

\bibitem{ATLAS:2013hta} 
  ATLAS Collaboration,
  arXiv:1307.7292 [hep-ex].
  
\bibitem{CMS:2013xfa} 
  CMS Collaboration,
  arXiv:1307.7135 [hep-ex].

\bibitem{Dawson:2013bba} 
  S.~Dawson, A.~V.~Gritsan, H.~Logan, J.~Qian, C.~Tully, R.~Van Kooten {\it et al.},
  arXiv:1310.8361 [hep-ex].

\bibitem{Behnke:2013xla} 
  T.~Behnke, J.~E.~Brau, B.~Foster, J.~Fuster, M.~Harrison, J.~M.~Paterson, M.~Peskin and M.~Stanitzki {\it et al.},
  arXiv:1306.6327 [physics.acc-ph].
  
\bibitem{Koratzinos:2013ncw} 
  M.~Koratzinos, A.~P.~Blondel, R.~Aleksan, O.~Brunner, A.~Butterworth, P.~Janot, E.~Jensen and J.~Osborne {\it et al.},
  arXiv:1305.6498 [physics.acc-ph].


\bibitem{Butterworth:2008iy} 
  J.~M.~Butterworth, A.~R.~Davison, M.~Rubin and G.~P.~Salam,
  Phys.\ Rev.\ Lett.\  {\bf 100}, 242001 (2008)
  [arXiv:0802.2470 [hep-ph]].


\bibitem{Shu:2013uua} 
  J.~Shu and Y.~Zhang,
  Phys.\ Rev.\ Lett.\  {\bf 111}, 091801 (2013) [arXiv:1304.0773 [hep-ph]].


\bibitem{Grzadkowski:1992sa} 
  B.~Grzadkowski and J.~F.~Gunion,
  Phys.\ Lett.\ B {\bf 294}, 361 (1992)
  [hep-ph/9206262].


  \bibitem{pdg} 
Particle Data Group, J. Beringer {\it et al.},
  Phys.\ Rev.\ D {\bf 86}, 010001 (2012).


\bibitem{Campbell:2011bn} 
  J.~M.~Campbell, R.~K.~Ellis and C.~Williams,
  JHEP {\bf 1107}, 018 (2011)
  [arXiv:1105.0020 [hep-ph]].


\bibitem{Kauer:2012hd} 
  N.~Kauer and G.~Passarino,
  JHEP {\bf 1208}, 116 (2012)
  [arXiv:1206.4803 [hep-ph]].

\bibitem{1307.4935} 
  F.~Caola and K.~Melnikov,
  Phys.\ Rev.\ D {\bf 88}, 054024 (2013) [arXiv:1307.4935 [hep-ph]].


\bibitem{Avery:2012um} 
  P.~Avery {\it et al.},
  Phys.\ Rev.\ D {\bf 87}, 055006 (2013)
  [arXiv:1210.0896 [hep-ph]].
  

%
\bibitem{cteq1}
  J.~Pumplin, D.~R.~Stump, J.~Huston, H.~L.~Lai, P.~M.~Nadolsky and W.~K.~Tung,
  JHEP {\bf 0207}, 012 (2002)
  [hep-ph/0201195].

\bibitem{cteq2}
P.~M.~Nadolsky {\it et al.},
Phys.\ Rev.\  D {\bf 78}, 013004 (2008).
  [arXiv:0802.0007 [hep-ph]].

\bibitem{thesis-nhan} 
N. Tran, Ph. D. thesis, Johns Hopkins University, CERN-THESIS-2011-127 (2011). 

\bibitem{Chatrchyan:2011ya} 
  CMS Collaboration, S.~Chatrchyan {\it et al.},
  Phys.\ Rev.\ D {\bf 84}, 112002 (2011)
  [arXiv:1110.2682 [hep-ex]].
  
\end{thebibliography}
\end{document}